\documentclass[11pt]{article}
\usepackage{latexsym,amsthm,amssymb,amsfonts,amsbsy,amsmath,multibox,calrsfs,stmaryrd}

\usepackage[dvips]{graphicx,epsfig}

\csname @addtoreset\endcsname{equation}{section}

\def\mso{\mathfrak{so}}
\def\ms{\mathfrak{s}}
\def\miso{\mathfrak{iso}}
\def\msu{\mathfrak{su}}
\def\msl{\mathfrak{sl}}

\def\msp{\mathfrak{sp}}
\def\mosp{\mathfrak{osp}}

\def\mho{\mathfrak{ho}}
\def\me{\mathfrak{e}}
\def\mg{\mathfrak{g}}
\def\ml{\mathfrak{l}}
\def\mpe{\mathfrak{p}}
\def\mm{\mathfrak{m}}
\def\mh{\mathfrak{h}}
\def\mD{\mathfrak{D}}
\def\mR{\mathfrak{R}}
\def\mV{\mathfrak{V}}
\def\mW{\mathfrak{W}}

\def\mC{\mathfrak{C}}
\def\mN{\mathfrak{N}}\def\mM{\mathfrak{M}}
\def\mI{\mathfrak{I}}

\def\Real{{\mathbb R}}
\def\Comp{{\mathbb C}}
\def\integ{{\mathbb Z}}
\def\1{1\hspace{-4pt}1}
\def\j1{\widetilde{1\hspace{-4pt}1}}

\def\Tr{{\mathrm{Tr}}}
\def\Str{{\mathrm{Str}}}

\def\adj{{\mathrm{ad}}}
\def\ac{{\mathrm{ac}}}
\def\Ad{{\mathrm{Ad}}}

\def\bec{\begin{center}}
\def\ec{\end{center}}
\def\a{\alpha} \def\ad{\dot{\a}} 
\def\unA{\underline A}\def\unB{\underline B}
\def\b{\beta} \def\bd{\dot{\b}} 
\def\c{\gamma} 
\def\C{\Gamma}
\def\d{\delta} 
\def\D{\Delta}
\def\e{\epsilon}

\def\k{\kappa}
\def\l{\lambda}
\def\L{\Lambda}
\def\m{\mu}
\def\n{\nu}
\def\r{\rho}
\def\s{\sigma}

\def\t{\tau}

\def\x{\xi}
\def\y{\eta}

\def\O{\Omega}
\def\o{\omega}
\def\pb{{\bar\pi}}

\def\sb{{\bar\s}}

\def\cF{{\cal F}}

\def\cA{{\cal A}}

\def\cR{{\cal R}}
\def\cS{{\cal S}}

\def\cA{{\cal A}}

\def\yb{{\bar y}}

\def\ts{\tilde{\s}}

\def\nn{\nonumber}
\newcommand{\eq}[1]{(\ref{#1})}

\newcommand{\w}[1]{\\[0.#1cm]}
\def\be{\begin{equation}}
\def\ee{\end{equation}}
\def\bea{\begin{eqnarray}}
\def\eea{\end{eqnarray}}
\def\ba{\begin{array}}
\def\ea{\end{array}}

\def\mx#1#2#3#4{\left#1\begin{array}{#2} #3 \end{array}\right#4}

\def\ft#1#2{{\textstyle{{\scriptstyle #1}
\over {\scriptstyle #2}}}}

\def\ket#1{|#1\rangle}
\def\bra#1{\langle#1|}
\def\scs#1{\section{\sc \large #1}}
\def\scss#1{\subsection{\sc #1}}
\def\scsss#1{\subsubsection{ \small #1}}

\def\ad{\dot\alpha}
\def\bd{\dot\beta}
\def\sb{\bar\sigma}

\def\cross{{}_\times\!\!\!\!{}^\times}
\def\ssum{\subsetplus}
\def\ssumr{\supsetplus}
\begin{document}

\begin{center}


{\Large\sc A Fiber Approach to Harmonic Analysis of \\[10pt] Unfolded Higher-Spin Field Equations}


\vspace{40pt}

C.~Iazeolla\\[15pt]
{\it{Dipartimento di Fisica, Universit\`{a} di Roma ``Tor
Vergata"}}\\
     \small{\it{INFN, Sezione di Roma ``Tor Vergata" }}\\
      \small{\it{Via della Ricerca Scientifica 1, 00133 Roma, Italy }}\\
     \small{\it{and\\
     Scuola Normale Superiore and INFN\\
Piazza dei Cavalieri 7, 56100 Pisa, Italy
\vspace{5pt}\footnote{Present address.} }}

\vspace{20pt}

P.~Sundell\\[15pt]
{\it\small Scuola Normale Superiore and INFN\\
Piazza dei Cavalieri 7, 56100 Pisa, Italy} \vspace{5pt}


\vspace{35pt} {\sc\large Abstract}\end{center}

In Vasiliev's unfolded formulation of higher-spin dynamics the
standard fields are embedded on-shell into covariantly constant
master fields valued in Lorentz-covariant slices of the
star-product algebra ${\cal A}$ of functions on the singleton
phase space. Correspondingly, the harmonic expansion is taken over
compact slices of ${\cal A}$ that are unitarizable in a rescaled
trace-norm rather than the standard Killing norm. Motivated by the
higher-derivative nature of the theory, we examine indecomposable
unitarizable Harish-Chandra modules consisting of standard
massless particles plus linearized runaway solutions. This
extension arises naturally in the above fiber approach upon
realizing compact-weight states as non-polynomial analytic
functions in ${\cal A}$.

\setcounter{page}{1}

\pagebreak

\tableofcontents


\scs{Introduction}\label{Sec:Intro}


\scss{Masslessness and compositeness}


Free massless spin-$s$ particles in four space-time dimensions
fall into infinite-dimensional representations $\mD(s+1;(s,s))$ of
the conformal group $\mso(4,2)$ with restrictions
$\mD(s+1;(s,s))|_{\miso(3,1)}\simeq \mD(m^2=0;|\l|=s)$ and
$\mD(s+1;(s,s))|_{\mso(3,2)}\simeq \mD(s+1;(s))$. Conformal
symmetry may be broken by Lorentz invariant interactions down to
$\miso(3,1)$, $\mso(3,2)$ or $\mso(4,1)$ depending on the
cosmological constant $\L$. One should distinguish between
ordinary and exotic gauge theories depending on whether the
infrared cutoff set by $\L$ is immaterial for local interactions
or not, and adopt a unified approach starting with non-vanishing
$\L$ \cite{Frons-dal}. If the (effective) field theory is ordinary
then $\L$ may as well be kept finite as long as one describes
processes at energy scales much larger than $\sqrt{|\L|}$, which
is where quantum field theory has been tested so far. At any such
local rate, massless particles in four space-time dimensions
exhibit compositeness: the lowest-energy representations
$\mD(s+1;(s))|_{\mso(3,2)}$ are squares of a more fundamental
representation namely Dirac's supersingleton $\mD(\ft12;(0))\oplus
\mD(1;(\ft12))$ \cite{Dirac:1963ta}, as summarized by the
remarkable Flato-Fronsdal formula \cite{Flato:1978qz}:
\bea [\mD(\ft12;(0))\oplus \mD(1;(\ft12))]^{\otimes 2}&=&
\bigoplus_{2s=0,1,2,\dots}\left[\mD(s+1;(s))\oplus
\mD(s+1+\d_{s0};(s))\right]\ .\label{FF1}\eea
Strictly speaking, the generators of $\mso(4,2)/\mso(3,2)$, and
hence $\mso(4,1)$, act on the left-hand side in a non-tensorial
split (see comment in Appendix \ref{Sec:4D}) and
$\mD(m^2=0;|\l|=s)|_{\miso(3,1)}$ is recovered at scales much
larger than $|\L|$ \cite{Angelopoulos:1980wg}, as mentioned above.

Compositeness is directly related to the higher-spin Lie-algebra
extension $\mho(D+1;\Comp)$ of $\mso(D+1;\Comp)$ (see, for
example, \cite{Konstein:1988yg, mishasalg, Engquist:2005yt} and
references therein) that: i) acts transitively on the
$D$-dimensional scalar singleton $\mD(\e_0;(0))$,
$\e_0=\ft12(D-3)$, which thus functions as its fundamental
lowest-energy representation; and ii) arranges the $D$-dimensional
Flato-Fronsdal tower $[\mD(\e_0;(0)))]^{\otimes
2}=\bigoplus_{s=0}^\infty \mD(s+2\e_0;(s))$ into an irrep which
one may refer to as the massless higher-spin multiplet\footnote{By
definition, a massless lowest-weight space of $\mso(D+1;\Comp)$
results from factoring out a non-trivial ideal, corresponding to
the longitudinal modes of a gauge field, from a generalized Verma
module (see Appendix \ref{App:B}). For $D>4$ one needs to
distinguish between conformal, composite and partial masslessness
\cite{Angelop,BMV,FronsdalFerrara,Deser} (see also
\cite{Skvortsov:2006at,Francia:2008hd} and references therein). In
this paper we focus on composite masslessness.}. As we shall show
below (see Section \ref{Sec:Adj}), the adjoint representation
itself is also composite and isomorphic to the tensor product of a
singleton (lowest-energy representation) and its negative energy
counterpart, an anti-singleton (highest-energy representation):
\bea \mho(3,2)&\simeq& [\mD^+(\ft12;(0))\otimes
\mD^-(-\ft12;(0))]\oplus [\mD^-(-\ft12;(0))\otimes
\mD^+(\ft12;(0))]\ .\eea

There appears to be a clear separation between interactions that
break and preserve higher-spin gauge symmetry. In particular, the
latter are exotic in a sense to be discussed next.

\scss{Interlude: Canonical vs unfolded higher-spin equations}

In the standard approach to field theory, based on kinetic terms
built using the metric, perturbative higher-spin gauge
interactions necessarily contain higher derivatives. Removing
unnecessary derivatives by perturbative field redefinitions leads
to standard vertices with minimal required numbers of derivatives,
that one may refer to as the canonical vertices. The number of
derivatives in the canonical spin $s_1-s_2-s_3$ vertices grows
linearly with $s_1+s_2+s_3$ \cite{pos, Metsaev, FV1,
Boulanger:2006gr, Boulanger:2008tg}. For $\L=0$ these vertices are
ordinary and hence weakly coupled at sufficiently low energies. In
particular, in the case of $s_1=2$ and $s_2=s_3=s>2$ there are
minimal non-abelian vertices with $2s-2$ derivatives,
Chern-Simons-like vertices (albeit parity preserving) with $2s$
derivatives, and Born-Infeld-like vertices with $2s+2$
derivatives.  However, the standard gravitational cubic spin
$2-s-s$ coupling (with two derivatives) is plagued by a classical
anomaly, and hence the non-linear completion based on non-abelian
higher-spin symmetries seems problematic in an expansion around
flat spacetime (although some of the above-mentioned cubic
vertices do pass certain consistency tests at the quartic level
\cite{Boulanger:2005br}).

On the other hand, if $\L\neq 0$ there exists an exotic cancelation
mechanism found by Fradkin and Vasiliev \cite{FV1}, whereby the
``non-gravitational'' minimal spin $2-s-s$ vertex from flat space,
which has $2s-2$ derivatives, is completed with lower derivatives
all the way down to the standard gravitational two-derivative
coupling, with dimensionful parameters fixed in units of $\L$.
Moreover, there exists a fully non-linear completion in the form
of Vasiliev's unfolded master-field equations (\cite{Vasiliev:en,
Vasiliev:2003ev}, see also \cite{Vasiliev:1999ba, Bekaert:2005vh}
for reviews and more references) which are manifestly higher-spin
gauge invariant and locally homotopy invariant, \emph{i.e.}
referring to the base manifold only via differential forms and the
exterior derivative $d$, as well as manifestly background
independent in the sense that their salient features do not rely
on the expansion around any specific solution nor on the reference
to a (non-degenerate) metric. The equations actually assume a
remarkably simple form: they describe a covariantly constant
zero-form and a flat connection taking their values in a deformed
``fiber''. The crux of the matter lies, of course, in what
constitutes a good set of observables.

In the sub-sector where the Weyl zero-form is
``weak''\footnote{This notion can be formulated in more precise
terms using a set of local zero-form observables which are given
by traces of algebraic powers of the zero-form
\cite{Sezgin:2005pv, Vasiliev:2005zu}.}, the master-field
equations imply generally covariant and locally Lorentz invariant,
albeit non-canonical, standard field equations for perturbatively
defined dynamical components of the master fields, that one may
refer to as the microscopic fields. In particular, there appear a
microscopic vielbein and spin-connection defining what one may
refer to as the Vasiliev frame, to be related to the canonical
Einstein frame by a (highly non-local) field redefinition. The
microscopic standard field equations (with box-like kinetic terms)
are obtained by eliminating auxiliary fields after first having
expanded the master-field equations in weak curvatures and then
around ``large'' Vasiliev-frame metric while treating the
remaining microscopic fields as ``small''. The result contains two
parameters: i) a dimensionless AdS-Planck constant $g^2\equiv (\l
\ell_p)^{D-2}$ that counts the order in the perturbative
weak-field expansion, where $\ell_p$ enters via the normalization
of the effective standard action and we are working with
dimensionless physical fields; and ii) a massive parameter $\l$
that simultaneously iia) sets the infrared cutoff via $\L\sim
\l^2$ and critical masses $M^2 \sim\l^2$ for the dynamical fields;
and iib) dresses the derivatives in the interaction vertices thus
enabling the Fradkin-Vasiliev (FV) mechanism.

At fixed order in $g$, that one may always take to be a small
parameter, the non-canonical microscopic interactions are given by
Born-Infeld-like series expansions in $\l^{-2}\nabla\nabla$, that
one may refer to as tails (see \cite{Kristiansson:2003xx} for a
computation of the quadratic scalar-field contributions to the
microscopic stress tensor). On general grounds, these microscopic
tails should be related to the canonical vertices via non-local,
potentially divergent, field redefinitions. Thus one has the
following scheme:
\bea \ba{c}\mbox{Unfolded}\\[-3pt]\mbox{master-field}\\[-3pt]
\mbox{equations}\ea\quad& \stackrel{\tiny\ba{c}{\rm weak}\\{\rm
fields}\ea}{\leftrightarrows} &\quad \ba{c}\mbox{Standard-exotic}\\
[-3pt]\mbox{microscopic}\\[-3pt]\mbox{field equations}\ea\qquad \stackrel{\tiny\ba{c}{\rm non-local}\\
{\rm field\, redef.}\ea}{\rightleftarrows}\qquad \ba{c}\mbox{Standard-exotic}\\
[-3pt]\mbox{canonical}\\[-3pt]\mbox{field equations}\ea\eea

We stress that what makes higher-spin theory exotic is the dual
purpose served by $\l$ within the FV mechanism whereby positive
and negative powers of $\l$ appear in mass terms and
vertices\footnote{This brings to mind the transition from the
standard perturbative regime of closed string theory in 10D flat
space where physical, string and Planck lengths $\ell$, $\ell_s$
and $\ell_p=g_s^{\ft14}\ell_s$ obey $\ell_p\ll\ell_s\ll\ell$, to a
high-energy regime where $\ell_p\ll \ell\ll \ell_s$, and thus
$\ell_s$ switches role from a stringy UV cutoff to a stringy IR
cutoff (\emph{e.g.} in one-loop amplitudes where the fundamental
domain can be chosen such that $\ell/\ell_s,{\rm Im}\t\in[0,1]$).
Sending $\ell_s$ to infinity, which in string-field language means
taking $\langle e^{\phi_{\rm dil}}\rangle =g_s\rightarrow 0$ keeping
$\ell_p$ fixed, may lead to an unbroken phase with stringy IR
cutoff set by $\l$ \cite{Fradkin:1991mw}.}, respectively. Thus, at
each order of the canonical expansion scheme, the local bulk
interactions -- and in particular the standard minimal
gravitational two-derivative couplings -- are dominated by
strongly coupled ``top vertices'' going like finite positive
powers of (energy scale)/(IR cutoff). On the other hand, in the
microscopic expansion scheme each order is given by a potentially
divergent Born-Infeld tail, suggesting that classical solutions as
well as amplitudes should be evaluated directly within the
master-field formalism, which offers transparent methods based on
requiring associativity of the operator algebra for setting up and
assessing regularized calculational schemes.

We also emphasize that the metric-dependent standard symplectic
structure differs from the background-independent unfolded
symplectic structure, which treats all the derivatives of the
physical fields as \emph{a priori} independent variables
\cite{action}. Besides providing a more systematic handling of the
initial data formulation in the presence of higher derivatives,
the unfolded structure leads to a Gaussian path-integral measure
that suppresses higher-derivative fluctuations.

The above-mentioned issues of regularization and off-shell
formulation go beyond the present scope of this paper, although
they serve as key motivations for our work. Similar considerations
are also made in reference \cite{Boulanger:2008tg}.

\scss{Phase-space quantization}

The higher-spin master-field formalism of \cite{Vasiliev:sa,
Vasiliev:1989yr} has been conjectured in \cite{Engquist:2005yt} to
be directly connected to the geometric Cattaneo-Felder formulation
\cite{Cattaneo:1999fm} of the phase-space quantization
\cite{Bayen:1977ha} of singletons in terms of a two-dimensional
first-order parent action, whose BV quantization leads to an
embedding of the algebra ${\cal A}$ of functions on the singleton
phase space into an infinite-dimensional quantum version
$\widehat\C$ governed by a ``topological'' BRST operator $\widehat
q$. One is led to attempt to formulate a complete theory with: i)
a first-order total Lagrangian for locally defined master fields
with ``kinetic'' terms built from $d+\widehat q$; ii) globally
defined observables that are given by integrals over (cycles in)
the space-time base manifold times (cycles in) the quantum phase
space; iii) two coupling constants $g$ and $\l^{-1}$ corresponding
to the perturbative expansions of the base manifold and of the
internal sigma model\footnote{The flat and infinitely curved
limits $\l\rightarrow 0$ and $\l\rightarrow\infty$ should thus
correspond to the strongly and weakly coupled limits of the sigma
model, respectively.}, respectively; and iv) a limit in which the
master fields are classical differential forms on the base
manifold taking their values in $\widehat \C$ (\emph{c.f.}
classical string-field theory).

Vasiliev's fiber algebra, that we shall denote by $\widehat{\cal
A}$, arises as a truncation of $\widehat\C$ \cite{Engquist:2005yt}
that is formally associative, and equipped with trace operations
given by integrations over suitable cycles in $\widehat{\cal A}$
\cite{Engquist:2005yt, Sezgin:2005pv, Vasiliev:1986qx}. Thus
Vasiliev's full master fields take their values in $\widehat{\cal
A}$, and the simply-looking full unfolded master-field equations
makes use only of the ordinary associative $\star$-product, or
2-product, on $\widehat {\cal A}$. This algebra contains ${\cal
A}$ as a subalgebra. Correspondingly, there are reduced master
fields taking their values in ${\cal A}$ and obeying highly
non-linear reduced unfolded master-field equations making use of
$n$-products with $n=2,3,\dots$ given perturbatively in the
weak-field expansion.

These equations admit a free limit on maximally symmetric spaces,
such that the properties of free standard higher-spin fields are
``dual'' to properties of ${\cal A}$ viewed as an
$\mho(D+1;\Comp)$-module. As such ${\cal A}$ is isomorphic to the
enveloping algebra ${\cal U}[\mso(D+1;\Comp)]$ modulo the ideal
consisting of all elements that are trivial in the scalar
singleton module. The ideal is generated by the covariant scalar
singleton equations of motion\footnote{An affine extension of the
singleton equations of motion, related to discretized tensionless
$p$-branes in $AdS$, has been examined in \cite{Engquist:2007pr}
at the level of lowest-weight modules.} \cite{Angelop} (see
Section \ref{sec:FT} for notation):
\bea \frac{1}{2} M_{\{A}{}^C \star M_{B\}C} \ \approx \ 0 \ ,
\qquad M_{[AB}\star M_{CD]} \ \approx \ 0 \ . \label{singeom}\eea
In \cite{Angelop} it was found that the first condition, when
imposed on lowest-weight spaces, selects the singletons, further
restricted to be scalars (or spinors, in $D=4$) by the second
condition. Moreover, somehow in the spirit of the deformation
quantization program \cite{Bayen:1977ha}, in the following we
shall exploit the fact that the standard one-particle states can
be mapped to non-polynomial Weyl-ordered phase-space functions
corresponding to similar special functions in the
enveloping-algebra realization of ${\cal A}$. For example, in
$D=4$ we have
\bea \ba{c}\mbox{Free one-particle}\\\mbox{lowest-weight state}\\[5pt]
\ket{1;(0)}=\ket{\ft12;(0)}\otimes
\ket{\ft12;(0)}\ea\quad&\leftrightarrows&\quad
\ba{c}\mbox{Phase-space}\\\mbox{function}\\[5pt]
4[\exp(-2\bar a^i a_i)]_{\rm Weyl}\ea\qquad
\leftrightarrows\qquad \ba{c}\mbox{Enveloping-algebra}\\
\mbox{element}\\[5pt] 4\exp(-4E) \ .\ea\label{unfoldedparticle}\eea
The incorporation of lowest-weight and highest-weight spaces into
${\cal A}$ is a manifestation of compositeness (and it applies
equally well to finite-dimensional lowest-and-highest-weight
spaces). However, the enveloping-algebra approach, which does not
refer \emph{a priori} to lowest-weight states, is more general,
and it indeed gives rise to additional representations, some of
which do not contain any lowest nor highest-weight state. We
stress that, while these representations arise in a dual fashion
on the standard free field-theory side, we are mainly interested
in the enveloping-algebra side which seems more relevant for
computing the exotic higher-spin interactions, as discussed above.
We shall explore the possibility of using such a fiber approach to
naturally incorporate the massless modules $\mD(s+1;(s))$ and
linearized runaway solutions of higher-spin gauge theories in a
bigger, indecomposable module.

\scss{Runaway solutions and extended
compositeness}\label{Sec:Runaway}

In an ordinary field theory one may consider three distinct
classes of solutions: particles, runaway solutions and solitons.
For a free massless scalar field in flat spacetime one has: i)
positive and negative energy mass shells $\mD^\pm(0;0)$ with
continuous spectrum for the translation generator $P_a=(P_0,P_r)$,
with $a=0,1,...,D-1$; ii) the $\miso(D-1,1)$ orbit $\mW(\ell)$ of
the static runaway spin-$\ell$ solution $r^{\ell}Y_\ell(\widehat
n)$, $\ell=0,1,\dots$, where $r$ is the radial coordinate and
$Y_\ell$ are spherical harmonics obeying
$(\nabla^2_{|S^{D-2}}+\ell(\ell+2\e_0))Y_\ell=0$ with $\e_0=
\ft12(D-3)$; and iii) the $\miso(D-1,1)$ orbit
$\widetilde\mW(\ell)$ of the static singular spin-$\ell$ solution
$r^{-\ell-2\e_0}Y_\ell(\widehat n)$. The mass shells can be given
in a basis $\{\ket{e;(\ell)}\}$, with $e\in \Real_+$ and
$\ell=0,1,2,...$, where each energy level forms a
finite-dimensional irrep of the $\mso(2)$ generated by $P_0$ and
the $\mso(D-1)$ generated by the spatial rotations $M_{rs}$, which
one refers to as an $(\mso(2)\oplus \mso(D-1))$-type. In this
basis the plane-waves with energy $e\neq 0$
are given by $\ket{(e,e\hat p_r)}=\sum_\ell \frac{(ie)^\ell}{\ell!}
c_\ell\ket{e;(\ell)}$ for coefficients $c_\ell$ that are non-vanishing for all $\ell$ and with the $e=0$ limit $\ket{(0,0)}=\ket{0;(0)}$, while the static spin-$\ell$ runaways are given by $\ket{0;(\ell)}$. The set of runaway modules $\{\mW(\ell)\}$,
which are simply the spaces of traceless polynomials of degree
$\ell$ in the cartesian coordinates $x^a$, forms an indecomposable
chain (see Appendix \ref{App:B}) in which $\mW(\ell)\subset\mW(\ell+1)$, and
$\{\widetilde\mW(\ell)\}$ exhibits the ``dual'' structure
$\widetilde\mW(\ell)\supset \widetilde\mW(\ell+1)$. Massless
static fields $\phi$ have conformal weight $\e_0$ in the $D-1$
spatial dimensions, and consequently the runaway and singular spin-$\ell$
solutions are related by
\bea \ba{c}\mbox{runaway}\\ \phi_\ell=r^\ell Y_\ell(\hat n)\ea
\quad&\stackrel{r\leftrightarrow \ft1{r}}{\rightleftarrows}& \quad
\ba{c}\mbox{singular}\\
\widetilde\phi_\ell=r^{-\ell-2\e_0}Y_\ell(\hat n)\ea\
.\label{r1/r}\eea
One also notes that, whereas $\mD(0;(0))\simeq \mD'(0;(0))$ with
$\mD'(e;(0))$ being the compact weight-space module consisting of
states $\ket{\psi}\in\mD(0;(0))$ of the form $\ket{\psi}=
\int_{e>0} de \,\psi(e)\,\ket{e;(\ell)}$, the action of $P_0$
cannot be diagonalized in $\mW(\ell)$ and $\widetilde\mW(\ell)$
which can therefore not be decomposed in terms of
$(\mso(2)\oplus\mso(D-1))$-types.

The effects of $\L$ and/or mass terms show up in the
large-$r$/infrared limit, while they do not alter the
small-$r$/ultraviolet limit. For $\L<0$ and unitary masses, the
free-field fluctuations that are normalizable in the Killing norm
$\parallel\cdot\parallel_{\rm can}$, induced from the standard
canonical action, form a lowest-weight representation $\mD$
containing the one-particle states \cite{Breitenlohner:1982jf}.
Among the linearized fields with divergent
$\parallel\cdot\parallel_{\rm can}$ are the runaway and singular
solutions. These may have non-linear completions depending on the
details of the interactions. In ordinary field theories some of
the singular solutions may become solitonic objects with finite
energy and localizable energy density. In exotic field theories,
on the other hand, one may expect that the perturbative
descriptions of localized soliton (or particle) solutions involve
derivative expansions that are coupled more strongly than those of
corresponding runaway solutions. One might entertain the idea that
the Born-Infeld tails of higher-spin gauge theory lead to a
restoration of the duality \eq{r1/r} at the full level. Another
noteworthy feature of higher-spin gauge theory is that all even
one-particle and runaway $(\mso(2)\oplus\mso(D-1))$-types, of all
fields, are generated by the $\mho(D-1,2)$ action from the static
runaway mode $\phi_{0;(0)}$ of the scalar field obeying
$(\nabla^2+4\e_0)\phi=0$; for example in $AdS_4$ with metric
$ds^2=\frac{1}{\cos^2\xi}(-dt^2+d\xi^2+\sin^2\xi d\O^2_{S^2})$
one has
\bea \ba{c}\mbox{Static, $\ell=0$}\\\mbox{runaway field}\\
\phi_{0;(0)}(\xi)\ =\ \frac{\xi}{\tan\xi}\ea\quad&\stackrel{\rm
unfolding}{\leftrightarrows}&\quad \ba{c} \mbox{Static,
$\ell=0$}\\\mbox{analytic ${\cal A}$ function}\\ \frac{\sinh 4E}{
4E}\ea\ .\label{stgs}\eea

The above reasoning and the fashion in which the degeneracies at
the infra-red ``apices'' of the flat-space massless mass-shells
are resolved by the deformation by $\Lambda<0$, whereby particles
and runaway solutions fall into distinct
$\mso(2)\oplus\mso(D-1)$-types, suggest a unification in
higher-spin gauge theory of the one-particle states in $\mD$ and
the runaway solutions ${\cal W}$ into a single module ${\cal M}$
that: i) has the indecomposable structure ${\cal M}=\mD\ssum {\cal
W}$ (runaway solutions are allowed to emit particles into the
bulk), where the definition of the semi-direct sum symbol $\ssum$
is recalled in Appendix \ref{App:B}; ii) is factorizable,
\emph{i.e.} realized via special functions in ${\cal A}$; iii) is
generated by the $\mso(D-1,2)$ action starting from the
$D$-dimensional generalization of the 4D static ground state
\eq{stgs}; iv) is unitarizable (so that the duality \eq{r1/r} may
make sense); and v) has an associative structure that lifts to
$\widehat{\cal A}$ (existence of perturbatively defined full
master fields).

In this paper we shall focus on (i)--(iv), leaving (v) for a
future publication \cite{companion}.


\scs{Enveloping-Algebra Realization of Higher-Spin Master
Fields}\label{sec:FT}


The unfolded formulation of minimal bosonic higher-spin gauge
theory in $D\geqslant 4$ space-time dimensions makes use of two
(reduced) master fields: a one-form $A$ and a zero-form $\Phi$
taking their values, respectively, in the adjoint and
twisted-adjoint representations $\mho_0$ and ${\cal T}_+$ of the
minimal higher-spin extension of $\mso(D+1;\Comp)$. In this
section, we describe how these representations arise in the
associative algebra ${\cal A}$ obtained by factoring out the
singleton annihilator ${\cal I}[V]$ from the universal enveloping
algebra ${\cal U}[\mso(D+1;\Comp)]$. The ideal consists of
monomials containing at least two contracted or three
anti-symmetrized $(D+1)$-dimensional vector indices, so that
${\cal A}$ consists of rectangular $\mso(D+1;\Comp)$ tensors of
height two. The adjoint and twisted-adjoint $\mho_0$ actions are
induced, respectively, via the commutator and a twisted version
that represents the Lorentz translations $P_a$ by the
anti-commutator $\{P_a,\cdot\}$. As a result, $\mho_0$ and ${\cal
T}_+$ decompose under $\mso(D+1;\Comp)$ into irreducible levels
containing, respectively, the finite sets of gauge fields and
infinite sets of Weyl-tensor/matter-field zero-forms required
within the unfolded formalism for describing one spin-$s$ degree
of freedom (where $s$ is related to the level).

In the leading order of the weak-field expansion, the zero-forms
obey a covariant constancy condition and source the one-forms via
a Chevalley-Eilenberg cocycle of the form $F_{(1)}=e^a \wedge e^b
J_{(1)ab}$, where $F_{(1)}$ is the linearized adjoint curvature
two-form, $e^a$ is the vielbein, and $J_{(1)ab}$ is linear in
$\Phi$. Thus, the local propagating degrees of freedom are carried
by $\Phi$ while $A$ may carry zero-modes (that are solutions with
$\Phi=0$ that cannot be gauged away) \cite{Iazeolla:2007wt}. As we
shall see, the cocycle has a weight-space analog in the form of a
short exact sequence connecting the adjoint levels to the
corresponding composite massless lowest-weight spaces.

Next, we define a nontrivial trace operation $\Tr_\cA$ on the
algebra $\cA$ that will endow the latter with a nondegenerate
inner product, and related \emph{reflector} states that possess a
series of useful properties. These tools will be useful for
establishing the state/operator correspondence mentioned in the
Introduction. For example, as we shall examine later, the
reflectors enable presentations of the master fields as elements
in left bimodules. The definition of $\Tr_\cA$ generalizes the
supertrace operation first given in \cite{Vasiliev:1986qx} for the
case of the 4D oscillator realization of higher-spin
superalgebras, as shown in Appendix \ref{Sec:4D}.

Finally, we shall define harmonic expansions of $\Phi$ in
maximally symmetric backgrounds as maps between the
Lorentz-covariant and maximally compact slicings of the
infinite-dimensional twisted-adjoint modules. This formalism will
then be examined in greater detail in the cases of de Sitter and
anti-de Sitter geometry.


\scss{Associative quotient algebra}\label{assocquot}


The universal enveloping algebra ${\cal U}[\mg]$ of a Lie algebra
$\mg$ is the associative algebra consisting of the unity $\1$ and
arbitrary polynomials in the generators $M_\a$ of $\mg$ modulo the
commutation rules, \emph{viz.}\footnote{The symmetrized monomials
are linearly independent according to the Poincar\'e-Birkhoff-Witt
theorem.} ${\cal U}[\mg]=\bigoplus_{n=0}^\infty {\cal U}_n\
,\qquad {\cal U}_n\ \ni\ X_n=x^{\a(n)}M_{\a(n)}$, where
$x^{\a(n)}\in \Comp$ and the basis consists of the symmetrized
monomials
\bea M_{\a(n)} &=&M_{\a_1}\cdots M_{\a_n}\ \equiv\
\left\{\ba{ll}{1\over n!}\sum_{\pi\in{\cal S}_n}
M_{\a_{\pi(1)}}\star\cdots \star
M_{\a_{\pi(n)}}&\mbox{for $n=1,2,\dots$.}\ ,\\[5pt] \1 &\mbox{for $n=0$}\ea
\right.\ ,\label{calU}\eea
where $\star$ and juxtaposition, respectively, denote the
non-commutative product of ${\cal U}[\mg]$ and the commutative
product given by symmetrization of the $M_\a$, so that $X_m\star
X_n=X_m X_n+\sum_{p\leqslant m+n-1}X'_p$. The algebra ${\cal
U}[\mg]$ has the involutive anti-automorphism\footnote{A map
$\phi$ from an associative algebra to itself is an automorphism if
$\phi(X\star Y)=\phi(X)\star \phi(Y)$ and an anti-automorphism if
$\phi(X\star Y)=\phi(Y)\star \phi(X)$. An (anti)-automorphism is
said to be involutive if $\phi^2=1$.} given by the
``transposition''
\bea \tau(M_{\a})&=&-M_{\a}\ ,\qquad \tau(X_n)\ =\ (-1)^n X_n\
.\label{taumap}\eea
The adjoint and anti-commutator actions of ${\cal U}[\mg]$ on
itself defined by $\adj_X(Y)=[X,Y]_\star$ and
$\ac_X(Y)=\{X,Y\}_\star$ obey the closure relation
$[\ac_X,\ac_Y]_\star=\adj_{[X,Y]_\star}$.

The Lie algebra $\mg=\mso(D+1;\Comp)$ has generators $M_{AB}$
obeying
\bea [M_{AB},M_{CD}]_\star&=& M_{AB}\star M_{CD}-M_{CD}\star
M_{AB}\ =\ 4i\eta_{[C|[B}M_{A]|D]}\ ,\label{soD+1}\eea
where $A=(\sharp,a)$, $a=(\sharp',r)$, $r=1,\dots,D-1$, and
$\eta_{AB}={\rm diag}(-\sigma,\eta_{ab})$, $\eta_{ab}={\rm
diag}(-\sigma',\d_{rs})$ with $\sigma,\sigma'=\pm 1$. We write
$\sharp=0'$ and $\sharp=D+1$ when $\s=+1$ and $\s=-1$,
respectively, and $\sharp'=0$ and $\sharp'=D$ when $\s'=+1$ and
$\s'=-1$, respectively, and refer to $\s'=1$ and $\s'=-1$,
respectively, as Lorentzian and Euclidean signatures. Splitting
$M_{AB}\rightarrow (M_{ab},P_a)$, $P_a=M_{\sharp a}$, the
commutation rules read
\bea [M_{ab},M_{cd}]_\star&=&4i\eta_{[c|[b}M_{a]|d]}\ ,\qquad
[M_{ab},P_c]_\star\ =\ 2i\eta_{c[b}P_{a]}\ ,\qquad
[P_a,P_b]_\star\ =\ i\s M_{ab}\ .\eea
We let $\mm\simeq\mso(D;\Comp)$ and $\ms\simeq\mso(D-1;\Comp)$
denote the subalgebras generated by $M_{ab}$ and $M_{rs}$,
respectively, and refer to them as the Lorentz and spin algebras.
The energy operator $E$ is defined by $[E,\ms]=0$ and we write
$\me=\Comp\otimes E$. The real forms $\mso(-\s,-\s',D-1)$,
$\mso(-\s',D-1)$ and $\mso(D-1)$ of $\mg$, $\mm$ and $\ms$,
respectively, are defined by
$(M^\Real_{AB})^\dagger=M^\Real_{AB}$. The corresponding maximally
symmetric spaces of radius $\l^{-1}$ are
\bea S(\sigma,\s')&=& {SO(-\s,-\s',D-1)\over SO(-\s',D-1)}\ =\
\left\{X\in \Real^{D+1}\ :\quad \l^2 X^A X^B \eta_{AB}\ =\ -\sigma
\ \right\}\ ,\label{maxsymm}\eea
\emph{i.e.}, $S(+,+)=AdS_D$, $S(+,-)=H_D$, $S(-,+)=dS_D$ and
$S(-,-)=S^D$. One splits $\mso(-\s,-\s',D-1)=\mh\ssum\ml$ where
$\mh\simeq \mso(p)\oplus \mso(D+1-p)$ is the maximal compact
subalgebra and $[\mh,\ml]=\ml$ and $[\ml,\ml]=\mh$.

The monomials $X_n$ with $n\geqslant 2$ decompose under
$\adj_{\mg}$ into $\eta_{AB}$-traceless and Young-projected
tensors. The trace parts and shapes with more than two rows form
the ideal \cite{Engquist:2007pr}
\bea {\cal I}[V]&=& \left\{ X\ =\ V\star X'\ \mbox{for $X'\in{\cal
U}[\mg]$}\right\}\ ,\qquad V=\l^{AB}V_{AB}+\l^{ABCD}V_{ABCD}\
,\label{idealV}\eea
where $\l^{AB},\l^{ABCD}\in\Comp$ and
\bea V_{AB}&\equiv & \frac12 M_{(A}{}^C M_{B)C}-{1\over
2(D+1)}\eta_{AB} M^{CD} M_{CD}\label{VAB}\ ,\qquad V_{ABCD}\
\equiv\ M_{[AB} M_{CD]}\ ,\label{VABCD}\eea
absorbing the trace parts and more-than-two-row shapes,
respectively. Defining ${\cal U}'[\mg]={\cal U}[\mg]\setminus\1$,
one has the chain ${\cal U}[\mg]\supset {\cal U}'[\mg]\supset{\cal
I}[V]$ of proper ideals. Factoring out ${\cal I}[V]$ yields the
infinite-dimensional unital associative algebra\footnote{Let
$V=\l^{A_1,\dots,A_n}V_{A_1,\dots,A_n}$ with
$\l^{A_1,\dots,A_n}\in \Comp$ and $V_{A_1,\dots,A_n}$ given by an
$\mg$-irreducible, \emph{i.e.} Young projected and traceless,
monomial in $M_{AB}$ and $\eta_{AB}$, thus obeying
$[M_{BC},V_{A_1,\dots,A_n}]_\star=2i\sum_{i=1}^n
\eta_{A_i[C}V_{\dots,A_{i-1},|B],A_{i+1},\dots,A_n}$. It follows
that if $X\in {\cal U}[\mg]$ then $X\star V=V\star X'$ for some
$X'\in {\cal U}[\mg]$. Thus ${\cal I}[V]\equiv V\star {\cal
U}[\mg]={\cal U}[\mg]\star V$ is a two-sided ideal and $[X\star
V\star X']=0$ in ${\cal U}[\mg]/{\cal I}[V]$, that is $X\star
V\star X'\approx 0$ for all $X,X'\in{\cal U}$. The above holds
true also if $V$ is $\mg$-reducible.}
\bea {\cal A}&\equiv & {{\cal U}[\mg]\over {\cal I}[V]}\ =\
\bigoplus_{n=0}^\infty {\cal A}_n\ ,\qquad {\cal A}_n\ \ni \ X_n\
\approx\ x^{A(n),B(n)} M_{A(n),B(n)}\
,\label{calA}\label{calA1}\eea
where $X\approx X'$ means $X-X'\in{\cal I}[V]$. A basis for ${\cal
A}$ consists of the traceless type-$(n,n)$ tensors\footnote{We
refer to tensors with the symmetry of the Young diagram with $n_i$
cells in the $i$th row ($i=1,\dots,\nu$) as
type-$(n_1,\dots,n_\nu)$ tensors, and work with normalized and
mostly symmetric projectors
$$\mathbf P_{n_1,n_2,\dots,n_\nu}={1\over
\prod_{\mbox{cells}}(\mbox{hook-lengths})} \prod_{\mbox{rows $i$}}
\mathbf S_i \prod_{\mbox{columns $j$}}\mathbf A_j\ ,$$ where
$\mathbf S_i$ and $\mathbf A_j$ are row symmetrizers and column
anti-symmetrizers, respectively. Thus a type-$(n_1,\dots,n_\nu)$
tensor has $\nu$ groups of symmetric indices $A^i(n_i)=A^i_1\dots
A^i_{n_i}$ subject to the over-symmetrization rule
$M_{\cdots,(A^i_1\dots A^i_{n_i},A^{i+1}_1)A^{i+1}_2\dots
A^{i+1}_{n_{i+1}},\cdots}=0$, $i=1,\dots,\nu-1$. Young-projected
index blocks are enclosed by $\langle\cdots\rangle$ and by
$\{\cdots\}$ if they are also traceless. With the exception of
\eq{calU}, indices distinguished by subindexation are always
assumed to be symmetrized; \emph{e.g.} $M_{A_1}{}^{B_1} \cdots
M_{A_n}{}^{B_n }$$=$$M_{(A_1}{}^{(B_1}\cdots
M_{A_n)}{}^{B_n)}$$=$$M_{(A_1}{}^{(B_1}\star \cdots\star
M_{A_n)}{}^{B_n)}$, or equivalently, $M_{A_1B_1} \cdots M_{A_nB_n
}$$=$$ M_{\langle A_1B_1}\cdots M_{A_nB_n\rangle }$$=$$M_{\langle
A_1B_1}\star \cdots\star M_{A_nB_n\rangle }$.}
\bea M_{A(n),B(n)}&=&M_{\{ A_1B_1} \cdots M_{A_nB_n\}}\ =\ M_{\{
A_1B_1}\star \cdots\star M_{A_nB_n\}}\nn\\[5pt] &=&\sum_{k=0}^{[n/2]}\k_{n;k} \eta_{\langle
A_1A_2}\eta_{B_1B_2}\cdots
\eta_{A_{2k-1}A_{2k}}\eta_{B_{2k-1}B_{2k}}M_{ A_{2k+1}B_{2k+1}}
\cdots M_{A_{n}B_{n}\rangle}\ ,\qquad\qquad\label{TAnBn}\eea
where $\k_{n;k}$ are fixed by $\eta^{CD} M_{A(n),B(n-2)CD}=0$ and
$\k_{n;0}=1$, and the trace parts contain monomials of rectangular
shapes. For example, if $n=2$ one has
\bea M_{A(2),B(2)}&=& M_{A_1 B_1}\star M_{A_2 B_2}-{2\over
D(D+1)}(\eta_{A_1 A_2}\eta_{B_1 B_2}-\eta_{A_1 B_1}\eta_{A_2 B_2})
C_2[{\cal S}]\ ,\eea
where $C_{2}[{\cal S}]$ is the quadratic Casimir operator
$\frac{1}{2}M^{AB}\star M_{AB}$ given by \eq{C2e0} in the left
$\cA$-module $\cal S$ that we shall define below. For our
purposes, we need
\bea\adj_{M_{AB}}(M_{C(n),D(n)})&=&2in\left(
\eta_{\{C_1|[B}M_{A]|C(n-1),D(n)\}}+\eta_{\{D_1|[B|}M_{|C(n)|,|A]|D(n-1)\}}\right)\
,\eea
and (see also \eq{TPa} and \eq{MabT})
\bea \ac_{M_{AB}}(M_{C(n),D(n)}) &=& 2\D_n
M_{[A|\{C(n),|B]D(n)\}}+ 2\l_n
\eta_{\{C_1|[A}\eta_{B]|D_1}M_{C(n-1),D(n-1)\}}\
,\label{AcMAB}\eea
where $\D_n={2(n+1)\over n+2}$ is fixed by the Young
projection\footnote{The traceless type-$(n,n)$ projection implies
that $\mathbf P(M_{AB}M_{C(n),D(n)})=\D_n\mathbf P
M_{[A|\{C(n)|,|B]|D(n)\}}=\D_n M_{AC(n),BD(n)}$ with $\mathbf
P\equiv \mathbf{P}_{\langle AC(n),BD(n)\rangle}$, and (suppressing
the anti-symmetry on $AB$)
\bea &&\eta_{\{C_1|[A}\eta_{
B]|D_1}M_{C(n-1),D(n-1)\}} \ =\
\eta_{AC_1}\eta_{BD_1}M_{C(n-1),D(n-1)}\nn\\[5pt]&&+\beta_n
\left(\eta_{C_1
C_2}\eta_{BD_1}M_{AC(n-2),D(n-1)}+\eta_{AC_1}\eta_{D_1D_2}M
_{C(n-1),BD(n-2)}\right.\nn\\[5pt]&&+\left.\eta_{C_1 D_1}
\eta_{C_2A}M_{BC(n-2),D(n-1)}+\eta_{C_1D_1}\eta_{D_2B}M_{C(n-1),AD(n-2)}\right)\nn\\[5pt]
&&+\a_n\left(\eta_{C_1 C_2}\eta_{D_1
D_2}M_{AC(n-2),BD(n-2)}-\eta_{C_1D_1}\eta_{C_2D_2}M_{AC(n-2),BD(n-2)}\right)\
,\eea
with $\a_n={1\over 4}{(n-1)^2\over(n+\e_0-1)(n+\e_0-\ft12)}$ and
$\b_n=-{1\over 2}{n-1\over n+\e_0-1}$.} while the contraction
coefficient
\bea \l_n&=&-{1\over 2}{n(n+1)(n+\e_0-1)\over n+\e_0+\ft12}\
,\label{lambdan}\eea
can be computed by imposing either {(i)} the trace condition using
$V\approx 0$; or {(ii)} the closure relation
$[\ac_{M_{AB}},\ac_{M_{CD}}]M_{E(n),F(n)}\equiv
2i\eta_{BC}\adj_{M_{AD}}M_{E(n),F(n)}-(A\leftrightarrow B)$. In
(i) it suffices to substitute \eq{TAnBn} to order ${\cal
O}(\eta^2)$ and ${\cal O}(\eta)$ on the left-hand and right-hand
sides, respectively, after which contraction by $\eta^{BD_1}$ and
usage of $V\approx0$ yields $\l_n$ (for fixed $n$). In (ii) the
closure requirement yields an inhomogeneous first-order recursion
relation for $\l_n$ with initial datum $\l_0=0$. The latter method
relies on the $\mg$-covariance of the entire procedure of
factoring out ${\cal I}[V]$ and does therefore not require any
explicit usage of $V\approx 0$.

The separate left and right actions of ${\cal A}$ on itself induce
a left ${\cal A}$-module ${\cal S}$ and a dual right ${\cal
A}$-module ${\cal S}^\ast$ characterized by $V\star {\cal S}={\cal
S}^\ast\star V=0$ and with pairing $X(Y)=(X,Y)_{\cal A}=
\Tr[X\star Y]$ where the trace is defined in \eq{Trprime}. We note
that since $X\star X$ contains a unit component for all $X\in{\cal
A}$ it follows that ${\cal S}_X\equiv {\cal A}\star X= {\cal
S}_{\1}={\cal S}$ and ${\cal S}^\ast_X\equiv X\star {\cal A}=
{\cal S}^\ast_{\1}={\cal S}^\ast$. Thus, as a two-sided module
\bea {\cal A}&\simeq& ({\cal S}\otimes {\cal S}^\star)/\sim\
,\qquad (X\star X')\otimes X''\ \sim\ X\otimes (X'\star X'')\
,\eea
where the isomorphism follows from ${\cal A}\otimes {\cal A}\sim
\1\otimes ({\cal A}\star {\cal A})=\1\otimes {\cal A}\simeq {\cal
A}$. The values\footnote{If $C_{2n}[\mg]$ assumes a fixed value in
a representation $\mR$ we shall denote it by
$C_{2n}[\mR]=C_{2n}[\mg|\mR]$.} $C_{2n}[{\cal S}]$ of the Casimir
operators $C_{2n}[\mg]=\ft12 M_{A_1}{}^{A_2}\star
M_{A_2}{}^{A_3}\star\cdots\star M_{A_{2n}}{}^{A_1}$ can be
expressed in terms of $C_2[\cS]$ using $V_{AB}\approx 0$ which
implies $M_A{}^B\star M_{BC}\approx{i(D-1)\over
2}M_{AC}+\mu^2\eta_{AC}$ with $\mu^2\equiv -{2C_2[{\cal S}]\over
D+1}$. For example $C_4[{\cal S}]={2\over
D+1}C_2[\cS]^2+{(D-1)^2\over 4}C_2[{\cal S}]$. Using also
$V_{ABCD}=M_{[AB}\star M_{CD]}=M_{[AB}\star
M_{C]D}-i\eta_{D[A}M_{BC]}$ one finds $M_A{}^B\star
V_{BCDE}\approx \left(\mu^2-\e_0\right)\star\eta_{A[C}M_{DE]}$,
from which one deduces that
\bea C_2[{\cal S}]&=& -\e_0(\e_0+2)\ ,\quad C_4[{\cal S}]\ =\
(\e_0^2+\e_0+1)C_2[{\cal S}]\ ,\qquad \e_0\ \equiv\ {D-3\over 2}\
.\label{C2e0}\label{C4e0}\eea
These values equal those of the scalar singleton $\mD_0\equiv
\mD(\e_0;(0))$, as can be seen by using \eq{C2lhws}, and one can
show more generally that $C_{2n}[{\cal S}]=C_{2n}[\e_0;(0)]$ for
all $n$. One can also show that ${\cal I}[V]$ is isomorphic to the
scalar-singleton annihilator\footnote{The elements of an
associative algebra ${\cal A}$ that act trivially in an ${\cal
A}$-module $\mR$ generate an ideal ${\cal I}[\mR]$ referred to as
the $\mR$ annihilator. Eqs. \eq{annI} and \eq{annID4} follow the
lemma that if $X\in {\cal U}'[\mg]$ belongs to an adjoint $\mg$
irrep then $X\in {\cal I}[\mD]$ where $\mD$ is a lowest-weight
representation iff $X$ annihilates the lowest-weight state. If $D$
is even then $V_{AB}\ket{e_0;\overrightarrow{s}_0}=0$ admits only
$\ket{\e_0,(0)}$ and $\ket{\e_0+\ft12,(\ft12)}$. The condition
$V_{ABCD}\approx 0$ rules out the spinor for $D>4$. Indeed,
$C_2[\mg|\e_0;(0)]= -\e_0(\e_0+2)$ and
$C_2[\mg|\e_0+\ft12;(\ft12)]= -\frac12
\left(\e_0+\frac12\right)(\e_0+2)$ are equal iff $D=4$. }
\bea {\cal I}[V]&\simeq&{\cal I}[\mD_0]\ .\label{annI}\eea
In $D=4$, also the spinor singleton $\mD_{\ft12}\equiv
\mD(1;(\ft12))$ is annihilated by ${\cal I}[V]$, that is
\bea D=4&:& {\cal I}[V]\ \simeq\ {\cal I}[\mD_0]\ \simeq\ {\cal
I}[\mD_{\ft12}]\ . \label{annID4}\eea
%


\scss{Adjoint and twisted-adjoint $\mso(D+1;\Comp)$
modules}\label{adjtwadj}


The space ${\cal A}$ also provides reducible modules ${\cal
T}^{(\mm_l)}$ for the generalized adjoint $\mg$ actions
\bea \adj^{(\mm_l)}_{M_{AB}}X &=& M_{AB}\star X-X \star
\pi_{(\mm_l)}(M_{AB})\ ,\label{adjt}\eea
where $\pi_{(\mm_l)}$ are the involutive ${\cal A}$-automorphisms
$\pi_{(\mm_l)}(M_{AB})= 2\left({\bf
P}_{(\mm_l)}\right)_{AB}{}^{CD} M_{CD} - M_{AB}$ with ${\bf
P}_{(\mm_l)}$ being the projector onto the generalized Lorentz
subalgebras $\mm_l\simeq\mso(D+1-l;\Comp)\oplus \mso(l;\Comp)$ of
$\mg$ for $l=0,1,2,\dots$. Thus, $\mg=\mm_l\ssum \mpe_l$ with
$[\mm_l,\mm_l]_\star=\mm_l$, $[\mm_l,\mpe_l]_\star=\mpe_l$ and
$[\mpe_l,\mpe_l]=\mm_l$, and $\pi_{(\mm_l)}(X)=X$ for $X\in \mm_l$
and $\pi_{(\mm_l)}(X)=-X$ for $X\in \mpe_l$. The modules ${\cal
T}^{(\mm_l)}$ decompose into $\mg$ irreps ${\cal
T}^{(\mm_l)}_{\ell}$ that we shall refer to as the levels of
${\cal T}^{(\mm_l)}$. If $\mm_1\simeq \mm$ we write
\bea {\cal L}&=& {\cal T}^{(\mg)}\ ,\qquad {\cal T}\ =\ {\cal
T}^{(\mm)}\ ,\eea
and define the twisted-adjoint action $\widetilde{\adj}\equiv
\adj^{(\mm)}$ by
\bea \widetilde{\adj}_{M_{AB}} X&=&\widetilde M_{AB} X\ =\
M_{AB}\star
X-X\star \pi(M_{AB})\ ,\qquad \pi\ =\ \pi_{(\mm)}\\[5pt] \pi(M_{ab})&=&M_{ab}\ ,\quad \pi(P_a)\ =\ -P_a\ ,\qquad P_a\ =\ M_{\sharp a}\ .\ \label{pimap}\eea
The automorphism $\pi$, which is outer in ${\cal A}$, becomes
inner\footnote{The inner automorphisms of a unital associative algebra ${\cal A}$ are the adjoint maps $\Ad_g$ defined for arbitrary invertible elements $g\in{\cal A}$. We note that the latter elements form a group inside ${\cal A}$, that one may view as the non-minimal higher-spin group in the case that ${\cal A}$ is defined as in \eq{calA1}. } in the enlarged algebra\footnote{The operator $k$
intertwines the extended adjoint and twisted-adjoint modules
$\mho_0\oplus ({\cal T}_+\star k)$ and ${\cal T}_+\oplus
(\mho_0\star k)$ defined in \eq{mho0} (although this does not
imply the agreement \eq{c2ell} and \eq{c4ell} between the values
of the Casimir operators on the irreducible subspaces). We recall
that a linear map $f: V\rightarrow W$ is an intertwiner between
two $\mg$ modules $V$ and $W$ if $f$ commutes with $\mg$ action,
that is $Xf(v)=f(X(v))$ for all $v\in V$ and $x\in \mg$. If $f$ is
a vector space isomorphism, then $V$ and $W$ are isomorphic as
modules. }
\bea \cA_k&=& \cA\oplus(\cA\star k)\ ,\label{enlalg}\eea
where by definition $k\star X=\pi(X)\star k$, $k\star k=\1$,
$\tau(k)=\pi(k)=k$ and $k$ acts as $\pi$ on lowest-weight spaces.
As shown in Appendix \ref{App:VAB}, the ideal ${\cal I}[V]$ has
the Lorentz-covariant presentation
\bea P^a\star P_a&\approx &\s\e_0\ ,\qquad P_{[a}\star P_b\star
P_{c]}\ \approx\ 0\ ,\label{papa}\eea
and we note the auxiliary trace constraints $P^a\star
M_{ab}\approx M_{ba}\star P^a\approx i(\e_0+1)P_b$ and
$M_{(a}{}^c\star M_{b)c}\approx \s P_{(a}\star
P_{b)}-\e_0\eta_{ab}$, and that the value of $C_2[\mm]=\ft12
M_{ab} \star M^{ab}$ in ${\cal S}$ (left action) is given by
$C_2[\mm|{\cal S}]=-\e_0(\e_0+1)$. Thus the elements of ${\cal A}$
have the Lorentz-covariant expansions
\bea X&\approx &\sum_{n\geqslant m\geqslant 0} X^{a(n),b(m)}
T_{a(n),b(m)}\ ,\label{Xod}\eea
where the traceless type-$(n,m)$ basis elements\footnote{The
non-monomial basis elements in \eq{TAnBn} and \eq{Tambn2} are
strongly equal, \emph{i.e.} as elements in ${\cal U}[\mg]$, and
obey $\tau(M_{A(n),B(n)})= (-1)^n M_{A(n),B(n)}$ and
$\pi(T_{a(n),b(m)})=(-1)^{n-m}T_{a(n),b(m)}$}
\bea T_{a(n),b(m)}&\equiv & M_{\{ a(n),b(m)\}
\sharp(n-m)} \ =\ M_{\{a_1b_1}\cdots M_{a_mb_m}P_{a_{m+1}}\cdots P_{a_n\}}\nn\\[5pt]&=& \sum_{k=0}^{[m/2]}\k_{n,m;k} M_{\langle
a(n),b(m-2k) \sharp(n-m+2k)}\eta_{ b_1b_2}\cdots
\eta_{b_{2k-1}b_{2k}\rangle}\ ,\label{Tambn2}\eea
where $M_{A(n),B(n)}$ are given by \eq{TAnBn} and $\k_{n,m;k}$ are
fixed by the requirement that $T_{a(n),b(m)}$ is traceless. The
resulting \emph{Lorentz-covariant adjoint and twisted-adjoint
modules} read
\bea {\cal L}|_{\mg}&=&
\bigoplus_{\ell\in\{-\ft12,0,\ft12,1,\dots\}}^\infty {\cal
L}_\ell\ ,\qquad {\cal T}|_{\mg}\ =\
\bigoplus_{\ell\in\{-1,\ft12,0,\ft12,\dots\}}^\infty {\cal
T}_\ell\ ,\label{covmod}\eea
respectively, where the $\ell$th levels take the form
\bea Q_\ell|_{\mm}&=&Q^{A(2\ell+1),B(2\ell+1)}
M_{A(2\ell+1),B(2\ell+1)}\ ,\quad \ell=-\ft12,0,\ft12,1,\dots\ ,\label{Lell}\\[5pt]
S_{\ell}|_{\mm} &=&\sum_{k=0}^\infty {i^k\over k!}S^{a(s+k),b(s)}
T_{a(s+k),b(s)}\ ,\quad \ell=-1,-\ft12,0,\ft12,\dots\ ,\quad
s=2\ell+2\ .\label{Tell}\eea
As shown in Appendix \ref{App:Cas}, the adjoint and
twisted-adjoint levels with $\ell\geqslant -\ft12$ have equal
quadratic and quartic Casimirs, namely
\bea C_2[\mg|{\cal L}_\ell]&=&C_2[\mg|{\cal T}_\ell]\ =\ C_2[\ell]\ =\ 2(s-1)(s+2\e_0)\ ,\label{c2ell}\\[5pt]
C_4[\mg|{\cal L}_\ell]&=&C_4[\mg|{\cal T}_\ell]\ =\ C_4[\ell]\ =\
(s^2+(2\e_0-1)s+2\e_0^2-\e_0+1)C_2[\ell]\ .\label{c4ell}\eea
Moreover, as can be seen using \eq{C2lhws} and \eq{C4lhws}, these
values coincide with those of the composite-massless lowest-weight
spaces $\mD(s+2\e_0;(s))$, \emph{i.e.}
\bea C_2[\ell]&=& C_2[\mg|s+2\e_0;(s)]\ ,\qquad C_4[\ell]\ =\
C_4[\mg|s+2\e_0;(s)]\ .\label{c2c4ell}\eea
As we shall see, these agreements follow from direct relationships
between ${\cal L}_{\ell}$, ${\cal T}_\ell$ and
$\mD(\pm(s+2\e_0);(s))$ visible in the corresponding
$\mso(2)\oplus \mso(D-1)$-covariant weight-space modules. In
${\cal T}_\ell$, which is infinite-dimensional, the required
change of basis is non-trivial, and actually amounts to the
\emph{harmonic expansion} of the linearized Weyl tensors. One
therefore distinguishes between the Lorentz-covariant slicing
\eq{covmod}, where the twisted-adjoint elements are arbitrary
polynomials in ${\cal A}$, and the corresponding
\emph{twisted-adjoint compact-weight modules}
\bea {\cal M}(\s,\s')|_{\mg}&=& \bigoplus_{s}{\cal
M}_{(s)}(\s,\s')\ ,\qquad {\cal M}_{(s)}(\s,\s')|_{\mh}\ =\
\bigoplus_{\bf \k} \Comp\otimes T^{(s)}_{\bf \k}\
,\label{compmod}\eea
where the elements by definition are arbitrary polynomials in
compact basis elements $T^{(s)}_{\bf \k}$ that belong to
finite-dimensional (unitary) representations of $\mh$. The latter
are given by series expansions in $T_{a(s+k),b(s)}$
($k=0,1,\dots)$ except for ${\cal M}_{(s)}(+,-)$, in which case
$\mm=\mh$.


\scss{Adjoining the adjoint and composite-massless
representations}\label{Sec:adjoin}

\begin{figure}[!h]
\begin{center}
\unitlength=.6mm
\begin{picture}(120,200)(0,-80)
\put(0,0){\vector(1,0){150}} \put(0,-80){\vector(0,1){200}}
\put(150,-10){$j_1$}
\put(-10,110){$E$}
\put(0,0){\line(1,1){50}}\put(0,0){\line(1,-1){50}}\put(50,-50){\!\!$\bullet$}\put(50,50){\line(0,-1){100}}
\put(47,-5){{\small s-1}}\put(-10,-50){{\small
1-s}}\put(-10,50){{\small s-1}}\multiput(0,50)(6,0){8}{-}
\multiput(0,-50)(6,0){8}{-} \put(-10,-42){{\small
2-s}}\put(58,-42){\line(0,1){130}}\put(58,-42){\!$\bullet$}\multiput(0,-42)(6,0){10}{-}
\put(58,-42){\line(1,1){50}}
\put(58,70){\line(0,1){40}}\put(58,-42){\!$\bullet$}\multiput(0,70)(6,0){10}{-}
\put(58,70){\line(1,1){40}}\put(58,70){\!$\bullet$}\put(-20,70){{\small
$s+2\epsilon_0$}}


\end{picture}
\end{center}
\caption{{\small The adjoint module $\mD(-(s-1);(s-1))$ is
connected via the conjugate massless module $\mD(-(s-2);(s))$ to
the massless module $\mD(s+2\e_0;(s))$.}} \label{wfig1}
\end{figure}
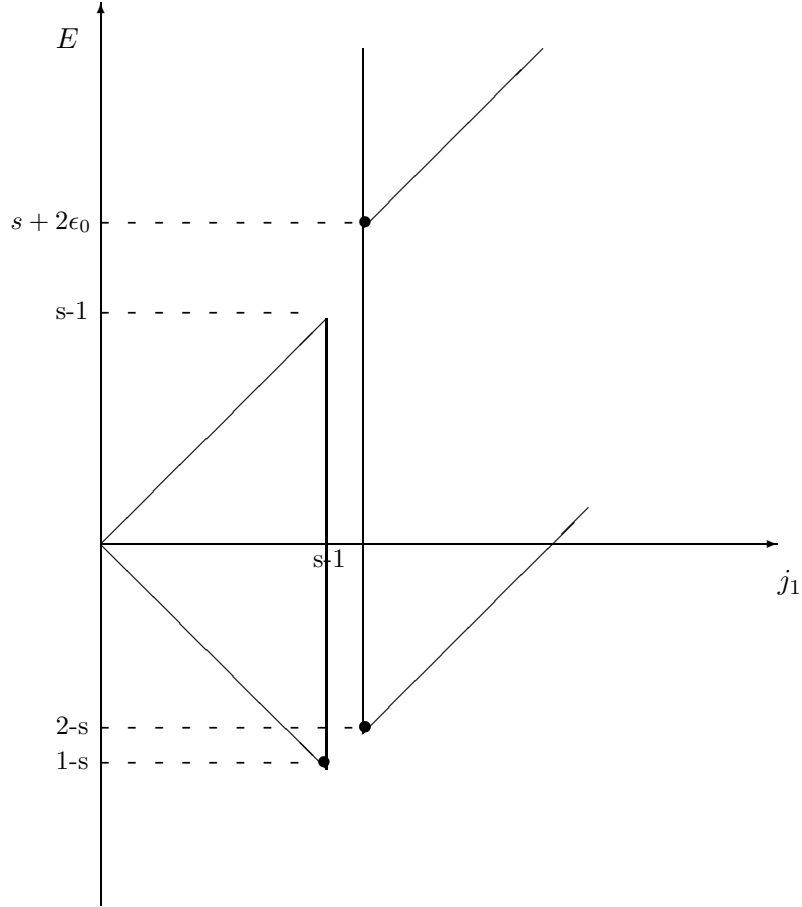

We propose that the Casimir relations \eq{c2ell}, \eq{c4ell} and
\eq{c2c4ell} follow from that ${\cal L}_\ell$, ${\cal
M}_\ell(+,+)$ and $\mD(s+2\e_0;(s))$, where $s=2\ell+2$, can
actually be adjoined in the $\mso(2)\oplus \mso(D-1)$ weight
space. As shown in Section \ref{Sec:LSM}, the space
$\mD(s+2\e_0;(s))$ arises upon harmonic expansion inside ${\cal
M}(+,+)$. Moreover, we propose that $\mD(s+2\e_0;(s))$ is adjoined
to the lowest-and-highest-weight space $\mD(-s+1;(s-1))\simeq
{\cal L}_\ell$ via the intermediate
conjugate-massless\footnote{The composite-massless and
conjugate-massless lowest-weight states are related by the
``conjugation'' $e_0\rightarrow D-1-e_0$, that leaves invariant
the values of the Casimir operators (see eqs. \eq{C2lhws} and
\eq{C4lhws}). Following a standard terminology (see, for example,
\cite{Metsaev:2008fs} and references therein), we will sometimes
refer to them as \emph{shadow} fields.} lowest-weight space
$\mD(-s+2;(s))$ (see Figure \ref{wfig1}) in such a way that the
Harish-Chandra modules
\bea 0\ \hookrightarrow\ \mC(-s+1;(s-1))&\longrightarrow
&\mC(-s+2;(s))\ \longrightarrow\ \mC(s+2\e_0;(s))\
\longrightarrow\ 0\ ,\label{ses}\eea
enter a short exact sequence for certain Young-projected actions
of $L^+_r$. First, the ground state of $\mC(-s+2;(s))$ has the
same quantum numbers as the singular vector
\bea \ket{-s+2;(s)}&=&L^+_{\{r_1}\ket{-s+1;(s-1)}_{r(s-1)\}}\ \in\
\mI(-s+1;(s-1))\ .\label{singvector}\eea
Next, we propose that the lowest-weight state of
$\mC(s+2\e_0;(s))$ has the same quantum numbers as the following
singular vectors in $\mI(-s+2;(s))$:
\bea D=4&:& \ket{s+1;(s)}_{r(s)}\ =\ \e_{r_1t_1u_1}\cdots \e_{r_s
t_s u_s}L^+_{u_1}\cdots L^+_{u_{s-1}}\ket{2;(s,1)}_{t(s),u_s}\
,\label{subsing2}\\[5pt]D\geqslant 5&:& \ket{s+2\e_0;(s)}_{r(s)}\ =\
(x^+)^{\ft{D-5}2}L^+_{t_1}\cdots
L^+_{t_s}\ket{2;(s,s)}_{r(s),t(s)}\ ,\label{subsing1} \eea
where $x^+\equiv L^+_r L^+_r$ and $\ket{2;(s,s)}$ and
$\ket{2;(s,1)}_{t(s),u_s}$ are the descendants of $\ket{-s+2;(s)}$
given by
\bea \hspace{-0.7cm} D\geqslant 5&:& \ket{2;(s,s)}_{r(s),t(s)}\ =\
L^+_{\{r_1}\cdots
L^+_{r_s}\ket{-s+2;(s)}_{t(s)\}}\ ,\label{wsweyl1}\\[5pt]
\hspace{-0.7cm} D=4&:& \ket{2;(s,1)}_{r(s),t}\ =\ \mathbf
P_{\{s,1\}} \left(\prod_{i=1}^{s-1}
\e_{r_iu_iv_i}L^+_{u_i}\right)\ket{-s+2;(s)}_{v(s-1)t}\
,\label{wsweyl2}\eea
where $\mathbf P_{\{s,1\}}$ denote the traceless type-$(s,1)$
projection of the index group $r(s),t$. We have checked that
\eq{subsing2} and \eq{subsing1} are singular vectors for $s=1$ and
$s=2$ in $D=4$ and $D=5$. In $D=6,8,\dots$ we define $\mg$ action
on $\sqrt{x^+}$ by extending
\bea L^-_r (x^+)^n&=&
(x^+)^nL^-_r+4n(x^+)^{n-1}\left(iL^+_sM_{rs}+ L^+_r
(E+n-\e_0-1)\right)\ ,\label{lminusxn}\eea
which is valid in ${\cal U}[\mg]$, to arbitrary differentiable
functions $f(x^+)$ as follows:
\bea [L^-_r ,f(x^+)]&=& 4{d\over dx^+}f(x^+)\left(iL^+_sM_{rs}+
L^+_r (E-\e_0)\right)+4x^+{d^2\over d(x^+)^2} f(x^+) L^+_r\
.\hspace{1cm}\label{lminusf}\eea
The singular vectors \eq{subsing1} and \eq{subsing2} vanish
formally if $\ket{-s+2;(s)}$ is substituted with \eq{singvector},
so that $\ket{-s+2;(s)}$ and $\ket{2;(s,s)})$ or $\ket{2;(s,1)}$
are weight-space analogs of an abelian gauge field and its Weyl or
Cotton tensor, respectively. In other words, the sequence \eq{ses}
is the weight-space counterpart of the Chevalley-Eilenberg cocycle
appearing in the linearized one-form constraint \eq{linoneform}.

\scss{Non-composite trace and reflectors}\label{Sec:Tr}


A trace on an associative algebra ${\cal A}$ is a linear map
$\Tr:{\cal A}\rightarrow\Comp$ that obeys $\Tr(X\star Y-Y\star
X)=0$. If ${\cal A}$ contains a Lie subalgebra $\ml$ and
decomposes under $\adj_{\ml}$ into finite-dimensional irreps
${\cal A}_\l$, that is ${\cal A}|_{\ml}=\bigoplus_{\l} {\cal
A}_\l$, ${\cal A}\ni X=\sum_{\l}T_\l X_\l$, and if the
non-singlets $T_{\l'}$ are expressible as $\star$-commutators,
then $\Tr[X]=\Tr_{{\cal A}|\ml}[X]=\sum_{\l_0\in\L} t_{\l_0}
X_{\l_0}$, provided the trace is finite, where $\l_0$ label the
$\ml$ singlets and $t_{\l_0}$ are complex numbers compatible with
cyclicity. If ${\cal A}[\mg|\mR]\equiv {\cal U}[\mg]/{\cal
I}[\mR]$ where $\mg$ is a finite-dimensional Lie algebra with
module $\mR$, then the \emph{non-composite trace}\footnote{ The
non-composite trace of the universal enveloping algebra ${\cal
U}[\mg]$ is trivial in the sense that $\Tr[X\star Y]=(X\star
Y)_{\1}=X_{\1}Y_{\1}$.} is defined as
$\Tr_{\cA[\mg|\mR]|\mg}[X]\equiv X_{\1}$ where $X_{\1}$ is the
coefficient of $\1\in {\cal A}[\mg|\mR]$ in a \emph{preferred
basis}. This trace is well-defined since $X\approx X'$ iff $X$ and
$X'$ have the same expansions in the preferred basis. The
cyclicity follows from $\Tr[\t(X)]=\Tr[X]$ where $\tau$ is the
anti-automorphism defined in \eq{taumap}, which implies that if
$X,Y\in\cA[\mg|\mR]$ then $\Tr[X\star Y]=\Tr[\t(X\star
Y)]=\Tr[\t(Y)\star\t(X))]=\Tr[Y\star X]$, as can be seen by
splitting $X=X_++X_-$ with $\t(X_\pm)=\pm X_\pm$, \emph{idem} $Y$,
and noting that $\Tr(X_\pm\star Y_\mp)=0$. If ${\cal A}[\mg|\mR]$
is infinite-dimensional, which requires $\mR$ to be
infinite-dimensional, the non-composite and composite
traces\footnote{ If $\mR$ has a finite dimension $|\mR|$ then the
algebra ${\cal A}[\mg|\mR]\subset GL(|\mR|;\Comp)$, and the
non-composite trace is equivalent to the composite trace
$\Tr_\mR[X]=\sum_{n}\bra{n^\ast}X\ket{n}$ with $\ket{n}$ and
$\bra{n^\ast}$ being basis elements for $\mR$ and $\mR^\ast$ with
normalization $\bra{m^\ast}n\rangle=\d_m^n$. The composite trace
can be written as
$\Tr_\mR[X]={}_{12}\bra{\1^\ast_\mR}X(1)\ket{\1}_{12}$ where
$\ket{\1_\mR}_{12}=\sum_n\ket{n}_1\otimes \ket{n}_2$ and
${}_{12}\bra{\1^\ast_\mR}=\sum_n{}_1\bra{n^\ast}\otimes
{}_2\bra{n^\ast}$ are referred to as composite reflectors of
$\mR$.} are in general not equivalent.

In the case of ${\cal A}={\cal A}[\mso(D+1;\Comp)|\mD_0]$ we
therefore distinguish between the composite trace $\Tr_{\mD_0}$
over the scalar singleton (see Appendix \ref{App:Blws}) and the
non-composite trace, and define
\bea \Tr[X]&=& X_0\ =\ X^{(0,0)}\ ,\label{Trprime}\eea
where $X_0$ is the $\mg$-singlet in \eq{calA} defined with respect
to the preferred basis elements \eq{TAnBn}, and $X^{(0,0)}$ is the
corresponding Lorentz singlet in \eq{Xod} defined with respect to
the preferred basis elements \eq{Tambn2}. These singlets are equal
since the preferred basis elements are equal (strongly) in ${\cal
U}[\mg]$, and the $(n,n)$-plet of $\mg$ contains the singlet of
$\mm$ iff $n=0$. The trace $\Tr$ extends uniquely to ${\cal A}_k$
defined in \eq{enlalg} since $\cA\star k$ does not contain any
singlet, which implies $\Tr[X\star k]=0$. As shown in Appendix
\ref{Sec:4D}, the $\Tr$ is equivalent in $D=4$ to the composite
supertrace
\bea D=4\ :\quad \Tr\ =\ \Tr_{\mD_0}-\Tr_{\mD_{1/2}}\
.\label{D4strace}\eea

The trace $\Tr$ can equivalently be written as
\bea \Tr[X]&=& _{12}{}\bra{\1^\ast} X(1)\ket{\1}_{12}\
,\label{trvev}\eea
where the \emph{non-composite reflector} $\ket{\1}_{12}\in {\cal
B}$ and its dual ${}_{12}\bra{\1^\ast}\in{\cal B}^\ast$ are
elements in the left and right ${\cal A}_k$-bimodules\footnote{A left bimodule is a left 2-module, where, in general, a left $P$-module $V$, $P\in\mathbb{N}$, is a linear space with the property that if $v\in V$ and $X,Y\in{\cal A}$, then $X(\xi)v\in V$ and $(X\star Y)(\xi)v=X(\xi)(Y(\xi)v)$ for $\xi=1,...,P$, and $X(\xi)Y(\eta)v=Y(\eta)X(\xi)v$ if $\xi\neq\eta$. We
recall that if $V$ is an ${\cal A}$ left module then $V^\ast$ is a
${\cal A}$ right module via $v^\ast(Xw)=(v^\ast X)(w)$ for
$v^\ast\in V^\ast$, $w\in V$ and $X\in{\cal A}$. This generalizes
straightforwardly to $P$-modules by defining $v^\ast(X(\xi)w)=(v^\ast X(\xi))(w)$. Finite-dimensional bimodules are
composite in the sense that their elements are sums of factorized
elements of the form $u_1\otimes v_2$, while the analog need not
hold for infinite-dimensional bimodules. We also note that since
$\pi(V)=V$ it follows that if $\ket{ X}_{12}\in {\cal B}$ and
${}_{12}\bra{ X^\ast}\in{\cal B}^\ast$, then $k(\xi)\ket{
X}_{12}\in {\cal B}$ and ${}_{12}\bra{ X^\ast}k(\xi)\in{\cal
B}^\ast$ for $\xi=1,2$.}
\bea {\cal B}&=&\left\{\ket{X}_{12}:\ V(\xi)
\ket{X}_{12}=0\right\}\ ,\qquad {\cal B}^\ast\ =\
\left\{{}_{12}\bra{X^\ast}:\ {}_{12}\bra{X^\ast}V(\xi)=0\right\}\
,\label{defB}\label{defBast}\eea
where $\xi=1,2$, obeying the normalization condition
$_{12}{}\bra{\1^\ast}\1\rangle_{12}=1$ and the overlap condition
\bea \left( X(1)-(\tau\circ\pi)( X)(2)\right)\ket{\1}_{12}&=&0\
,\qquad
{}_{12}\bra{\1^\ast}\left(X(1)-(\tau\circ\pi)(X)(2)\right)\ =\ 0\
,\qquad \label{overlapX}\eea
for all $X\in {\cal A}_k$ and where $\tau\circ\pi$ is the
anti-automorphism composed by $\tau$ defined in \eq{taumap} and
$\pi$ defined in \eq{pimap}. The overlap conditions and
$(\pi\tau)^2={\rm Id}$ imply that $ _{12}{}\bra{\1^\ast}(X\star
Y)(1)\ket{\1}_{12}$$=$$_{12}{}\bra{\1^\ast}( Y\star
X)(1)\ket{\1}_{12}$ so that
$_{12}{}\bra{\1^\ast}[M_{AB}(1),X(1)]_\star\ket{\1}_{12}=0$. Since
$X_n$ are $\mg$-commutators for $n>0$ it follows that
$_{12}{}\bra{\1^\ast}
X(1)\ket{\1}_{12}=_{12}{}\bra{\1^\ast}\1\rangle_{12}X_0=\Tr[X]$.
Moreover, taking the expectation value of $(P^a\star k\star
P_a-\s\e_0 k)(1)=0$ implies that
$_{12}{}\bra{\1^\ast}k(1)\ket{\1}_{12}=0$. From \eq{overlapX} it
follows that ${\cal O}_{12}\equiv _{23}\bra{\1^\ast}\star
\ket{\1}_{13}$ obeys $ X\star {\cal O}={\cal O}\star X$ for all
$X\in {\cal A}_k$ so that ${\cal O}=c\1$. Defining
\bea _{23}\bra{\1^\ast}\star\ket{\1}_{13}&=& \1_{12}\
,\label{def1}\eea
one has the sequence of \emph{reflection maps} ${\cal
B}_{12}\stackrel{R_2}{\longrightarrow}
(\cA_k)_{12}\stackrel{R_1}{\longrightarrow}{\cal B}^\ast_{12}$
given by
\bea | X\rangle_{12}&=&X(1)\ket{\1}_{12}\ ,\qquad
{}_{12}\bra{X^\ast}\ =\ {}_{12}\bra{\1^\ast} X(1)\
,\label{doubleton}\\[5pt] X_{12}&=&
_{23}\bra{\1^\ast}\star\ket{X}_{13}\ =\ {}_{23}\bra{
X^\ast}\star\ket{\1}_{13}\ .\label{invrefl}\eea
We note that the condition \eq{overlapX} implies, in particular,
that $\ket{\1}_{12}$ is a $\mm_{\rm diag}$-invariant element.

We also define the twisted non-composite reflectors
\bea \ket{\j1}_{12}&=&k(1)\ket{\1}_{12}\ =\ k(2)\ket{\1}_{12}\
,\qquad {}_{12}\bra{\j1^\ast}\ =\ {}_{12}\bra{\1^\ast}k(1)\ =\
{}_{12}\bra{\1^\ast}k(2)\ ,\label{twrefl}\eea
obeying the overlap conditions
$(X(1)-\t(X)(2))\ket{\j1}_{12}$$={}_{12}\bra{\j1^\ast}(X(1)-\t(X)(2))=0$,
which in particular imply that the twisted reflectors are
$\mg_{\rm diag}$-invariant. The twisted reflectors have the
normalizations ${}_{12}\langle{\j1^\ast}\ket{\j1}_{12}=1$,
${}_{12}\langle{\j1^\ast}\ket{\1}_{12}$$=$$
{}_{12}\langle{\1^\ast}\ket{\j1}_{12}$$=0$ and
${}_{23}\bra{\j1^\ast}\star \ket{\j1}_{12}=\1_{13}$,
${}_{23}\bra{\j1^\ast}\star\ket{\1}_{12}$$=$${}_{23}\bra{\1^\ast}\star\ket{\j1}_{12}$$=k_{13}$,
and $\Tr$ now admits the manifestly $\mg$-invariant form
\bea \Tr[X]&=&{}_{12}\bra{\j1^\ast}X(1)\ket{\j1}_{12}\ .\eea


\scss{The adjoint and twisted-adjoint higher-spin representations}


The adjoint and twisted-adjoint actions of ${\cal A}$ on itself
induces the minimal bosonic higher-spin algebra
$\mho_0=\mho_0(D+1;\Comp)$ and its even and odd twisted-adjoint
representations:
\bea \mho_0&\equiv&\{Q\in{\cal A}:\t(Q)\ =\ -Q\}\ ,\qquad {\cal
T}_\pm\ \equiv\ \{S_\pm\in{\cal A}:\t(S_\pm)\ =\ \pm \pi(S_\pm)\}\
,\label{mho0}\eea
where $\t$ and $\pi$ are given in \eq{taumap} and \eq{pimap},
respectively. Their level decompositions read
$\mho_0|_{\mg}=\bigoplus_{\ell=0}^\infty {\cal L}_\ell$ and ${\cal
T}_\pm|_{\mg}\ =\ \bigoplus_{\ell=-1}^\infty {\cal
T}_{\ell+\ft12(1\mp 1)}$ with levels ${\cal L}_\ell$ and ${\cal
T}_\ell$ given in \eq{covmod}. The $\mho_0$ transformations are
defined by
\bea \adj(Q)(Q')&=& [Q,Q']_\star\ =\ Q\star Q'-Q'\star Q\
.\label{adjrep}\\[5pt]
\widetilde{\adj}_Q(S)&=&[Q,S]_\pi\ =\ \widetilde Q S\ =\ Q\star
S-S\star \pi(Q)\ ,\label{twadjrep}\eea
and mix the levels as follows (see \cite{mishasalg,
Bekaert:2005vh, Sagnotti:2005ns, Eastwood} and also
\cite{Engquist:2007kz} for a more recent discussion):
$\widetilde{\adj}_{Q_\ell}(S_{\ell'})=\sum_{\tiny\ell''=
\max(-1,\ell'-\ell)}^{\tiny\ell+\ell'}S_{\ell''}$ and
$[Q_{\ell},Q_{\ell'}]_\star=\sum_{\tiny\ell''=|\ell-\ell'|}^{\tiny\ell+\ell'}Q_{\ell''}$,
where the lower bound on $\ell''$ in the twisted-adjoint case
follows from the contraction rules $V_{AB}\approx V_{ABCD}\approx
0$. The algebra $\mho_0$ is a minimal Lie-algebra extension of
$\mg$ in the sense that if ${\cal L}'$ is a Lie subalgebra obeying
$\mg\subseteq {\cal L}'\subseteq \mho_0$ then either ${\cal
L}'=\mg$ or ${\cal L}'=\mho_0$. The minimal set-up can be extended
in various ways by enlarging the underlying associative algebra
$\cA$, \emph{e.g.} to $\cA_k$ which corresponds to adding extra
auxiliary fields (see, for example, \cite{Vasiliev:1999ba} and
references therein), and/or by tensoring with internal associative
algebras which adds Yang-Mills-like sectors, and fermionic
generators \cite{Konstein:1988yg, mishasalg, Vasiliev:1986qx,
Engquist:2002vr}. The in some sense simplest extension is to add
the half-integer levels, which leads to the non-minimal bosonic
higher-spin algebra
\bea \mho&\equiv & {\cal L}\ ,\qquad \mho|_{\mg}\ =\
\bigoplus_{\ell=-\ft12,0,\ft12,1,\dots} {\cal L}_\ell\
,\label{algmho}\eea
acting irreducibly in ${\cal T}={\cal T}_+\oplus{\cal T}_-$, such
that if $\tau(Q_\pm)=\mp Q_\pm$ and $\t(S_\pm)=\pm \pi(S_\pm)$
then $\adj_{Q_\pm}(Q_\e)=Q_{\pm\e}$ and $\widetilde
\adj_{Q_\pm}(S_\e)=S_{\pm\e}$ for $\e=\pm$.

The trace operation $\Tr$ induces the $\mho$-invariant bilinear
forms on $\mho$ and ${\cal T}$:
\bea (Q,Q')_{\mho}&=&-\Tr[Q\star Q']\ =\ -_{12}\langle{Q}|
Q'\rangle_{12}\
,\label{hoinnerproduct}\\[5pt]
(S,S')_{\cal T}&=& \Tr[\pi(S)\star S']\ =\ _{12}\langle{\pi(S)}|
S'\rangle_{12}\ ,\label{Tinnerproduct}\eea
where \eq{doubleton} has been used to define
$\ket{Q_\pm}_{12}=Q_\pm(1)\ket{\1}_{12}= \ft12
(Q_\pm(1)\mp\pi(Q_\pm)(2))\ket{\1}_{12}$ and $\ket{S_\pm}_{12}=
S_\pm(1)\ket{\1}_{12}=\ft12 (S_\pm(1)\pm S_\pm(2))\ket{\1}_{12}$,
and ${}_{12}\bra{Q_\pm}={}_{12}\bra{\1^\ast}Q_\pm(1)=\ft12
{}_{12}\bra{\1^\ast}(Q_\pm(1)\mp\pi(Q_\pm)(2))$ and
${}_{12}\bra{\pi(S_\pm)}$$= {}_{12}\bra{\1^\ast}\pi(S_\pm)(1)$$=
\ft12{}_{12}\bra{\1^\ast} (\pi(S_\pm)(1)\pm \pi(S_\pm)(2))$. These
states carry the higher-spin representations
\bea \ket{\adj_Q Q'}_{12}\ =\ \widetilde Q\ket{Q'}_{12}\ ,\qquad
\ket{\widetilde{\adj}_Q S}_{12}\ =\ Q\ket{S}_{12}\ ,\eea
and ${}_{12}\bra{\adj_Q Q'}= -{}_{12}\bra{ Q'}\widetilde Q$,
${}_{12}\bra{\pi(\widetilde\adj_Q S)}=-{}_{12}\bra{\pi(S)}Q$,
where
\bea \widetilde Q \ket{X}_{12}&\equiv
&(Q(1)+\pi(Q)(2))\ket{X'}_{12}\ ,\qquad Q\ket{S}_{12}\ \equiv\
(Q(1)+Q(2))\ket{S}_{12}\ .\eea
The $\mho$ action is (anti-)self-adjoint with respect to the inner
products \eq{hoinnerproduct} and \eq{Tinnerproduct}, that is
\bea ( \adj_Q Q',Q'')_{\mho}&=&-(Q',\adj_Q Q'')_{\mho}\ ,\qquad (
\widetilde \adj_Q S,S')_{{\cal T}}\ =\ -(S,\widetilde\adj_Q
S')_{\cal T} \ .\label{selfadjoint}\eea
The real forms of $\mho$ and ${\cal T}$ are given by
\bea Q^\dagger&=&-Q\ ,\qquad S^\dagger\ =\ \pi(S)\ ,\eea
where $\dagger$ acts as $(M_{AB})^\dagger=M_{AB}$ and as standard
complex conjugation of coefficients.


\scss{Adjoint and twisted-adjoint master fields}\label{Sec:adj}


The unfolded formulation of a field theory (see, for example,
\cite{Vasiliev:1999ba,Bekaert:2005vh} and references therein)
involves a space $\cR=\bigoplus_{\a}\O^{[p_\a]}\otimes \cR^\a$ of
locally defined rank-$p_\a$ differential forms $X^\a$ taking their
values in $\mg$ modules $\cR^\a$ and obeying generalized curvature
constraints $R^\a\equiv dX^\a+f^\a(X^\b)=0$, where the functions
$f^\a$ (that are sums of multi-linear products) obey generalized
Jacobi identities $f^\b\partial_\b f^\a\equiv 0$ assuring gauge
and diffeomorphism invariance (for arbitrary base manifolds). The
corresponding operator $Q=f^\a\partial_\a$ is nilpotent and, upon
expansion around backgrounds, its linearization induces a (locally
defined) cohomology in the space
$\cR^{[\ast]}=\bigoplus_{\a}\O^{[\ast]}\otimes \cR^\a$ containing
the standard gauge parameters, gauge fields, and field equations,
while the non-closed and trivial elements encode St\"uckelberg
symmetries, auxiliary fields and Bianchi identities
\cite{Bekaert:2005vh, Vasiliev:2005zu}.

In Vasiliev's frame-like formulation of higher-spin gauge theory,
the basic building blocks are master fields in $\O^{[0]}$ and
$\O^{[1]}$. The \emph{full} master fields are valued in
representations of an extended higher-spin algebra $\widehat{\cal
A}$, and obey full constraints of a remarkably simple form
\cite{Vasiliev:en, Vasiliev:2003ev, Sezgin:2002ru,
Sagnotti:2005ns}. The full fields can be expanded perturbatively
in terms of \emph{reduced} master fields, namely a one-form $A$
and a zero-form $\Phi$, referred to as the Weyl zero-form. In the
minimal bosonic model the reduced fields are
\bea A&\in &\O^{[1]}\otimes \mho_0\ ,\qquad \Phi\ \in\
\O^{[0]}\otimes {\cal T}_+\ ,\eea
where $\mho_0$ and ${\cal T}_+$ are defined in \eq{mho0}. More
explicitly,
\bea A&=&\sum_{s=2,4,6,\dots} A_{(s)}\ ,\quad A_{(s)}\ =\
-i\sum_{t=0}^{s-1}dx^M A_{M,a(s-1),b(t)}(x^N)M^{a_1
b_1}\cdots M^{a_tb_t}P^{a_{t+1}}\cdots P^{a_{s-1}}\ ,\hspace{1.5cm}\label{A}\label{As}\\[5pt]
\Phi&=&\sum_{s=0,2,4,\dots} \Phi_{(s)}\ ,\quad \Phi_{(s)}\ =\
\sum_{k=0}^\infty{i^k\over k!}
\Phi^{a(s+k),b(s)}(x^M)M_{a_1b_1}\cdots
M_{a_sb_s}P_{a_{s+1}}\cdots P_{a_{s+k}}\ ,\label{Phis}\eea
where $x^M$ are the coordinates of the base manifold. The
perturbative constraints have the form
\bea F+ \sum_{n=1}^\infty J_{(n)}(A,A;\Phi,\dots,\Phi)&=&0\
,\qquad D\Phi+\sum_{n=2}^\infty P_{(n)}(A;\Phi,\dots,\Phi)\ =\ 0\
,\label{mfe}\eea
where $F\equiv dA+A\star A$ and $D\Phi\equiv d\Phi+[A,\Phi]_\pi$,
and $J_{(n)}:(\mho_0)^{\otimes 2}\otimes ({\cal T}_+)^{\otimes
n}\rightarrow \mho_0$ and $P_{(n)}:\mho_0\otimes ({\cal
T}_+)^{\otimes n}\rightarrow {\cal T}_+$ are multi-linear products
that can be computed using the oscillator realizations (see
\cite{Sezgin:2002ru} for discussions). The standard formulation in
$D$-dimensional spacetime follows upon splitting $A=A_{(2)}+W+K$
where
\bea A_{(2)} &=&-i(e^aP_a +\ft12 \omega^{ab}M_{ab})\ =\ -i dx^\mu
\left(e_\mu{}^a P_a+\ft12 \omega_{\mu}{}^{ab} M_{ab}\right)\ ,
\label{spin2gaugeefields}\eea
contains the microscopic Vasiliev-frame vielbein $e_\m{}^a$ and
Lorentz connection $\o_\m{}^{ab}$; $W=\sum_{s=4,6,\dots} A_{(s)}$;
and $K$ is a field redefinition such that $e_\m{}^a$ and the
component fields in $W$ and $\Phi$ transform as tensors under the
canonical Lorentz transformations \cite{Vasiliev:1999ba,
Sezgin:2002ru, Vasiliev:1992av}. If $A_{(2)}$ is treated exactly
while $W$ and $\Phi$ are treated as weak fields, then the
expansion of \eq{mfe} yields a set of manifestly diffeomorphism
and locally Lorentz invariant constraints, where thus $\o$ appears
only via $\nabla\equiv d+\o$ or $R\equiv d\o+\o^2$. If $e_\m{}^a$
is invertible, then exhaustion of the resulting algebraic
constraints and St\"uckelberg symmetries leaves a set of dynamical
microscopic fields: a \emph{scalar field}, a \emph{metric} and a
tower of \emph{doubly-traceless symmetric rank-$s$ tensor gauge
fields} ($s=4,6,\dots$) given by\footnote{There are exceptions
already in the minimal case, such as the chiral models in
Euclidean and Kleinian signatures in $D=4$ \cite{Iazeolla:2007wt},
whose metric and tensor-gauge fields are half-flat, say with
left-handed curvatures, leaving the right-handed components in
$\Phi$ as independent dynamical fields.}:
\bea \phi&=&\Phi_{(0)}|_{P_a=0}\ , \quad g_{\mu\nu} \ =\ e_\mu{}^a
e_{\nu a}\ ,\quad \phi_{a(s)}\ =\ (e^{-1})_{(a}{}^\mu
W_{\mu,a(s-1))}\ .\label{dybfields}\eea
Among the auxiliary zero-forms are the \emph{generalized spin-$s$
Weyl tensors} $\Phi_{a(s+k),b(s)}$ with $s=0,2,4,\dots$ and
$k=0,1,2,\dots$, given on-shell for $s\neq 2$ by $s$ curls and $k$
gradients of $\phi_{a(s)}$ plus higher-order corrections in weak
fields, and for $s=2$ by the $k$th gradient of the full spin-2
Weyl tensor plus weak-field corrections. At first order in weak
fields
\bea {\cal R}+\nabla
W-ie^a\{P_a,W\}_\star&=&-\ft{i}2\sum_{s=2,4,6,\dots} e^a\wedge e^b
\Phi_{ac(s-1),bd(s-1)}M^{c_1 d_1}\cdots M^{c_{s-1}d_{s-1}}\ ,\hspace{1cm}\label{linoneform}\\[5pt]
\nabla\Phi-ie^a\{P_a,\Phi\}_\star&=&0\ ,\label{linzeroform}\eea
where ${\cal R}=-i(T^a P_a +\ft12 (R^{ab}+\sigma e^a e^b)M_{ab}$,
$T^a=\nabla e^a=de^a+\o^{ab}e_b$,
$R^{ab}=d\o^{ab}+\o^{ac}\o_{c}{}^b$ and $\nabla=d-\ft{i}2 \o^{ab}
M_{ab}$. Since ${\cal D}\equiv d+A_{(2)}=\nabla-i e^a P_a$ obeys
${\cal D}^2\equiv {\cal R}$, which starts as a weak field, it
follows that \eq{linoneform} and \eq{linzeroform} are consistent
in the first order\footnote{The consistency at higher orders in
weak fields depends crucially on the contributions from $J_{(n)}$
with $n\geqslant 1$ and $P_{(n)}$ $n\geqslant 2$.}: the
consistency of \eq{linoneform} requires $e^a\wedge e^b \wedge e^c
\nabla_a \Phi_{bd(s-1),ce(s-1)}=0$, which follows from
\eq{linzeroform}, which is in its turn consistent. Combining
\eq{Phis} and \eq{linzeroform} with \eq{TPa} yields the following
component form of the zero-form constraint (see Appendix
\ref{App:T}):
\bea \nabla_c \Phi_{a(s+k),b(s)}-2k\D_{s+k-1,s} \eta_{c\{
a}\Phi_{a(s+k),b(s)\} }+ {2\l^{(s)}_{k+1}\over k+1} \Phi_{c\{
a(s+k),b(s)\}}&=&0\ .\label{DPhicomponents}\eea
Using $\Phi_{\langle c\langle a(s+k),b(s)\rangle\rangle}=\Phi_{c
a(s+k),b(s)}$ one finds that
\bea \Phi_{a(s+k),b(s)}&=& {(-1)^k k!\over 2^k \prod_{l=1}^k
\l^{(s)}_{l}} \nabla_{\{ a_1}\cdots
\nabla_{a_k}\Phi_{a(s),b(s)\}}\ ,\qquad k=1,2,\dots\
.\label{Phiauxiliary}\eea
Other projections of \eq{DPhicomponents} yield the following
Bianchi identities and mass-shell conditions:
\bea \nabla_{[\mu} \Phi_{\nu |a(s+k-1),|\rho]b(s-1)}&=&0\ ,\qquad s\geqslant 1\ ,\label{BI}\\[5pt]
(\nabla^2-M^2_{s,k})\Phi_{a(s+k),b(s)} &=&0\ ,\qquad s\geqslant 0\
,\label{massshell}\eea
with critical masses given by
\bea M^2_{s,k}&=&-4\l_{k}^{(s)}-4\l_{k+1}^{(s)}\D_{s+k,s}
{(k+s+2\e_0)(k+s+\e_0+\ft32)(k+2s+2\e_0+1)\over (k+s+1)
(k+2s+2\e_0)(k+s+\e_0+\ft12)}\nn\\[5pt] &=& -\s\left(4\e_0+2s+(k+2s+2\e_0+1)k\right)\ .\label{msk}\eea
These masses can be computed without the explicit usage of the
expression for $\l_k^{(s)}$, by first rewriting \eq{linzeroform}
as $\nabla_a\Phi_{(s)}=i\ac_{P_a}\Phi_{(s)}$ and then using the
Casimir relation $C_2[\mg]=C_2[\mm]-2\s P^a\star P_a$, that is
$-\ac_{P^a} \ac_{P_a}\Phi_{(s)}= \ft\s
2(\widetilde{\adj}_{M_{AB}}\widetilde{\adj}_{M^{AB}}-
\adj_{M_{ab}}\adj_{M^{ab}})\Phi_{(s)}$. This yields
\bea \nabla^2 \Phi_{a(s+k),b(s)}&=&\s
\left(C_2[\mg|\ell]-C_2[\mm|(s+k,s)]\right)\Phi_{a(s+k),b(s)}\
,\label{casrel}\eea
with $C_2[\mm|(s+k,s)]=(s+k)(s+k+D-2)+s(s+D-4)$. Inserting the
value of $C_2[\mg|\ell]$ given by \eq{c2ell} then gives \eq{msk}.


\scss{Harmonic expansion of the Weyl zero-form}


The maximally symmetric coset geometries \eq{maxsymm} can be
embedded into Vasiliev's equations as exact solutions given by
\bea \Phi&=&0\ ,\qquad A\ =\ A_{(2)}\ =\ \O \ \equiv\ L^{-1}\star
dL\ ,\label{maxsymmspace}\eea
where $\O$ is thus a flat $\mg$-connection parameterized by a
coset element\footnote{For example, in Lorentz-covariant
stereographic coordinates one may take $e^a=-{2\l dx^a\over h^2}$
and $\omega^{ab}=-{4\s\l^2 x^{[a}dx^{b]}\over h^2}$ with
$h=\sqrt{1-\s \l^2 x^2}$, which arise from $L=\exp_\star (4i \xi
\l x^a P_a)= f\exp(i\l g x^\mu \d_\mu^a P_a)$ with $\xi=\frac{{\rm
artanh}\sqrt{\frac{1-h}{1+h}}}{\sqrt{1-h^2}}$ and $f=f(h)$ and
$g=g(h)$. In $D=4,6$, where ${\cal T}_{(0)}$ is conformal, one has
$f=\left[\frac{2h}{1+h}\right]^{\e_0+\ft12}$ and $g=\frac{4}{1+h}$
(see Appendix \ref{App:T0}).} $L=L(x)$. The leading order of the
weak-field expansion around these solutions is a self-consistent
set of linearized field equations that are invariant under abelian
gauge transformations. The linearized zero-form constraint
\eq{linzeroform} is solved by
\bea \Phi_{(s)}&=& L^{-1}\star S_{(s)}\star \pi(L)\ ,\quad
dS_{(s)}\ =\ 0\ .\label{linPhi}\eea
Expanding $S_{(s)}$ in the twisted-adjoint compact-weight module
${\cal M}_{(s)}(\s,\s')$,
\bea S_{(s)}&=& \sum_{\bf \k} S^{(s)}_{\bf \k} T^{(s)}_{\bf \k}\
,\quad S^{(s)}_{\bf \k}\in \Comp\ ,\eea
yields the \emph{harmonic expansion} of the linearized Weyl
tensors
\bea \Phi_{(s)}&=& \sum_{\bf \k} S^{(s)}_{\bf \k} L^{-1}\star
T^{(s)}_{{\bf \k}} \star \pi(L)\ =\ \sum_{k=0}^\infty
T_{a(s+k),b(s)} \sum_{\bf \k} S^{(s)}_{\bf \k}
D^{(s);a(s+k),b(s)}_{\bf \k}\ ,\eea
where the \emph{generalized harmonic functions}
\bea D^{(s);a(s+k),b(s)}_{\bf \k}={\cal N}_{s,k}^{-1}
~\Tr\left[T^{a(s+k),b(s)} \star L^{-1}\star T^{(s)}_{\bf \k}\star
\pi(L)\right]={\cal N}_{s,k}^{-1} {}_{12}\bra{T_{a(s+k),b(s)}}
L^{-1}\ket{\k}_{12}\ .\quad\eea
The first equality follows from \eq{trtabtcd}, while the second
from \eq{trvev}, \eq{doubleton} and the overlap condition
\eq{overlapX}, which implies that
$\pi(L)(1)\ket{\1}_{12}=L^{-1}(2)\ket{\1}_{12}$ so that $L^{-1}$
acts in the diagonal representation where
$M_{AB}=M_{AB}(1)+M_{AB}(2)$ and
\bea \ket{\k}_{12}&=& T^{(s)}_{\bf \k}(1)\ket{\1}_{12}\ ,\qquad
\ket{T_{a(s+k),b(s)}}\ =\ T_{a(s+k),b(s)}(1)\ket{\1}_{12}\
.\label{test}\eea
The harmonic functions obey the Bianchi identity \eq{BI} and the
mass-shell condition \eq{massshell}.

In Euclidean signature, the bilinear forms $(\cdot,\cdot)_{\mho}$
and $(\cdot,\cdot)_{\cal T}$ are \emph{positive definite} for
$\s=-1$ and $\s=+1$, respectively. Thus, the $\mm$-covariant
expansion of $\Phi$ around $H_D$ provides a unitarizable
$\mho$-module. In Lorentzian signature, the unitarizable
$\mho$-modules arise in $\Phi$ for both $\s=+1$ ($AdS_D$) and
$\s=-1$ ($dS_D$) upon going to the corresponding compact bases of
${\cal T}$. The former case will be examined next and the latter
case is examined in Section \ref{Sec:WS}.

\scss{Unitarity of the harmonic expansion on $dS_D$}

The the real form $\mso(1,D)$ with Lorentz algebra $\mso(1,D-1)$
and transvections obeying $[P_a,P_b]_\star=-iM_{ab}$, splits into
$\mh\ssum \ml$ where the maximal compact subalgebra $\mh=\mso(D)'$
is generated by $J_{mn}=(M_{rs},P_r)$ obeying
$[J_{mn},J_{pq}]_\star=4i\delta_{[p|[n}J_{m]|q]}$, and $\ml={\rm
span}_\Real K_m$ where $K_m=(M_{r0},P_0)$ obeys
$[K_m,K_n]_\star=iJ_{mn}$. We can define the twisted-adjoint
compact-weight module ${\cal M}(-,+)={\cal M}^{(+)}(-,+)\oplus
{\cal M}^{(-)}(-,+)$, where ${\cal M}^{(\pm)}(-,+)$ are
$\mho$-irreps that decompose under $\mso(D,1)$ into
\bea {\cal M}^{(\pm)}(-,+)&=& \bigoplus_{s=0}^\infty {\cal
M}^{(\pm)}_{(s)}(-,+)\ ,\qquad {\cal M}^{(\pm)}_{(s)}(-,+)\ =\
\bigoplus_{k=0}^\infty {\cal M}^{(\pm)(s)}_{(s+k,s)}\ ,\eea
where ${\cal M}^{(\pm)}_{(s)}(-,+)$ consists of generalized
elements in ${\cal T}_\ell$, $s=2\ell+2$, given by regular series
expansions, and ${\cal M}^{(\pm)(s)}_{(s+k,s)}$ are traceless
type-$(s+k,s)$ tensors of $\mso(D)'$. These decompose further
under $\ms$ as ${\cal
M}^{(\pm)(s)}_{(s+k,s)}=\bigoplus_{s+k\geqslant j_1\geqslant
s\geqslant j_2\geqslant0} {\cal M}^{(\pm)(s)}_{(s+k,s|j_1,j_2)}$
where the traceless type-$(j_1,j_2)$ basis elements
\bea T^{(\pm)(s)}_{(s+k,s|j_1,j_2)}&=& \sum_{n=0}^\infty
f^{(\pm)(s)}_{(s+k,s|j_1,j_2);n}T^{(s)}_{(j_1,j_2);n}\ ,\quad
\left[T^{(s)}_{(j_1,j_2);n}\right]_{r(j_1),t(j_2)}\ =\ T_{0(n)\{
r(j_1),t(j_2)\} 0(s-j_2)}\ ,\qquad\label{Tsj1j2}\eea
with $T_{a(m),b(n)}$ defined in \eq{Tambn2}
and$f^{(\pm)(s)}_{(s+k,s|j_1,j_2);n}\in\Real$ determined by the
above embedding conditions\footnote{An $\mso(D)'$-tensor of
type-$(s_1,s_2)$ decomposes under $\ms$ into type-$(j_1,j_2)$
tensors with $s_1\geqslant j_1\geqslant s_2\geqslant j_2\geqslant
0$, and can thus be a (generalized) element of ${\cal T}_{\ell}$
for $s=2\ell+2$ with expansion of the form given in \eq{Tambn2}
only if $s_2=s$. For $D=4$ the type-$(j_1,j_2)$ tensors with
$j_2\geqslant 2$ are trivial, since an irreducible
$\mso(N;\Comp)$-tensor is trivial if the sum of the heights of the
first two columns in its Young diagram exceeds $N$.}. For example,
using
\bea \widetilde P_{\{r|} T_{0(n)|r(k)\}}&=& 2T_{0(n)\{r(k+1)\}}+
{n(n-1)(n+k+1)(n+k+2\e_0-1)\over 8
(n+k+\e_0-\ft12)(n+k+\e_0+\ft12)}T_{0(n-2)\{r(k+1)\}}\ ,\qquad\eea
for $s=0$ one finds that the generating functions
$f^{(\pm)(0)}_{(k|k)}(z)=\sum_{n=0}^\infty
f^{(\pm)(s)}_{(k|k);n}z^n$ of the ``top'' elements
$T^{(\pm)(0)}_{(k|k)}$ are given up to overall constants by
\bea f^{(+)(0)}_{(0|0)}(z)&=& {}_2F_3\left[\ft{k+\e_0+\ft32}2,\ft{k+\e_0+\ft52}2;\ft12,\ft{k+3}2,\ft{k+2\e_0+1}2;-4z^2\right]\ ,\\[5pt] f^{(-)(0)}_{(0|0)}(z)&=& z\, {}_2F_3\left[\ft{k+\e_0+\ft32}2,\ft{k+\e_0+\ft52}2;\ft32,\ft{k+4}2,\ft{k+2\e_0+2}2;-4z^2\right] \ ,\eea
that simplify in $D=4$ to $f^{(+)(0)}_{(k|k)}(z)=\cos 4z$ and
$f^{(-)(0)}_{(k|k)}(z)=\ft14 \sin 4z$. The $\widetilde\mho$ action
on $T^{(\pm)(0)}_{(0|0)}$ fill out the spaces ${\cal
M}^{(\pm)}(-,+)$. By a choice of normalization, the
twisted-adjoint $\mso(D,1)$ representation matrix in ${\cal
M}^{(\pm)}_{(s)}(-,+)$ takes the twisted-adjoint form $\widetilde
K T^{(\pm)(s)}_{(s+k|s)}=2T^{(\pm)(s)}_{(s+k+1|s)}+2\l_k^{(s)}
\mathbf P_{(s+k,s)}\left[\delta\otimes
T^{(\pm)(s)}_{(s+k-1|s)}\right]$ with $\l_k^{(s)}$ given by
\eq{lambda}. These $\mso(D,1)$ modules are unitarizable in the
bilinear inner product
\bea (S,S')_{{\cal M}^{(\pm)}_{(s)}(-,+)}&\equiv & {1\over {\cal
N}^{(\pm)}_{(s)}}\Tr(\pi(S)\star S')\ ,\eea
defined by the analog of the prescription that we shall give in
detail later on, below \eq{inner}, with
$\Tr(\pi(T^{(\pm)(s)}_{(s,s|s,s)})\star
T^{(\pm)(s)}_{(s,s|s,s)})={\cal N}^{(\pm)}_{(s)}\mathbf
P_{(s,s)}$. For example, for $s=0$ it follows from
$\pi((T^{(\e)(0)}_{(0|0)})^\dagger)=\pi(T^{(\e)(0)}_{(0|0)})=\e
T^{(\e)(0)}_{(0|0)}$ and $\pi(\widetilde\adj_{Q}
S)^\dagger)=-\widetilde\adj_{Q^\dagger}\pi(S^\dagger)$ that the
real forms are given by $S^{(\e)}=\sqrt{\e}\sum_{k}i^k
S^{(\e)(0)}_{(k)}T^{(\e)(0)}_{(k)}$ with
$S^{(\e)(0)}_{(k)}\in\Real$, and hence\footnote{The harmonic
expansion of $(\nabla^2-M^2)\phi=0$ in $dS_D$ yields a
compact-weight module ${\cal M}_{(0)}(M^2)$ with
$C_2[\mso(D,1)|M^2]=-M^2$ and representation matrix where
$\l_k^{(0)}(M^2)={k\over
8(k+\e_0+\ft12)}(k^2+2\e_0k-1-2\e_0+M^2)$. For $M^2>0$ this module
is irreducible and unitarizable. If $M^2=M^2_p\equiv
-(p+2\e_0+1)(p-1)$ then it follows from $\l_{k}^{(0)}(M^2_p)>0$
for $k>p$, and $\l_{p}^{(0)}(M^2_p)=0$, that ${\cal
M}_{(0)}(M^2_p)$ contains the unitarizable invariant subspace
${\cal M}_{(0)}(M^2_p)'=\bigoplus_{k\geqslant p}{\cal
M}^{(0)}_{(k)}(M^2_p)$.}
\bea (S,S')_{{\cal M}^{(\pm)}_{(0)}(-,+)}&=&\pm\sum_{p,q}\left[S^{(\pm)(0)}_{(p)}\right]^{m(p)}{2^{-p}\over {\cal N}^{(\pm)}_{(0)}}\Tr[S^{(\pm)(0)}_{(0)}\star (\widetilde K)^p_{m(p)}(\widetilde K)^q_{n(q)} S^{(\pm)(0)}_{(0)}]\left[S^{(\pm)(0)\prime}_{(q)}\right]^{n(q)}\nn\\[5pt]
&=&\pm \sum_p N_p ~S^{(\pm)}_{(p)}\cdot S^{(\pm)(0)\prime}_{(p)}\
,\qquad N_p\ =\ \prod_{k=0}^p \l_k^{(0)}\ ,\eea
with $(\widetilde K)^p_{m(p)}=\widetilde K_{\{m_1}\cdots
\widetilde K_{m_p\}}$, that are manifestly positive or negative
definite.


\scs{On the Harmonic Expansion on $AdS_D$}\label{Sec:WS}


We first define the twisted-adjoint compact-weight modules ${\cal
M}(+,+|\mu)$ of $\mso(2,D-1)$ consisting of basis elements
$T^{(s)}_{e;(j_1,j_2)}$ with energy $e\in\integ+\mu$ and spin
$(j_1,j_2)$, given by \emph{regular} series expansions in the
Lorentz covariant basis, and corresponding to harmonic functions
that are \emph{finite} in the interior of $CAdS_D$. In the case of
$\mu=0$ we propose that the even and odd submodules ${\cal
M}^{(\pm)}(+,+|0)$ are generated by $\widetilde\mho$ from the
\emph{static ground states} $T^{(0)}_{0;(0)}$ and
$T^{(0)}_{0;(1)}$. Then, in terms of \emph{one-sided} ${\cal
A}$-modules ${\cal S}^{(s)}_{e;(j_1,j_2)}$ and their duals, we
identify the Flato-Fronsdal factorization of a subsector of ${\cal
M}(+,+|0)$, and also the complete factorization of ${\cal
M}^{(+)}(+,+|0)$ in terms of ${\cal S}^{(0)}_{0;(0)}$ and its
dual. These fill wedges in compact-weight space and we shall refer
to them as \emph{angletons}, as opposed to the singletons that
fill single lines. The angletons contain (one-sided)
$\mso(D-1)$-submodules with \emph{negative spin}, and the
factorization is given by a direct product modulo an equivalence
relation (see \eq{statsing}). We then proceed to identifying the
standard composite-massless spaces $\mD^\pm(\pm(s+2\e_0);(s))$ as
invariant subspaces of ${\cal M}_{(s)}(+,+|0)$. Their complements
${\cal W}_{(s)}$ are (composite-massless) \emph{lowest-spin
modules} in the sense that they are unbounded in ordinary weight
space, \emph{i.e.} the energy and spin eigenvalues are not bounded
from above nor below, while the spins $(j_1,j_2)$ obey
$j_1\geqslant j_2\geqslant 0$. We propose that the $\Tr$ on ${\cal
A}$ induces norms (\emph{i.e.} bilinear forms with definite
signatures) on ${\cal W}_{(s)}$, as we shall verify explicitly in
the case of ${\cal W}^{(+)}_{(0)}$. Finally, we describe the map
from ${\cal M}^{(s)}(+,+)$ to ${\cal T}_\ell$ as a decomposition
of the reflector $\ket{\1}_{12}$. As by-product, we find a natural
generalization of the Flato-Fronsdal formula whereby the adjoint
representation $\mho$ is identified as the direct product between
singletons and anti-singletons (see \eq{FFformulatwisted}).


\scss{Twisted-adjoint compact-weight module}\label{Sec:LSM}


The twisted-adjoint $\mso(2)\oplus \mso(D-1)$-covariant modules
${\cal M}(+,+|\mu)$, $\mu\in[0,1]$, are defined by
\bea {\cal M}(+,+|\mu)&=&\bigoplus_{s=0}^\infty {\cal
M}_{(s)}(+,+|\mu)\ ,\qquad
{\cal M}_{(s)}(+,+|\mu)\ =\ \!\!\!\!\!\!\!\!\bigoplus_{\tiny\ba{c}e-\mu\in \integ\\
j_1\geqslant s\geqslant j_2\geqslant
0\ea}\!\!\!\!\!\!\!\!\Comp\otimes T^{(s)}_{e;(j_1,j_2)}\
,\label{calMs}\eea
where the $\mg$ submodules ${\cal M}_{(s)}(+,+|\mu)$ consist of
\emph{regular} elements $T^{(s)}_{e;(j_1,j_2)}$ with spin
$(j_1,j_2)$ and energy $e\in\Real$, that is
\bea \left[T^{(s)}_{e;(j_1,j_2)}\right]_{r(j_1),t(j_2)}&=&
\sum_{n=0}^\infty f^{(s)}_{e;(j_1,j_2);n}
\left[T^{(s)}_{(j_1,j_2);n}\right]_{r(j_1),t(j_2)}\
,\label{Ts}\eea
where $T^{(s)}_{(j_1,j_2);n}\in{\cal A}$ are the same as in
\eq{Tsj1j2}, and the generating functions
$f^{(s)}_{e;(j_1,j_2)}(z)=\sum_{n=0}^\infty
f^{(s)}_{e;(j_1,j_2);n} z^n$ are determined \emph{uniquely} from
\bea \ac_E T^{(s)}_{e;(j_1,j_2)}&=&e~T^{(s)}_{e;(j_1,j_2)}\
,\qquad f^{(s)}_{e;(j_1,j_2);0}\ =\ 1\ ,\label{TE}\eea
as can be seen from \eq{TPa} which implies $\ac_E
(T^{(s)}_{(j_1,j_2);n})=\l^{(s)}_{(j_1,j_2);n}T^{(s)}_{(j_1,j_2);n+1}+
\l^{\prime(s)}_{(j_1,j_2):n}T^{(s)}_{(j_1,j_2);n-1}$ where the
coefficients are non-vanishing except
$\l^{\prime(s)}_{0;(j_1,j_2)}=0$. It also follows that
$f^{(s)}_{e;(j_1,j_2);n}\in\Real$ which together with
$(T^{(s)}_{(j_1,j_2);n})^\dagger=T^{(s)}_{(j_1,j_2);n}$ implies
that $(T^{(s)}_{e;(j_1,j_2)})^\dagger= T^{(s)}_{e;(j_1,j_2)}$.
Moreover, $\pi(T^{(s)}_{e;(j_1,j_2)})=
(-1)^{j_1-s}T^{(s)}_{-e;(j_1,j_2)}$, that is $\pi:~{\cal
M}_{(s)}(+,+|\mu)\ \longrightarrow {\cal M}_{(s)}(+,+|1-\mu)$, and
$f^{(s)}_{-e;(j_1,j_2)}(z)= f^{(s)}_{e;(j_1,j_2)}(-z)$. The ideal
relations imply that
\bea \widetilde L^\pm_{r}
\left[T^{(s)}_{e;(s,j_2)}\right]_{rt(s-1),u(j_2)}&=&0\qquad
\mbox{for $j_1=s\geqslant 1$ and $j_2<s$}\ ,
\label{Mdiv}\\[5pt] \mathbf P_{\{j_1,j_2,1\}}\left[\widetilde L^\pm_{u}
\left[T^{(s)}_{e;(j_1,j_2)}\right]_{r(j_1),t(j_2)}\right]&=&0\qquad
\mbox{for $j_2\geqslant 1$}\ ,\label{Mcurl}\eea
and from $\widetilde
C_{2n}[\mg]\left(T^{(s)}_{e;(j_1,j_2)}\right)=\sum_{p=0}^\infty
f^{(s)}_{e;(j_1,j_2)} \ft12 \widetilde M_{A_1}{}^{A_2}\cdots
\widetilde M_{A_{2n}}{}^{A_1}\left(T^{(s)}_{p;(j_1,j_2)}\right)$
it follows that
\bea \widetilde C_{2n}[\mg|{\cal M}_{(s)}]&=& C_{2n}[\ell]\
,\label{C2nMell}\eea
where $s=2\ell+2$ and $C_2[\ell]$ and $C_4[\ell]$ are given in
\eq{c2ell}. The space ${\cal M}(+,+|\mu)$ and its subspaces ${\cal
M}_{(s)}(+,+|\mu)$ decompose under the $\widetilde \mho$ and
$\widetilde \mg$ actions into \emph{even and odd submodules}
${\cal M}(+,+|\mu)={\cal M}^{(+)}(+,+|\mu)\oplus {\cal
M}^{(-)}(+,+|\mu)$ where
\bea {\cal M}^{(\pm)}(+,+|\mu)&=&\!\!\!\!\bigoplus_{\tiny\ba{c}e;(j_1,j_2)\\
e-\mu+j_1+j_2\\=\ft12 (1\mp 1)\ \mbox{mod
$2$}\ea}\!\!\!\!\Comp\otimes T^{(s)}_{e;(j_1,j_2)}\ .\eea
The Weyl zero-form $\Phi$ (obeying $\Phi^\dagger=\pi(\Phi)$) can
be expanded as
\bea \Phi&=&\int d\mu ~\Phi(\mu)\ ,\qquad \Phi(\mu)\ =\ \sum_s \Phi_{(s)}(\mu)\ ,\\[5pt]\Phi_{(s)}(\mu)&=&\!\!\!\!
\sum_{\tiny \ba{c}e-\mu\in\integ\\j_1\geqslant s\geqslant
j_2\geqslant 0\ea }\sum_{k=j_1-s}^\infty T_{a(s+k),b(s)}
S^{(s)}_{e;(j_1,j_2)}(\mu) D^{(s);a(s+k),b(s)}_{e;(j_1,j_2)}\
,\label{harmexp}\eea
where $S^{(s)}_{e;(j_1,j_2)}(\mu)\in\Comp$ obey the reality
condition
\bea
\left(S^{(s)}_{e;(j_1,j_2)}(\mu)\right)^\ast&=&(-1)^{j_1-s}S^{(s)}_{-e;(j_1,j_2)}(1-\mu)\
,\eea
and the generalized harmonic functions ($k\geqslant j_1-s$)
\bea
\left[D^{(s);a(s+k),b(s)}_{e;(j_1,j_2)}\right]_{r(j_1),t(j_2)}&=&
{\cal N}_{s,k}^{-1} ~\Tr\left[T^{a(s+k),b(s)}\star L^{-1}\star
\left[T^{(s)}_{e;(j_1,j_2)}\right]_{r(j_1),t(j_2)}\star
\pi(L)\right]\nn\\[5pt]&=&
{\cal N}_{s,k}^{-1} {}_{12}\bra{T_{a(s+k),b(s)}} L^{-1}
\ket{(s);e;(j_1,j_2)}_{12;r(j_1),t(j_2)}\ ,\eea
where $\ket{(s);e;(j_1,j_2)}_{12}=
T^{(s)}_{e;(j_1,j_2)}(1)\star\ket{\1}_{12}$. At $L=1$ the overlaps
are finite and given by
\bea
\left[D^{(s);a(s+k),b(s)}_{e;(j_1,j_2)}\right]_{r(j_1),t(j_2)}|_{L=1}=
\d^{\{a(s+k),b(s)\}_D}_{\{0(n)\{r(j_1),t(j_2)\}_{D-1}0(s-j_2)\}_D}
f^{(s)}_{e;(j_1,j_2);n}\ ,\quad n=s+k-j_1\ ,&&\qquad\eea
where $\{\cdots\}_{D}$ and $\{\cdots\}_{D-1}$, respectively,
denote $\mso(D)$ and $\mso(D-1)$ traceless Young projections.

In what follows we shall focus on the case $\mu=0$, and we shall
therefore write ${\cal M}\equiv{\cal M}(+,+|0)$, ${\cal
M}^{(\pm)}\equiv{\cal M}^{(\pm)}(0)$ and ${\cal
M}^{(\pm)}_{(s)}\equiv{\cal M}^{(\pm)}_{(s)}(0)$.


\scss{Static ground states}


We propose that ${\cal M}_{(s)}^{(\pm)}$ are generated by
$\widetilde {\cal U}[\mg]$ from the elements with $e=0$ and
minimal $j_1+j_2$, namely the \emph{static ground states}
\bea s=0&:&\quad T^{(0)}_{(\pm)}\ =\ T^{(0)}_{0;(\sigma_\pm)}\
;\qquad\qquad s>0\ :\quad T^{(s)}_{(\pm)}\ =\
T^{(s)}_{0;(s,\sigma_\pm)}\
,\label{scalarstatic}\label{hsstatic}\eea
where $\sigma_\pm=(1\mp 1)/2$. Furthermore, we propose that
$T^{(s)}_{(\pm)}$ with $s>0$ are generated by $\widetilde {\cal
U}[\mho]$ action from $T^{(0)}_{(\pm)}$, so that
\bea {\cal M}_{(s)}^{(\pm)}&=& {\cal U}[\widetilde\mg]
T^{(s)}_{(\pm)}\ ,\qquad {\cal M}^{(\pm)}\ =\ {\cal
U}[\widetilde\mho] T^{(0)}_{(\pm)}\ .\label{tilde}\eea
As shown in Appendix \ref{App:T0}, the generating functions of
$T^{(0)}_{(\pm)}$ are given by
\bea f^{(0)}_{0;(0)}(z)&=&\sum_{p=0}^\infty
{(4z)^{2p}(\e_0+\ft32)_{2p}\over (2)_{2p}(2\e_0+1)_{2p}}\ =\ {}_2
F_3\left({2\e_0+3\over 4},{2\e_0+5\over 4};
\ft32,\e_0+\ft12,\e_0+1;4z^2\right)\ ,\label{scalarf000plus}\\[5pt] f^{(0)}_{0;(1)}(z)&=&
\sum_{p=0}^\infty {(\e_0+\ft52)_{2p}\,z^{2p}\over
p!(2)_{p}(\e_0+1)_{p}(\e_0+2)_{p}}\ =\ {}_2 F_3\left({2\e_0+5\over
4},{2\e_0+7\over 4}; 2,\e_0+1,\e_0+2;4z^2\right)\
.\hspace{1cm}\label{scalarf000minus} \eea
In $D=4$ these functions are\footnote{The function
$f^{(0)}_{0;(0)}(z)$ corresponds to the twisted-adjoint element
representing the static and rotationally invariant scalar-field
profile discussed in \cite{Sezgin:2005pv}.}
\bea f^{(0)}_{0;(0)}(z)&=& {\sinh 4z\over 4z}\ ,\qquad
f^{(0)}_{0;(1)}(z)\ =\ {3\over 16 z^2}\left(\cosh 4z-{\sinh
4z\over 4z}\right)\ . \eea
The $\widetilde\mg$-action in ${\cal M}$ obeys relations of the
form
\bea &&\left(\widetilde L^\pm_t\widetilde L^\mp_t-{\cal
\mu}^{(s)}_{\pm e;(j_1,j_2)}\right) T^{(s)}_{e;(j_1,j_2)}\ =\
\left(\widetilde x^\pm\widetilde x^\mp - {\cal
\mu}^{\prime(s)}_{\pm e;(j_1,j_2)}\right)T^{(s)}_{e;(j_1,j_2)}\ =\ 0\ ,\label{deg1}\\[5pt]
&& \left(\widetilde x^\pm \widetilde L^\mp_{\{r_1}\widetilde
L^\mp_{r_2\}} -{\cal \mu}^{\prime\prime(s)}_{\pm
e;(j_1,j_2)}\widetilde L^\pm_{\{r_1}\widetilde
L^\mp_{r_2\}}\right)T^{(s)}_{e;(j_1,j_2)}\ =\ 0\ ,\qquad
\widetilde x^\pm\ =\ \widetilde L^\pm_r\widetilde L^\pm_r\
,\label{deg3}\eea
where ${\cal \mu}^{(s)}_{e;(j_1,j_2)}$, ${\cal
\mu}^{\prime(s)}_{e;(j_1,j_2)}$,${\cal
\mu}^{\prime\prime(s)}_{e;(j_1,j_2)}$$\in\Real$. To factor out
these relations we first write ${\cal M}^{(+)}_{(s)}$$=$${\cal
M}^{(+)>}_{(s)}\cup{\cal M}^{(+)0}_{(s)}\cup {\cal
M}^{(+)<}_{(s)}$ with
\bea {\cal M}^{(+)~{}^{>}\!\!\!\!\!{}_{<}}_{(s)}&=&
\{T^{(s)}_{e;(j_1,j_2)}:\ \pm e
>j_1+j_2-s\}\ ,\quad {\cal M}^{(+)0}_{(s)}\ =\ \{T^{(s)}_{e;(j_1,j_2)}:\
|e|\leqslant j_1+j_2-s\}\ .\qquad\label{M><}\eea
Thus, according to our proposal, there exist \emph{non-vanishing}
coefficients ${\cal C}^{(s)}_{e;(j_1,j_2)}$ such that
\bea {\cal M}^{(+)~{}^{>}\!\!\!\!\!{}_{<}}_{(s)}&:&
\left[T^{(s)}_{e;(j_1,j_2)}\right]_{r(j_1),t(j_2)}\ =\ {\cal
C}^{(s)}_{e;(j_1,j_2)} (\widetilde x^\pm)^p \widetilde
L^{\pm(j_1-s)}_{\{r(j_1-s)}\widetilde L^{\pm(j_2)}_{t(j_2)}
\left[T^{(s)}_{0;(s,0)}\right]_{r(s)\}}\
,\label{stategeneration1}\eea
for $p_\pm=\ft12(\pm e+s-j_1-j_2)$ and where $\widetilde
L^{\pm(n)}_{\{r(n)}=\widetilde L^+_{\{r_1}\cdots \widetilde
L^+_{r_n\}}$, and such that
\bea{\cal M}^{(+)0}_{(s)}&:&
\left[T^{(s)}_{e;(j_1,j_2)}\right]_{r(j_1),t(j_2)}\ =\ {\cal
C}^{(s)}_{e;(j_1,j_2)} (\widetilde L^{+(q_+)}\widetilde
L^{-(q_-)})_{\{r_1\cdots r_{j_1-s} t_1\cdots t_{j_2}}
\left[T^{(s)}_{0;(s,0)}\right]_{r(s)\}}\
,\label{stategeneration2}\eea
for $q_\pm=\ft12(j_1+j_2-s\pm e)$ and where $(\widetilde
L^{+(m)}\widetilde L^{-(n)})_{r_1\cdots r_{m+n}}=\widetilde
L^+_{\{r_1}\cdots \widetilde L^+_{r_m}\widetilde
L^-_{r_{m+1}}\cdots \widetilde L^-_{r_{m+n}\}}$. In particular,
for $s=0$ one has
\bea T^{(0)}_{e+2;(0)}&=&{\cal C}_{e}\widetilde x^+~
T^{(0)}_{e;(0)}\ ,\qquad
{\cal C}_e\ =\ -{1\over (e+2\e_0)(e+2)}\ ,\label{Ce}\\[5pt]
\widetilde L^{-}_r T^{(0)}_{e;(0)}&=&{\cal C}'_e \widetilde L^+_r
T^{(0)}_{e-2;(0)}\ ,\qquad {\cal C}'_e\ =\ {1\over {\cal
C}'_{2-e}}\ =\ -{(e-2\e_0)(e-2)\over e(e+2\e_0-2)}\
,\label{tildeCe}\eea
which implies $\mu^{(0)}_{e;(0)}=(e-2\e_0)(e-2)$,
$\mu^{\prime(0)}_{e;(0)}=(e-2\e_0)(e-2)e(e+2\e_0-2)$, and
\bea (\widetilde L^+_r+\widetilde L^-_r) f(\widetilde x^+)
T^{(0)}_{e;(0)}\ =\ (D_e f)(\widetilde x^+) T^{(0)}_{e;(0)}\
,\label{lpluslminus}\eea
for differentiable functions $f(\widetilde x^+)$ and with
\bea D_e&=& 4\widetilde x^+{d^2\over d(\widetilde
x^+)^2}+4(e-\e_0){d\over d\widetilde x^+}+1+{(e-2\e_0)(e-2)\over
\widetilde x^+}\ .\label{De}\eea
For even $s=2p\geqslant 2$ the static ground states are of the
form
\bea \left[T^{(2p)}_{0;(2p)}\right]_{r(2p)}&=& \sum_{n=0}^{p-1}
\x_{2p;n} \widetilde L^+_{\{r_1}\widetilde L^-_{r_2}\cdots
\widetilde L^+_{r_{2n-1}}\widetilde L^-_{r_{2n}}\widetilde
Q_{r(2p-2n)\}} T^{(0)}_{0;(0)}\
,\label{gen1}\\[5pt]\left[T^{(2p)}_{0;(2p,1)}\right]_{r(2p),s}&=&
\sum_{n=0}^{p-1} \x'_{2p;n} \widetilde L^+_{\{r_1}\widetilde
L^-_{r_2}\cdots \widetilde L^+_{r_{2n-1}}\widetilde
L^-_{r_{2n}}\widetilde
Q_{r(2p-2n)}\left[T^{(0)}_{0;(1)}\right]_{s\}}\ ,\label{gen2}\eea
for some $\x_{2p;n},\x'_{2p;n}\in \Real$ and $Q_{r(2n)}=
L^+_{\{r_1}\star L^-_{r_2}\star\cdots \star L^+_{r_{2n-1}}\star
L^-_{r_{2n}\}}\in\mho$. Similarly, for odd spin $s=2p+1\geqslant
3$ one has
\bea \left[T^{(2p+1)}_{0;(2p+1)}\right]_{r(2p+1)}&=&
\sum_{n=0}^{p-1} \x_{2p+1;n} \widetilde L^+_{\{r_1}\widetilde
L^-_{r_2}\cdots \widetilde L^+_{r_{2n-1}}\widetilde
L^-_{r_{2n}}\widetilde Q_{r(2p-2n)}
\left[T^{(1)}_{0;(1)}\right]_{r_{2p+1}\}}\
,\label{gen3}\\[5pt]\left[T^{(2p+1)}_{0;(2p+1,1)}\right]_{r(2p+1),s}&=&
\sum_{n=0}^{p-1} \x'_{2p+1;n} \widetilde L^+_{\{r_1}\widetilde
L^-_{r_2}\cdots \widetilde L^+_{r_{2n-1}}\widetilde
L^-_{r_{2n}}\widetilde
Q_{r(2p-2n)}\left[T^{(1)}_{0;(1,1)}\right]_{r_{2p+1},s\}}\
,\label{gen4}\eea
for some $\x_{2p+1;n},\x'_{2p+1;n}\in \Real$, and where
$T^{(1)}_{(\pm)}$ in their turn can be generated from the scalar
static ground states. For example, to generate $T^{(1)}_{0;(1)}$
from $T^{(0)}_{0;(1)}$ one may use $\widetilde
\adj_{EM_{rs}}\left[T^{(0)}_{0;(1)}\right]_{t}=
\left\{EM_{rs},\left[T^{(0)}_{0;(1)}\right]_{t}\right\}=
\d_{t[s}\left[T^{(1)}_{0;(1)}\right]_{r]}$, as follows from
$EM_{rs}=E\star M_{rs}=M_{rs}\star E$ and $E\star
\left[T^{(0)}_{0;(1)}\right]_{r}=-\left[T^{(0)}_{0;(1)}\right]_{r}\star
E\ =\ \ft12\adj_E \left[T^{(0)}_{0;(1)}\right]_{r}=-\ft{i}2
\left[T^{(1)}_{0;(1)}\right]_{r}$. The generation of
$T^{(1)}_{0;(1,1)}$ from $T^{(0)}_{0;(0)}$ is more involved since
$E\star T^{(0)}_{0;(0)}=\ft12\adj_E\, T^{(0)}_{0;(0)}=0$. For
example, two $\widetilde\mg$ transformations send
$T^{(0)}_{0;(0)}$ into $T^{(0)}_{0;(2)}$, which
$\widetilde\adj_{EM_{rs}}$ maps to $T^{(1)}_{0;(2)}$, from which
two $\widetilde\mg$ transformations lead down to
$T^{(1)}_{0;(1,1)}$.


\scss{Factorization in terms of singletons and
angletons}\label{Sec:Fact}


The element $T^{(s)}_{e;(j_1,j_2)}$ gives rise to a separate
${\cal A}$ left and right modules
\bea {\cal S}^{(s)}_{e;(j_1,j_2)}&=& {\cal A}\star
T^{(s)}_{e;(j_1,j_2)}\ ,\qquad {\cal S}^{(s)\ast}_{e;(j_1,j_2)}\
=\ T^{(s)}_{e;(j_1,j_2)}\star {\cal A}\ ,\label{calsc}\eea
which are subspaces of ${\cal M}^{(\pm)}(+,+|\mu)$ for
$e-\mu+j_1+j_2=\ft12(1\mp 1)$ mod $2$. According to \eq{tilde} an
element $S\in{\cal M}^{(\pm)}$ can be written as
$S=\sum_{X,X'\in{\cal A}} X\star T^{(0)}_{(\pm)}\star X'$. Thus,
the twisted-adjoint compact-weight modules can be factorized as
follows:
\bea {\cal M}^{(\pm)}&=& ({\cal S}^{(\pm)}\otimes {\cal
S}^{(\pm)\ast})/\sim\ ,\qquad {\cal S}^{(\pm)}\ =\ {\cal
S}^{(0)}_{0;(\s_\pm)}\ ,\quad {\cal S}^{(\pm)\ast}\ =\ {\cal
S}^{(0)\ast}_{0;(\s_\pm)}\ ,\label{statsing}\eea
where we shall refer to the factors as \emph{angletons}, and the
equivalence relation reads
\bea (X\star T^{(0)}_{(\pm)})\otimes (T^{(0)}_{(\pm)}\star Y)\sim
(X'\star T^{(0)}_{(\pm)})\otimes (T^{(0)}_{(\pm)}\star Y')~~
\Leftrightarrow~~ X\star T^{(0)}_{(\pm)}\star Y\ =\ X'\star
T^{(0)}_{(\pm)}\star Y'\ .\qquad\label{simrelation}\eea
The ideal relations $V_{AB}\approx 0$ and $C_2[\mg|{\cal
M}^{(\pm)}]\approx -\e_0(\e_0+2)$ imply that
\bea &&L^+_r\star L^+_r\ \approx \ L^-_r\star L^-_r\ \approx\ 0\ ,\qquad M_{rs}\star L^\pm_s\ \approx\ i(\e_0\pm E)\star L^\pm_r\ ,\\[5pt]
&&M_{\{r_1}{}^t\star M_{r_2\}t}\ \approx\ \ft12\{L^+_{\{r_1},L^-_{r_2\}}\}_\star\ ,\qquad L^+_{[r}\star L^-_{s]}\ \approx\ i(1-E)\star M_{rs}\ ,\label{tracem2}\\[5pt] &&\ft12 M^{rs}\star M_{rs}\ \approx\ E\star E-\e_0^2\ ,\qquad
\{L^+_r,L^-_r\}_\star\ \approx\ 4(E\star E+\e_0)\
,\label{d-1cas}\eea
which together with $V_{ABCD}\approx 0$ show that ${\cal A}$ has
the compact basis
\bea {\cal A}&=& \!\!\!\!\bigoplus_{\tiny
\ba{c}e\in\integ,~n\geqslant 0
\\ j_1\geqslant j_2\geqslant 0\\ |e|\leqslant j_1-j_2\ea}\!\!\!\!\Comp\otimes
M_{e;(j_1,j_2);n}\ ,\label{Xs}\eea
where $\left[M_{e;(j_1,j_2);n}\right]_{r(j_1),s(j_2)}=
L^{+(p_+)}_{\{r(p_+)} L^{-(p_-)}_{r(p_-)} M^{(j_2)}_{r(j_2)
s(j_2)\}} E^{n}$, $p_\pm=\ft12(j_1\pm e-j_2)$, has adjoint energy
$e$ and spin $(j_1,j_2)$, and $L^{\pm(p)}_{r(p)}\equiv
L^\pm_{r_1}\cdots L^\pm_{r_p}$, $M^{(p)}_{r(p)s(p)}\equiv M_{r_1
s_1}\cdots M_{r_p s_p}$. Since $T^{(0)}_{e;(0)}$ are series
expansions in $T_{0(n)}$, which are $\star$-polynomials in $E$ of
order $n$ (as follows from \eq{estaren}), it follows that
$[M_{0;(j_1,j_2);p},T^{(0)}_{e;(0)}]_\star=0$. In particular,
$\adj_E T^{(0)}_{e;(0)}=0$, which together with $\ac_E
T^{(0)}_{e;(0)}= e T^{(0)}_{e;(0)}$ and \eq{d-1cas} yields
\bea E\star T^{(0)}_{e;(0)}&=& T^{(0)}_{e;(0)}\star E\ =\ {e\over 2} T^{(0)}_{e;(0)}\ ,\label{Elemma}\\[5pt]
C_2[\ms]\star T^{(0)}_{e;(0)}&=& \nu_e(\nu_e+2\e_0) T^{(0)}_{e;(0)}\ ,\qquad \nu_e\ =\ \frac{|e|}2 -\e_0\ ,\label{Cslemma}\\[5pt]
\{L^+_r,L^-_r\}_\star \star T^{(0)}_{e;(0)}&=& \mu_e
T^{(0)}_{e;(0)}\ ,\qquad \mu_e\ =\ e^2+4\e_0\ .\label{mue}\eea
Thus ${\cal S}^{(0)}_{e;(0)}$ is spanned by the elements
$M_{e';(j'_1,j'_2);0}\star T^{(0)}_{e;(0)}$ as can be seen by
re-ordering $M_{e';(j'_1,j'_2);p}=\sum_{n=0}^p {\cal
C}^{p,n}_{e';(j'_1,j'_2)}M_{e';(j_1,j_2);0}\star E^{n}$, where
${\cal C}^{p,n}_{e';(j'_1,j'_2)}$ are finite coefficients, which
implies that $M_{e';(j'_1,j'_2);p}\star T^{(0)}_{e;(0)}=
\sum_{n=0}^p {\cal C}^{p,n}_{e';(j'_1,j'_2)} (e')^{n}
M_{e';(j'_1,j'_2);0}\star T^{(0)}_{e;(0)}$. In particular, the
even angleton
\bea {\cal S}^{(+)}&=& \bigoplus_{\tiny\ba{c} j_1\geqslant j_2\geqslant 0\\
|e|<j_1-j_2\ea} \Comp\otimes T_{e;(j_1,j_2)}\ , \qquad
T_{e;(j_1,j_2)}\ \equiv\ M_{e;(j_1,j_2);0}\star T^{(0)}_{0;(0)}\
,\eea
where the basis elements $T_{e;(j_1,j_2)}$ carry a representation
of the left action, \emph{viz.} $M_{e;(j_1,j_2);n}\star
T_{e';(j'_1,j'_2)}=\sum_{e'';(j''_1,j''_2)} {\cal
C}^{e'';(j''_1,j''_2)}_{e;(j_1,j_2);n|e';(j'_1,j'_2)}
T_{e'';(j''_1,j''_2)}$, and also twisted-adjoint energy and spin
$(e;(j_1,j_2))$, that is $T_{e;(j_1,j_2)}=\sum_{s=j_2}^{j_1} {\cal
C}^{(s)}_{e;(j_1,j_2)} T^{(s)}_{e;(j_1,j_2)}$. The
one-dimensionality of the compact weights $(0;(0))$, $(0;(1,1))$
and $(2;(1,1))$, implies that there exist finite coefficients
$\a$, $\a'$ and $\a''$ such that
\bea L^\pm_r\star T^{(0)}_{(+)}\star L^\pm_r&=&\a T^{(0)}_{(+)}\
,\quad
L^\pm_{[r}\star T^{(0)}_{(+)}\star L^\pm_{s]}\ =\ \a' M_{rs}\star T^{(0)}_{(+)}\ ,\\[5pt]
M_{rs}\star L^\pm_t \star T^{(0)}_{(+)}\star L^\pm_t&=&\a''
L^\pm_{[r}\star T^{(0)}_{(+)}\star L^\mp_{s]} \ ,\eea
where on the right-hand sides $M_{rs}\star T^{(0)}_{0;(0)}=
{4\e_0\over
(2\e_0+1)(2\e_0+2)}\left[T^{(1)}_{0;(1,1)}\right]_{r,s}$ and
$L^\pm_t \star T^{(0)}_{(+)}\star L^\pm_t=2\e_0 T^{(0)}_{2;(0)}$,
as can be seen using \eq{MrsT0n} and \eq{Ce}, respectively. The
remaining combinations involving one left and one right
$\star$-multiplication of ladder operators are non-degenerate, and
we conclude that the equivalence relation \eq{simrelation} is
generated by
\bea (M_{0;(j_1,j_2);0}\star T^{(0)}_{(+)})\otimes T^{(0)}_{(+)} &\sim& T^{(0)}_{(+)}\otimes (T^{(0)}_{(+)}\star M_{0;(j_1,j_2);0})\ ,\\[5pt]
(L^\pm_r\star T^{(0)}_{(+)})\otimes(T^{(0)}_{(+)}\star L^\pm_r)&\sim&\a T^{(0)}_{(+)}\otimes T^{(0)}_{(+)}\ ,\\[5pt]
(L^\pm_{[r}\star T^{(0)}_{(+)})\otimes(T^{(0)}_{(+)}\star L^\pm_{s]})&\sim&\a' (M_{rs}\star T^{(0)}_{(+)})\otimes T^{(0)}_{(+)}\ ,\\[5pt]
 (M_{rs}\star L^\pm_t \star T^{(0)}_{(+)})\otimes (T^{(0)}_{(+)}\star L^\pm_t)&\sim&\a''(L^\pm_{[r}\star T^{(0)}_{(+)})\otimes (T^{(0)}_{(+)}\star L^\mp_{s]})\ .\eea

The left $\mg$-module ${\cal S}^{(0)}_{e;(0)}$ contains the left
$\ms$-submodule
\bea {\cal S}(\nu_e)&\equiv& \bigoplus_{n=0}^\infty
\bigoplus_{p=0}^{\ft{n}2} \Comp\otimes{\bf
P}_{\{n,n-2p\}}M^{(n)}_{r(n)s(n)}\star T^{(0)}_{e;(0)}\ ,\qquad
\nu_e\ =\ \frac{|e|}2-\e_0\ ,\label{calSnu}\eea
where the traceless type-$(n,n-2p)$ projections ${\bf
P}_{\{n,n-2p\}}$ incorporate the ideal relations $C_2[\ms]\approx
C_2[\ms|\nu_e]= \nu_e(\nu_e+2\e_0)$ given in \eq{Cslemma} and
$\mathbf P_{\{2,1,1\}}M_{[r_1 s} M_{t]r_2}\approx 0$. It follows
that $\nu_e\in \integ+[\ft12(e+D)]$, and we say that ${\cal
S}(\nu_e)$ carries half-integer or integer one-sided spins,
respectively. Moreover, if $|e|<2\e_0$ then $C_2[\ms|\nu_e]<0$ and
say that ${\cal S}(\nu_e)$ carries negative one-sided spins.

From \eq{Tleft} it follows that
\bea \mbox{all $D$}&:&\mD^\pm\ =\ \mD^+\oplus \mD^-\ \equiv\ {\cal
A}\star T^{(0)}_{\pm 2\e_0;(0,0)}\star
{\cal A}\ ,\label{mD}\\[5pt]
D\neq 5&:&\mD^{\prime\pm}\ =\ \mD^{\prime+}\oplus\mD^{\prime-}\
\equiv \ {{\cal A}\star T^{(0)}_{\pm 2;(0,0)}\star {\cal A}\over
({\cal A}\star T^{(0)}_{\pm 2;(0,0)}\star {\cal A})\cap \mD^\pm}\
,\label{mDpr}\eea
are two-sided ${\cal A}$ submodules in ${\cal M}(+,+|0)$, and we
note that $({\cal A}\star T^{(0)}_{\pm 2;(0,0)}\star {\cal A})\cap
\mD^\pm$ is non-vanishing iff $D$ is odd. There is an analog of
$\mD'$ for $D=5$ to be defined below. In $\mD^+$ one has that if
$e=2\e_0+2n$ with $n=0,1,2,\dots$ then the one-sided spins are
positive integers. From \eq{Tleft} and the fact that ${\cal
C}_{2\e_0+2n}$ in \eq{Ce} are non-vanishing, it follows that
$L^-_{r_1}\star \cdots \star L^-_{r_{2n+1}}\star
T^{(0)}_{2\e_0+2n;(0)}=0$. Thus, viewed as a one-sided ${\cal A}$
module
\bea {\cal
S}^{(0)}_{2\e_0+2n;(0)}&\simeq&\mD(\e_0;(0))\quad\mbox{for
$n=0,1,2,\dots$}\ .\eea
Moreover, viewed two-sidedly, the element
$T^{(0)}_{2\e_0+2n;(0)}\simeq
\ket{\e_0+n;(n)}_{r(n)}{}^{r(n)}\bra{\e_0+n;(n)}$, which yields
the isomorphism
\bea \mD^+&\simeq&\mD_0\otimes \mD_0^\ast\ ,\label{envFF}\eea
underlying the enveloping-algebra analog of the Flato-Fronsdal
formula \eq{FFformula}, that we shall discuss in the next
Subsection. Likewise, in $D=4$ we identify the spinor singleton
\bea D=4&:& {\cal S}^{(0)}_{2+2n;(0)}\ \simeq\
\mD(1;(\ft12))\quad\mbox{for $n=0,1,2,\dots$}\ ,\eea
leading to the isomorphism $T^{(0)}_{2+2n;(0)}\simeq
\ket{1+n;(n+\ft12)}^{i,r(n)}{}_{i,r(n)}\bra{1+n;(n+\ft12)}$
underlying the enveloping-algebra version of the Flato-Fronsdal
formula for the 4D spinor singleton:
\bea D=4&:& \mD^{\prime+}\ \simeq\ \mD_{\ft12}\otimes
\mD_{\ft12}^\ast\ ,\eea
where King's rule rules out the elements with twisted-adjoint
energies between $2$ and the composite values. The scalar
singletons and the spinor singletons in $D=4$ have natural
extensions by negative spins described in Appendix
\ref{App:negspin} although their role in the factorization of
${\cal M}^{(\pm)}$ and its submodules is unclear at the moment.


\scss{Lowest-weight and lowest-spin submodules}\label{Sec:LWS}


\scsss{Admissibility analysis}

Lowest-weight and highest-weight $\widetilde\mg$ submodules in
${\cal M}_{(s)}$ correspond to the solutions of
\bea \widetilde L^-_r T^{(s)}_{e;(j_1,j_2)}&=& L^-_r\star
T^{(s)}_{e;(j_1,j_2)}-T^{(s)}_{e;(j_1,j_2)}\star L^+_r\ =\ 0\
.\label{candidatelws}\eea
If this holds, then $C_2[\mg]$ and $C_4[\mg]$ are on the one hand
given by \eq{C2lhws} and \eq{C4lhws}, and on the other hand by
\eq{C2nMell}. This leads to the necessary conditions
\bea x+y+z= x_0+y_0\ ,\quad x(x+\D)+y(y+\D')+z(z+\D'')=
x_0(x_0+\D)+y_0(y_0+\D')\ ,&&\qquad \label{c2c4conds}\eea
where $x=e(e-D+1)$, $y=j_1(j_1+D-3)$, $z=j_2(j_2+D-5)$ and
$\D=\ft12(D-1)(D-2)$, $\D'=\ft12 (D-3)(D-4)-1$ and
$\D''=\ft12(D-5)(D-6)-2$, and finally $x_0=e_0(e_0-D+1)$,
$y_0=s(s+D-3)$ and $e_0=s+D-3$. Moreover, combining
\eq{candidatelws} with \eq{Mdiv} yields
\bea e&=&s+D-3-{j_2\over s}\qquad\mbox{for $j_1=s\geqslant 1$ and
$j_2<s$}\ .\label{div}\eea
The combination of \eq{candidatelws} and \eq{Mcurl} yields yet
another necessary condition, valid for $j_2\geqslant 1$, which we
shall not need here.

If $j_2=0$ then $z=0$, and \eq{c2c4conds} has two roots:
$(x,y)=(x_0,y_0)$ and $(y_0+2-D,x_0+D-2)$. The first root
corresponds to $j_1=s$ and $e=s+D-3$ or $e=2-s$. The latter energy
level is ruled out for $s\geqslant 1$ due to the condition
\eq{div}. The second root corresponds to $j_1=s-1$, which is ruled
out for all $s$, and $j_1=4-D-s$, which is ruled for all $s$
except $s=0$ in $D=4$ where it coincides with the first root
(which is thus a double root in $D=4$). Thus, the admissible
lowest-weight states with $j_2=0$ are
\bea j_1&=& s\ ,\quad e\ =\ s+2\e_0\qquad\mbox{and}\quad j_1\ =\
s\ =\ 0\ ,\quad e\ =\ 2\ ,\eea
where we note the two-fold degeneracy of the root $j_1=s=0$, $e=2$
for $D=5$. If $j_2=s\geqslant 1$, which requires $D\geqslant 5$,
then the only admissible state is
\bea j_1\ =\ j_2\ =\ s\ =\ 1\ ,\qquad e\ =\ 2\ .\eea
In the remaining case of $D\geqslant 5$ and $s>j_2\geqslant 1$, we
have no conclusive statements to make, though we have not found
any admissible root here.

\scsss{Case of $s=0$}

The two admissible roots for $s=0$ are realized by the
lowest-weight states with generating functions
\bea f^{(0)}_{2\e_0;(0)}(z)&=& {}_1 F_1(\e_0+\ft32;2;-4z)\ ,
\qquad f^{(0)}_{2;(0)}(z)\ =\ {}_1 F_1(\e_0+\ft32;2\e_0;-4z)\
,\label{scalarf020minus}\label{scalarf020plus}\eea
taking the following particularly simple form in $D=4$:
\bea f^{(0)}_{1;(0)}(z)&=& e^{-4z}\ ,\qquad f^{(0)}_{2;(0)}(z)\ =\
(1-4z)e^{-4z}\ . \eea
For $D=2p+5$, $p=1,2,3,\dots$, the Harish-Chandra module
$\mC(2;(0))$ contains the singular vector $\ket{2\e_0;(0)}=(x^+)^p
\ket{2;(0)}$ and $\mC(2;(0))=\mD(2;(0))\ssumr \mD(2\e_0;(0))$
where $\mD(2;(0))=\mC(2;(0))/\mN(2;(0))$ is the scalar
$p$-lineton\footnote{For $e_0=\e_0-(p-1)$, $p=1,2,\dots$, the
Harish-Chandra module $\mC(e_0;(0))\supset \mN(e_0+2p;(0))$ with
singular vector given by $(x^+)^p| e_0;(0)\rangle$. Thus
$\mD(e_0;(0))$, that we shall refer to as the scalar $p$-lineton,
consists of $p$ lines in weight space, \emph{viz.}
$$\mD(e_0;(0))=\bigoplus_{k=0}^{p-1} \bigoplus_{n=0}^\infty |e_0+2k+n;(n)\rangle\ ,\quad |e_0+2k+n;(n)\rangle_{r(n)}=L^+_{r_1}\cdots L^+_{r_n}(x^+)^k\mid
e_0;(0)\rangle\ .$$ In particular, the $1$-lineton coincides with
the ordinary singleton.} and
$\mD(2\e_0;(0))=\mN(2;(0))=\mC(2\e_0;(0))$ is the
composite-massless lowest-weight space. In the enveloping-algebra
realization, the module $\mC(2;(0))\simeq \widetilde{\cal
U}[\mg]T^{(0)}_{2;(0)}$ and
\bea T^{(0)}_{2\e_0;(0)}&=& {\cal C}_{2\e_0-2}\cdots {\cal C}_2
(\widetilde x^+)^p\, T^{(0)}_{2;(0)}\quad\mbox{for $D=2p+5$,
$p=1,2,\dots$}\ .\eea
Thus, the scalar modules ${\cal M}_{(0)}^{(\pm)}$ contain the
ideals:
\bea \ba{lcl}D=4,6,\dots\!\!&:& {\cal I}^{(+)}_{(0)}\ =\ (1+\pi)\mD^+(2;(0))\ ,\quad {\cal I}^{(-)}_{(0)}\ =\ (1+\pi)\mD^+(2\e_0;(0))\ ,\\[10pt]
 D=5\!\!&:& {\cal I}^{(+)}_{(0)}\ =\ (1+\pi)\mD^+(2;(0))\ ,\\[10pt]
D=7,9,\dots\!\!&:&{\cal I}^{(+)}_{(0)}\ =\
(1+\pi)\left[\mD^+(2;(0)) \ssumr\mD^+(2\e_0;(0))\right]\ .\ea\eea

\scsss{Case of $s\geqslant 1$ and the Flato-Fronsdal formula}

From \eq{Tleft} it follows that $L^-_r(\xi)\ket{2\e_0;(0)}_{12}=
M_{rs}(\x)\ket{2\e_0;(0)}_{12}=0$ for $\x=1,2$ which yields an
enveloping-algebra analog of the Flato-Fronsdal formula:
\bea \ket{s+2\e_0;(s)}_{12;r(s)} &=& f_{(s)}
f_{r(s)}(1,2)\ket{2\e_0;(0)}_{12}\ ,\label{FFformula}\eea
where $f_{(s)}$ is a normalization fixed by \eq{TE} and $f_{r(s)}$
the composite operator\footnote{The condition
$(L^-_r(1)+L^-_r(2))\ket{s+2\e_0;(s)}_{12;r(s)} =0$ is equivalent
to $a_k f_{s;k}+a_{s-k+1}f_{s;k-1}=0$ with $a_k=2k(k+\e_0-1)$
implying $f_{s;k}=(-1)^sf_{s;s-k}=(-1)^k{a_{s-k+1}\cdots a_s\over
a_k\cdots a_1}f_{s;0}$, that becomes \eq{complw} for $f_{s;0}=1$.}
\bea f_{r(s)}(1,2)&=& (-1)^s f_{r(s)}(2,1)\ =\ \sum_{k=0}^s
f_{s;k}(L^+_{\{r_1}\cdots L^+_{r_{k}})(1)(L^+_{r_{k+1}}\cdots
L^+_{r_s\}})(2)\ ,\label{frs}\\[5pt] f_{s;k}&=&(-1)^s f_{s;s-k}\ =\ {s\choose k}{(1-s-\e_0)_k\over (\e_0)_k}\
.\label{complw}\eea
Indeed, applying $\bra{\1^\ast}_{23}$ to \eq{FFformula} yields the
lowest-weight elements:
\bea \left[T^{(s)}_{s+2\e_0;(s)}\right]_{r(s)}&=& f_{(s)}
\sum_{k=0}^s (-1)^{s-k} f_{s;k}L^+_{\{r_1}\star\cdots \star
L^+_{r_k}\star T^{(0)}_{2\e_0;(0)}\star L^-_{r_{k+1}}\star \cdots
\star L^-_{r_s\}}\ .\eea
Likewise, in $D=4$ the lowest-weight spaces $\mD(2;(0))$ and
$\mD(2;(1,1))\simeq \mD(2;(1))$ are the first two levels of the
tower of composite massless lowest-weight spaces
$\mD(s+1;(s,1))\simeq \mD(s+1;(s))$ contained in the tensor
product $\mD_{\ft12}\otimes \mD_{\ft12}$ of two spinor singletons.
We expect that all these lowest-weight spaces are realized in
${\cal M}^{(+)}$:
\bea D=4&:& \left[T^{(s)}_{s+1;(s)}\right]_{r(s)}=
f'_{(s)}\e_{tu\{r_{1} }\sum_{ k=1}^{s} f'_{s;k}
L^+_{r_2}\star\cdots\star L^+_{r_{k}}\star M_{tu}\star
T^{(0)}_{2;(0)}\star L^-_{r_{k+1}}\star\cdots\star L^-_{r_{s}\}}\
,\hspace{1cm}\eea
where $f'_{(s)}$ and $f'_{s;k}$ are fixed by \eq{TE} and
\eq{candidatelws}, respectively.

For $D\geqslant 6$ it follows from $C_2[\ms|(1-\e_0)]<0$ that the
$\ms$ submodule ${\cal S}(1-\e_0)$ defined in \eq{calSnu} is
infinite-dimensional, thus consisting of the elements $M_{\{r_1
t_1}\star \cdots M_{r_j t_j\}}\star T^{(0)}_{2;(0)}={\cal
C}^{(j)}_{2;(j,j)}\left[T^{(j)}_{2;(j,j)}\right]_{r(j),s(j)}$,
$j=0,1,2,\dots$, for non-vanishing ${\cal C}^{(j)}_{2;(j,j)}$,
since $M_{\{r_1}{}^t\star M_{r_2\}t}\star
T^{(0)}_{2;(0)}\approx\ft12\{L^+_{\{r_1},L^-_{r_2\}}\}_\star\star
T^{(0)}_{2;(0)}=0$. Thus $\widetilde L^-_u T^{(j)}_{2;(j,j)}=0$
for $j\geqslant 0$ and $D\geqslant 6$, and by analytical
continuation also for $D=5$, that is
\bea \widetilde L^-_u T^{(s)}_{2;(s,s)}&=&0\quad\mbox{for
$s=1,2,\dots$ and $D\geqslant 5$}\ ,\eea
while $D=4$ falls under King's rule. For example, ${\cal
M}^+_{(1)}$ contains the generalized Verma module
$\mC'(2;(1,1))={\mC(2;(1,1))\over{\cal I}[V]}$, which is
isomorphic to $\widetilde{\cal U}[\mg]T^{(1)}_{2;(1,1)}$ modulo
\eq{Mdiv} and \eq{Mcurl}. For $D=3+2p$, $p=1,2,\dots$, the
lowest-weight state of $\mD(1+2\e_0;(1))=\mD(1+2p;(1))$ is a
singular vector in $\mC'(2;(1,1))$, \emph{viz.}\footnote{At the
level of the Harish-Chandra module $\mC(2;(1,1))$, $$L^-_t x^{p-1}
L^+_s
\ket{2}_{s,u}=-4(p-1)x^{p-2}L^+_uL^+_s\ket{2}_{s,t}+2(p-1)(2p+5-D)
x^{p-2}L^+_tL^+_s\ket{2}_{s,u}+2(5-D)x^{p-1}\ket{2}_{t,u}\ ,$$
where $\ket{2}_{s,u}\equiv\ket{2;(1,1)}_{s,u}$. For $D=2p+3$ the
$(tu)$-projection vanishes and the $[tu]$-projection equals
$-6(D-5)L^+_s L^+_{[s}\ket{2}_{t,u]}$ that vanishes in
$\mC(2;(1,1))$ only if $D=5$, while it vanishes weakly in
$\mC'(2;(1,1))$ for all $D$.} $\widetilde L^-_t (\widetilde
x^+)^{p-1} \widetilde L^+_s T^{(1)}_{2;(1,1);u,s}\approx0$.
Factoring out the ideal $\mN'(2;(1,1))\simeq \mD(1+2\e_0;(1))$
from $\mC'(2;(1,1))$ yields the spin-1 $p$-lineton
\bea \mD'(2;(1,1))&=&{\mC'(2;(1,1))\over
\mN'(2;(1,1))}\quad\mbox{for $D=3+2p$, $p=1,2,\dots$}\ ,\eea
occupying $p$ lines in compact weight space. In $D=5$ it is the
ordinary spin-1 singleton, that is $\mD'(2;(1,1))=\mD(2;(1,1))$.

Thus, in summary, the sectors $\bigoplus_{s=1}^\infty {\cal
M}_{(s)}^{(\pm)}$ contain the ideals:
\bea D=4&:& {\cal I}^{(+)}\ =\ (1+\pi)\bigoplus_{s=1}^\infty \mD^+(1+s;(s,1))\ ,\\[5pt]
&&{\cal I}^{(-)}\ =\ (1+\pi)\bigoplus_{s=1}^\infty \mD^+(1+s;(s))\ ,\\[5pt]
D=5,7,\dots&:&{\cal I}^{(+)}\ =\ (1+\pi)\bigoplus_{s=1}^\infty [\mD^+(2\e_0+s;(s))\oplus \mD^+(2;(s,s))]\ ,\\[5pt]
D=6,8,\dots&:& {\cal I}^{(+)}\ =\ (1+\pi)\bigoplus_{s=1}^\infty \mD^+(2;(s,s))\ ,\\[5pt]
&&{\cal I}^{(-)}\ =\ (1+\pi)\bigoplus_{s=1}^\infty
\mD^+(2\e_0+s;(s)) \ .\eea

\scsss{Summary of the indecomposable structure of twisted-adjoint
module}

In summary, the compact twisted-adjoint module has the
indecomposable structure
\bea {\cal M}&=&{\cal W}\ssumr \mD' \ssumr\mD\
,\label{indecomposable}\eea
where $\mD=\mD^+\oplus \mD^-$ and
$\mD'=\mD^{\prime+}\oplus\mD^{\prime-}$ have the substructures
\bea \mbox{all $D$}&:& \mD\ \simeq \ (1+\pi)(\mD^+_0\otimes \mD^+_0)\ \simeq\ (1+\pi)\bigoplus_{s=0}^\infty \mD^+(s+2\e_0;(s))\ ,\\[5pt]D=4&:& \mD'\ \simeq\ (1+\pi)(\mD^+_{\ft12}\otimes \mD^+_{\ft12})\ \simeq\ (1+\pi)\bigoplus_{s=0}^\infty \mD^+(s+2\e_0+\d_{s,0};(s,1))\ ,\label{mDprime4}\\[5pt] D=5&:& \mD'\ \simeq\
(1+\pi)\bigoplus_{s=1}^\infty \mD^+(2;(s,s))\ ,\label{mDprime5}\\[5pt]
D\geqslant 6&:&\mD'\ \simeq\
(1+\pi)\left(\mD^+(2;(0))\oplus\bigoplus_{s=0}^\infty \mD^{\prime
+}(2;(s,s))\right)\ .\label{mDprime}\eea
Moreover, splitting into even and odd parts we have
\bea D=4,6,\dots&:& {\cal M}^{(+)}\ =\ {\cal W}^{(+)} \ssumr \mD'\ ,\qquad {\cal M}^{(-)}\ =\ {\cal W}^{(-)} \ssumr \mD\ ,\label{indec1}\\[5pt]
D=5,7,\dots&:& {\cal M}^{(-)}\ =\ {\cal W}^{(-)}\ ,\qquad {\cal
M}^{(+)}\ =\ {\cal W}^{(+)} \ssumr \mD'\ssumr \mD\
.\label{indec2}\eea
The lowest-spin spaces ${\cal W}^{(\pm)}$ do not contain any
lowest-weight nor highest-weight states. If ${\cal
I}^{(\pm)}_{(0)}$ is non-trivial then ${\cal W}^{(\pm)}$ consists
of the ``wedge'' ${\cal M}^{(\pm)0}$ defined in
\eq{stategeneration2} plus a finite number of energy levels in
${\cal M}^{(\pm)~{}^{>}\!\!\!\!\!{}_{<}}$ for each $j_1$. In
particular, one has
\bea {\cal W}^{(+)}_{(0)}&=&\bigoplus_{|e|\leqslant j}\Comp\otimes
T^{(0)}_{e;(j)}\ .\label{W0+}\eea
%


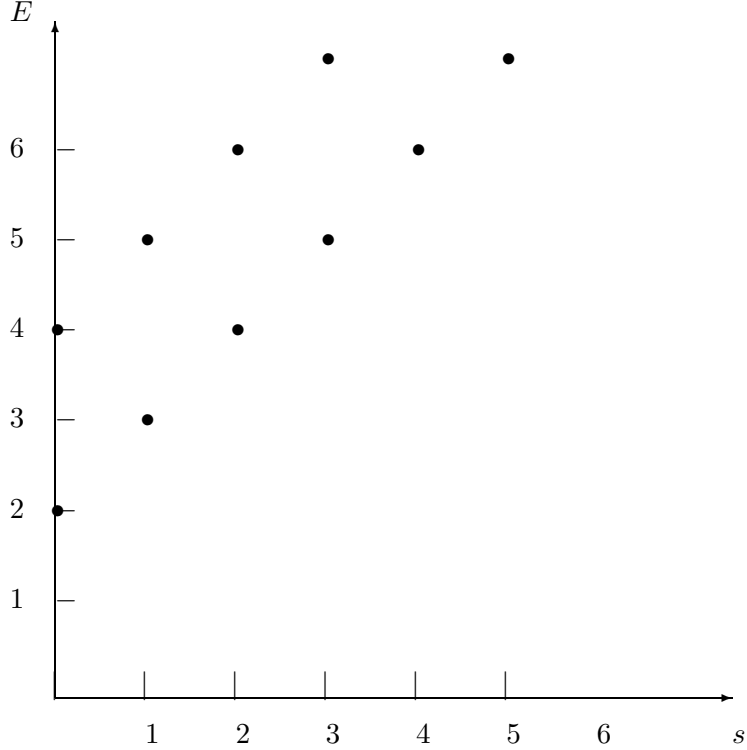
\begin{figure}[!h]
\begin{center}
\unitlength=.6mm
\begin{picture}(150,180)(0,-10)
\put(0,0){\vector(1,0){150}} \put(0,0){\vector(0,1){150}}
\put(150,-10){$s$} \put(-10,150){$E$}
\put(20,-10){1}\put(40,-10){2}\put(60,-10){3}\put(80,-10){4}\put(100,-10){5}\put(120,-10){6}
\put(-10,20){1}\put(-10,40){2}\put(-10,60){3}\put(-10,80){4}\put(-10,100){5}\put(-10,120){6}
\multiput(0,1)(20,0){6}{\!$|$} \multiput(1,20)(0,20){6}{\!$-$}
\multiput(0,40)(20,20){6}{\!$\bullet$}
\multiput(0,80)(20,20){4}{\!$\bullet$}
\end{picture}
\end{center}
\caption{{\small The $(\mso(2)\oplus\mso(D-1))$-types arising in
the scalar 2-lineton in $D=9$.}} \label{wfig2}
\end{figure}


\scss{Inner products, real forms and unitarity}\label{Sec:Inner}


The non-polynomial nature of the compact basis elements
$T^{(s)}_{e;(j_1,j_2)}$, viewed as generalized elements of ${\cal
T}$, together with the indecomposable structure of ${\cal M}$
imply that the bilinear inner product $(S,S')_{\cal
T}=Tr[\pi(S)\star S']$ induces inequivalent inner products in the
different sectors $\mV=\mD,\mD',{\cal W}^{(\pm)}$ of ${\cal M}$:
\bea (S,S')_{\mV}&=& {1\over {\cal N}_{\mV}} (S,S')_{\cal T}\
,\qquad {\cal N}_{\mV}\ =\ (T_{\mV},T_{\mV})_{\cal T}\
,\label{inner}\eea
where $T_{\mV}$ denotes $T^{(0)}_{\pm 2\e_0;(0)}$, $T^{(0)}_{\pm
2;(0)}$ and $T^{(0)}_{0;(\s_\pm)}$, respectively, and ${\cal
N}_{\mV}$ is factored out according to the following
\emph{prescription}: (i) expand $S$ and $S'$ in the bases
generated by the $\widetilde \mho$ action on $T_{\mV}$; (ii) use
\eq{selfadjoint}, \emph{i.e.} $\Tr[\pi(\widetilde \adj_Q S)\star
S']=-\Tr[\pi(S)\star \widetilde \adj_Q S']$ for $Q\in\mho$ and
$S,S'\in{\cal M}$ (which amounts to assuming $\Tr[X\star
S]=\Tr[S\star X]$ for $X\in{\cal A}$ and general $S\in{\cal M}$)
to write $(S,S')_{\cal T}=(T_{\mV},\widetilde Q(S,S')
T_{\mV})_{\cal T}$ for some $Q(S,S')\in \mho$; (iii) expand
$\widetilde Q(S,S') T_{\mV}=\sum_{s;e;(j_1,j_2)}{\cal
C}^{(s)}_{e;(j_1,j_2)}(S,S')T^{(s)}_{e;(j_1,j_2)}={\cal
C}_{\mV}T_{\mV}+\cdots$, where ${\cal
C}^{(s)}_{e;(j_1,j_2)}(S,S')\in\Comp$ are finite by construction,
and declare $(S,S')_{\mV}$ to be the coefficient of $T_{\mV}$,
\emph{i.e.}
\bea (S,S')_{\mV}&=&{\cal C}_{\mV}(S,S')\ .\eea
By construction $(\cdot,\cdot)_{\mV}$ is symmetric and
$\mho$-invariant; in particular, if $\widetilde E S_e
=\{E,S_e\}_\star=e S_e$ \emph{idem} $S'_{e'}$ then
\bea (S_e,S'_{e'})_{\mV}&=& \delta_{e+e',0}(S_e,S'_{e'})_{\mV}\
.\eea
Thus, on ${\cal M}^{(\pm)}$ we have the bilinear forms
$(S,S')_{{\cal M}^{(\pm)}}={\cal C}_{(\pm)}(S,S')$ where ${\cal
C}_{(\pm)}(S,S')$ are the coefficients of $T_{(\pm)}$ in
$-\widetilde Q(S)\widetilde Q (S') T_{(\pm)}$ with $\widetilde Q
(S)$ and $\widetilde Q (S')\in \mho$ defined by $S=\widetilde Q
(S)T_{(\pm)}$ and $S'=\widetilde Q (S')T_{(\pm)}$. These bilinear
forms split under\eq{indec1} and \eq{indec2} into a non-degenerate
inner product on ${\cal W}^{(\pm)}$ and trivial bilinear forms on
$\mD'$ and $\mD$, where the non-degenerate inner products are
instead defined by $(S,S')_{\mD'}=\ft1{{\cal N}_{2}}Tr[\pi(S)\star
S']$ and $(S,S')_{\mD}=\ft1{{\cal N}_{2\e_0}}Tr[\pi(S)\star S']$.
If $S$ and $S'$ obey the twisted-adjoint reality condition,
\emph{i.e.} $S^\dagger=\pi(S)$ \emph{idem} $S'$, then their inner
product is a real number, \emph{i.e.}
\bea (S,S')_{{\cal M}^{(\pm)}}&=&\ft1{{\cal N}_{(\pm)}}\Tr[
S^\dagger\star S']\ =\ \ft1{{\cal
N}_{(\pm)}}\Tr[( S^{\prime\dagger}\star S]\ =\ \left((S,S')_{{\cal M}^{(\pm)}}\right)^\ast\ ,\\[5pt]
( S, S^{\prime})_{\mD}&=&\ft1{{\cal N}_{2\e_0}}\Tr[ S^\dagger\star
 S^{\prime}]\ =\ \ft1{{\cal
N}_{2\e_0}}\Tr[( S^{\prime})^\dagger\star S]\ =\ \left(( S,
S^{\prime})_{\mD}\right)^\ast\ .\eea
To examine their signatures we use the following bases:
\bea S_{{\cal W}^{(\pm)}}&=& \sum_{s=0}^\infty \sum_{m,n=0}^\infty
S^{(\pm)(s)}_{m,n}(\widetilde
L^+)^m (\widetilde L^-)^n \widetilde Q_s T^{(0)}_{(\pm)}\ ,\label{wedgestate}\\[5pt]
S_{\mD}&=& \sum_{s=0}^\infty
\sum_{m}^\infty\left[S^{(s)}_{m}(\widetilde L^+)^m \widetilde R_s
T^{(0)}_{2\e_0;(0)}+\overline S^{(s)}_{m}(\widetilde L^-)^m
\pi(\widetilde R_s T^{(0)}_{2\e_0;(0)})\right]\
,\label{1pstate}\eea
where $S^{(s)}_{m,n},S^{(s)}_{m},\overline S^{(s)}_{m}\in\Comp$
and $Q_s,R_s\in\mho$ such that $\widetilde Q_s
T^{(0)}_{(\pm)}=T^{(s)}_{(\pm)}$ and $\widetilde R_s
T^{(0)}_{2\e_0;(0)}=T^{(s)}_{s+2\e_0;(s)}$. From
$(L^\pm_r)^\dagger=L^\mp_r$ and $\pi(\widetilde\adj_Q S)^\dagger=
-\widetilde \adj_{Q^\dagger} \pi(S^\dagger)$ it follows that
\bea {\cal W}^{(\pm)},\quad s=0&:& (S^{(\pm)(0)}_{m,n})^\ast\ =\
(-1)^{\ft{1\mp 1}2}
(-1)^{m+n}S^{(\pm)(0)}_{n,m}\ ,\\[5pt]
{\cal W}^{(\pm)},\quad s>0&:&(S^{(\pm)(s)}_{m,n})^\ast\ =\
(-1)^{m+n}S^{(\pm)(s)}_{n,m}\ ,\\[5pt]\mD,\quad s\geqslant 0&:&
(S^{(s)}_{m})^\ast\ =\ (-1)^{m}\overline S^{(s)}_m\ ,\eea
where the additional phase factor for $W^{(-)}_{(0))}$ arises from
$\pi((T^{(0)}_{(\pm)})^\dagger)= (-1)^{\ft{1\mp
1}2}T^{(0)}_{(\pm)}$. Thus
\bea ( S_{{\cal W}^{(\pm)}}, S^{\prime}_{{\cal W}^{(\pm)}})_{\cal
M}&=& \sum_{s=0}^\infty \sum_{m,n;m',n'=0}^\infty (
S^{(\pm)(s)}_{m,n})^\ast
N^{(\pm)(s)}_{m,n;m',n'} S^{(\pm)(s)\prime}_{m',n'}\ ,\\[5pt]
 ( S_{\mD}, S'_{\mD})_{\mD}&=& \sum_{s=0}^\infty
\sum_{m,m'=0}^\infty \left(( S^{(s)}_m)^\ast M^{(s)}_{m,m'}
 S^{(s)\prime}_{m'}+( S^{(s)\prime}_{m'})^\ast
M^{(s)}_{m',m} S^{(s)}_m\right)\ , \label{mDAdS}\eea
\bea\mbox{where}&& N^{(\pm)(s)}_{m,n;m',n'}\ =\
\left(T^{(s)}_{\pm},(\widetilde L^-)^m(\widetilde L^+)^n
(\widetilde L^+)^{m'}(\widetilde
L^-)^{n'}T^{(s)}_{\pm}\right)_{{\cal W}^{(\pm)}}\ ,\label{Ns}\\[5pt]&&M^{(s)}_{m,m'}\ = \left(T^{(s)}_{-(s+2\e_0);(s)},(\widetilde
L^-)^m(\widetilde L^+)^{m'} T^{(s)}_{s+2\e_0;(s)}\right)_{\mD}\
.\label{Ms}\eea
More explicitly, using \eq{stategeneration1} we have
$M^{(s)}_{m,n}((j_1,j_2)|(j'_1,j'_2))
=\d_{m,n}\d_{j_1,j'_1}\d_{j_2,j'_2}M^{(s)}((j_1,j_2);p))$ where
$p=\ft12(m+s-j_1-j_2)\geqslant 0$ and
$\left[M^{(s)}((j_1,j_2);p)\right]_{r(j_1),t(j_2)}^{r'(j_1),t'(j_2)}$
is given by
\bea \left(T^{(s)}_{-(s+2\e_0);(s);\{r(s)},(\widetilde
x^-)^{p}\widetilde L^{-(j_1-s)}_{r(j_1-s)} \widetilde
L^{-(j_2)}_{t(j_2)\}}(\widetilde x^+)^{p}\widetilde
L^{+(j_1-s)}_{\{r'(j_1-s)} \widetilde L^{+(j_2)}_{t'(j_2)}
T^{(s)}_{s+2\e_0;(s);r'(s)\}}\right)_{\mD}\ .&&\eea
This matrix and hence the inner product $(\cdot,\cdot)_{\mD}$ is
positive definite\footnote{The complex space
$\mD^+(s+2\e_0;(s))\oplus \mD^-(s+2\e_0;(s))$ has positive
definite antilinear-linear inner product
$\widehat{M}(e;(j_1,j_2)|e';(s'_1,s'_2))=
\left(\ket{e;(j_1,j_2)}\right)^\dagger
\ket{e';(s'_1,s'_2)}=\left[\ba{cc} M&0\\0&M\ea\right]$ with $M$
given by \eq{Ms}.}.

The matrix $N^{(+)(s)}$ can be expanded using
\eq{stategeneration1} and \eq{stategeneration2} as follows
\bea N^{(+)(s)}_{m,n;m',n'}((j_1,j_2)|(j_1',j_2'))&=&\!\!
\d_{m,m'}\d_{n,n'}\d_{m+n,j_1+j_2-s}\d_{j_1,j'_1}\d_{j_2,j'_2}N^{(+)(s)}
((j_1,j_2);p,q)\ ,\qquad\quad\eea
where $p=\ft12(m-n+s-j_1-j_2)$ with $n=0$ for $p>0$, and
$q=\ft12(n-m+s-j_1-j_2)$ with $m=0$ for $q>0$, and
$\left[N^{(+)(s)}((j_1,j_2);p,q)\right]_{r(j_1),t(j_2)}^{r'(j_1),t'(j_2)}$
is given by
\bea \left\{\ba{ll}\left(T^{(s)}_{\pm;\{r(s)},\left[(\widetilde
L^-)^{m} (\widetilde
L^+)^{n}\right]_{r(j_1-s),t(j_2)\}}\left[(\widetilde L^+)^{m}
(\widetilde
L^-)^{n}\right]_{\{r'(j_1-s),t'(j_2)}T^{(s)}_{\pm;r'(s)\}}\right)_{\cal
M}&\mbox{for $p,q\leqslant 0$}\\[10pt]
\left(T^{(s)}_{\pm;\{r(s)},(\widetilde x^-)^p(\widetilde
L^-)^{m}_{r(j_1-s),t(j_2)\}}(\widetilde x^+)^p (\widetilde
L^+)^{m}_{\{r'(j_1-s),t'(j_2)}T^{(s)}_{\pm;r'(s)\}}\right)_{\cal
M}&\mbox{for $p>0$ }\\[10pt] \left(T^{(s)}_{\pm;\{r(s)},(\widetilde x^+)^q(\widetilde
L^+)^{n}_{r(j_1-s),t(j_2)\}}(\widetilde x^-)^q(\widetilde
L^-)^{n}_{\{r'(j_1-s),t'(j_2)}T^{(s)}_{\pm;r'(s)\}}\right)_{\cal
M}&\mbox{for $q>0$ }\ea\right. \ .&&\qquad\nn\eea
For $s=0$ it follows from \eq{W0+} that
$N^{(+)(0)}_{m,n;m',n'}((j_1)|(j_1'))=\d_{m,m'}\d_{n,n'}\d_{,j_1}\d_{j_1,s'_1}N^{(+)(0)}
((j);p,q)$ where $j=m+n$ and $p,q\leqslant 0$, and hence
\bea \left[N^{(+)(0)}((j);p,q)\right]_{r(j)}^{r'(j)}=
\left(T^{(0)}_{(+)},\left[\widetilde
T_{n,m}\right]_{r(j)}\left[\widetilde
T_{m,n}\right]^{r'(j)}T^{(0)}_{(+)}\right)_{{\cal W}^{(+)}}=
\d_{\{r(j)\}}^{\{r'(j)\}}N^{(+)(0)}_{m,n}\ ,&&\qquad\eea
$$\left[\widetilde T_{m,n}\right]_{r(j)}\equiv \left[(\widetilde L^+)^{m} (\widetilde
L^-)^{n}T^{(0)}_{(+)}\right]_{\{r(j)\}}=\widetilde
L^+_{\{r_1}\left[\widetilde T_{m-1,n}\right]_{r(j-1)\}}=\widetilde
L^-_{\{r_1}\left[\widetilde T_{m,n-1}\right]_{r(j-1)\}}\ ,\qquad$$
where the matrix elements
\bea N^{(+)(0)}_{m,n}&=& {1\over \dim
(j)}\left(T^{(0)}_{(+)},\left[\widetilde
T_{n,m}\right]_{r(j)}\left[\widetilde
T_{m,n}\right]^{r(j)}T^{(0)}_{(+)}\right)_{{\cal W}^{(+)}}\ ,\eea
with $\dim (j)=\d_{\{r(j)\}}^{\{r(j)\}}$. These matrix elements
obey the recursion relation
\bea N^{(+)(0)}_{m,n}&=& {\dim (j-1)\over \dim (j)}D^+_{m,n}
N^{(+)(0)}_{m,n-1}\ =\ {\dim (j-1)\over \dim (j)}D^-_{m,n}
N^{(+)(0)}_{m-1,n}\ ,\eea
where the coefficients $D^\pm_{m,n}$ are defined by $$\widetilde
L^+_{s}\left[\widetilde T_{m,n}\right]_{s r(j-1)}\ =\
D^+_{m,n}\left[\widetilde T_{m,n-1}\right]_{r(j-1)}\ ,\quad
\widetilde L^-_{s}\left[\widetilde T_{m,n}\right]_{s r(j-1)}\ =\
D^-_{m,n}\left[\widetilde T_{m-1,n}\right]_{r(j-1)}\ ,$$and given
by
\bea D^+_{m,n}&=&D^-_{n,m}\ =\ {2n(n+\e_0-1)(m+n+2\e_0-1)\over
m+n+\e_0-1}\quad\mbox{for $m\geqslant 0$ and $n\geqslant 1$}\
,\eea
as can be seen by using ($n=0,1,2,\dots$)
\bea \widetilde L^+_r\widetilde L^-_r \widetilde T_{0,n}&=&\mu_{n}\widetilde T_{0,n}0\ ,\qquad \m_{n}\ =\ 2(n+1)(n+2\e_0)\ ,\\[5pt]
\widetilde x^+ \widetilde T_{0,n}&=&\mu'_{n}\widetilde
T_{1,n-1}+\widetilde T'\ ,\qquad \mu'_n\ =\ 4n(n+\e_0-1)\ ,\eea
where the first formula follows from $(\widetilde L^+_r\widetilde
L^-_r-4\e_0) T^{(0)}_{0;(0)}=0$, and $\widetilde T'$ is a
descendant of $\widetilde x^\pm T^{(0)}_{0;(0)}$ that decouples
from $(\cdot,\cdot)_{{\cal W}^{(+)}}$. Since $D^+_{m,n}>0$ for
$m\geqslant 0$ and $n\geqslant 1$ it follows that
$N^{(+)(0)}_{m,n}>0$ for all $m,n\geqslant 0$ and hence
$(\cdot,\cdot)_{{\cal W}^{(+)}}$ is positive definite in ${\cal
W}^{(+)}_{(0)}$.


\scss{Expansion of Lorentz tensors in compact
basis}\label{Sec:Refl}


\scsss{Decomposition of the identity}


Attempting to invert the harmonic map \eq{Ts} from ${\cal T}$ to
${\cal M}$ amounts to seeking a decomposition of the unity $\1$ in
the compact basis. From $[M_{ab},\1]_\star=0$, that is
\bea [M_{rs},\1]_\star&=&0\ ,\qquad [M_{0r},\1]_\star\ =\
(\widetilde L^+_r+\widetilde L^-_r)\1\ =\ 0\ ,\label{ltilde1}\eea
and from the nature of the indecomposable structure
\eq{indecomposable} of ${\cal M}_{(0)}$, it follows that
\bea \1_{\cal M}&=& (1+\pi)\sum_{\tiny \mV={\cal
W}^{(\pm)},\mD',\mD}~{\cal N}_{\mV}~\psi_{\mV}(\widetilde
x^+)T^{(0)}_{e_0^{(\mV)};(0)}\ ,\qquad \psi_{\mV}(0)\ =\ 1\
,\label{decunity}\eea
where $e_0^{(\mD)}=2\e_0$, $e_0^{(\mD')}=2$, $e_0^{(-)}=1$,
$e_0^{(+)}=0$; ${\cal N}_{\mV}$ are normalization constants; and
the embedding functions $\psi_{\mV}(\widetilde x^+)$ account for
the \emph{regular} contributions to $\1_{\cal M}$ from $\mV$ plus
eventual \emph{logarithmic} contributions from other modules that
arise if $\mV$ contains a finite number of $(\me\oplus\ms)$-types
with $\overrightarrow{s}=0$. Using \eq{Ce}, \eq{tildeCe},
\eq{lpluslminus} and \eq{De}, the condition \eq{ltilde1} can be
shown to imply that
\bea D_{2\e_0}\psi_{\mD}(x^+)&=&D_2\psi_{\mD'}(x^+)\ =\ 0\ ,\qquad
D_1\psi_{(-)}(x^+)\ =\ {2\e_0-1\over x^+}\ ,\\[5pt]
D_2\chi_{(+)}(x^+)&=&{4\e_0\over x^+}\ ,\qquad \psi_{(+)}(x^+)\ =\
1-{1\over 2\e_0} x^+ \chi_{(+)}(x^+)\ ,\eea
where
$D_e(x)=4x\ft{d^2}{dx^2}+4(e-\e_0)\ft{d}{dx}+1+(e-2\e_0)(e-2)x^{-1}$.
The transformations
\bea \psi_{\mD}(x^+)&=& y^{-\nu} J(y)\ ,\qquad \psi_{\mD'}(x^+)\ =\ y^{-\nu'} J(y)\ ,\\[5pt]
\psi_{(-)}(x^+)&=& 1-\sqrt{\pi}\left({y\over
2}\right)^{1-\nu'}J(y)\ ,\qquad \chi_{(+)}(x^+)\ =\ y^{-\nu'}J(y)\
,\eea
with $\nu=\e_0-1$ and $\nu'=1-\e_0$, lead to Bessel's differential
equation\footnote{ The Bessel, Struve, Neumann and modified Struve
functions, respectively, have the series expansions $$J_\nu(y)\ =\
\sum_{n=0}^\infty {(-1)^n\left({y\over 2}\right)^{\nu+2n}\over
n!\C(\nu+n+1)}\ ,\qquad {\bf H}_\nu(y)\ =\ \sum_{n=0}^\infty
{(-1)^n\left({y\over 2}\right)^{\nu+1+2n}\over
\C(n+\ft32)\C(\nu+n+\ft32)}\ ,$$$$\pi N_p(y)\ =\ {\bf
f}_{p}(y)-\sum_{n=0}^{p-1}{(p-n-1)!\over n!}\left({y\over
2}\right)^{2n-p}\ ,\qquad \widetilde {\bf H}_{-p-\ft12}(y)\ =\
{1\over p!}\left({\bf
f}_{-p-\ft12}(y)-\sum_{n=0}^{p-1}{(p-1-n)!\over
\Gamma(n+\ft32)}\left({y\over 2}\right)^{2n+\ft12-p}\right)\ ,$$
where $p\in\{0,1,2,\dots\}$ and
$$ {\bf f}_{\nu}(y)\ =\
\sum_{n=0}^\infty {(-1)^n(\log y^2-\psi(n+1)-\psi(n+\nu+1))\over
n!\Gamma(\nu+n+1)}\left({y\over 2}\right)^{\nu+2n}\ ,$$ where
$\psi(z)={d\over dz}\log \Gamma(z)$. These functions obey
$$ B_\nu J_\nu\ =\ B_p N_p\ =\ 0\ ,\quad B_\nu {\bf H}_\nu\ =\ {1\over \sqrt{\pi}\C(\nu+\ft12)}\left({y\over 2}\right)^{\nu-1}\ ,$$$$B_{-p-\ft12} \widetilde {\bf H}_{-p-\ft12}\ =\ {1\over \sqrt{\pi}}\left({y\over 2}\right)^{-p-\ft32}\ ,\quad B_\nu {\bf f}_\nu \ =\ {1\over\Gamma(\nu)} \left({y\over 2}\right)^{\nu-2}\ .$$}
\bea B_\nu J&=&K_{\mV}\ ,\qquad B_\nu\ =\ {d^2\over dy^2}+{1\over
y}{d\over dy}+1-{\nu^2\over y^2}\ ,\qquad \widetilde x^+\ =\ y^2\
,\label{Bessel}\eea
where the sources are given by
\bea K_{\mD}&=& K_{\mD'}\ =\ 0\ ,\qquad K_{(-)}\ =\
{1\over\sqrt{\pi}}\left({y\over 2}\right)^{\nu'-1}\ ,\qquad
K_{(+)}\ =\ 4\e_0y^{\nu'-2}\ .\eea
As a result, the regular embeddings are given by
\bea D\geqslant 4&:& \psi_{\mD}\ =\
\C(\e_0)\left(\frac{y}{2}\right)^{1-\e_0} J_{\e_0-1}(y)
 \ ,\label{psimD}\\[5pt]
 D=4,6,\dots&:&\psi_{\mD'}\ =\ \C(2-\e_0)\left(\frac{y}{2}\right)^{\e_0-1} J_{1-\e_0}(y) \
, \label{psimDprime}\\[5pt]D=5,7,\dots&:&
\psi_{{\cal W}^{(-)}}\ =\
1-\sqrt{\pi}\C(\ft32-\e_0)\left(\frac{y}2\right)^{\e_0}{\bf
H}_{1-\e_0}(y)\ , \label{psiwminus} \eea
where we note that $\mD={\cal W}^{(-)}$ in $D=4$ and $\mD=\mD'$ in
$D=5$. The irregular embeddings involving linear logarithms are
given by
\bea
 D=7,9,\dots&:& \psi_{\mD'}\ =\ -{\pi \over (\e_0-2)!}\left(\frac{y}2\right)^{\e_0-1} N_{\e_0-1}(y)
, \label{psimDprimelog}\\[5pt]
D=6,8,\dots&:& \psi_{{\cal W}^{(-)}}\ =\ 1-\sqrt{\pi}\left(\frac{y}2\right)^{\e_0}\widetilde {\bf H}_{1-\e_0}(y)\ ,\label{psiwminuslog}\\[5pt]
D=4,6,\dots&:& \psi_{{\cal W}^{(+)}}\ =\
1-2\Gamma(1-\e_0)\left(\frac{y}2\right)^{\e_0+1}{\bf
f}_{1-\e_0}(y)\ .\label{psiwpluseven} \eea
If $D=5,7,\dots$ the function $\psi_{(+)}$ contains quadratic
logarithms and this case is left out in this work. We note that
$[E,T^{(0)}_{e;(0)}]_\star=0$ and more generally that
\bea [E,f(\widetilde x^+,\widetilde
x^-)T^{(0)}_{e;(0)}]_\star&=&0\ \mbox{for differentiable $f$}\
,\label{Ef}\eea
which together with $[M_{ab},\1_{\cal M}]_\star=0$ implies that
$[P_a,\1_{\cal M}]_\star=0$. Conversely, the general solution to
$[M_{AB},{\bf M}]_\star=0$ is ${\bf
M}=\sum_{\mW=\mD^+,\mD^-,\mD^{\prime+},\mD^{\prime-},{\cal
W}^{(\pm)}}{\cal N}_{\mW} {\bf M}_{\mW}$, where ${\cal N}_{\mW}$
are constants and ${\bf M}_{\mD^\pm}=\psi_{\mD}(x^\pm)T^{(0)}_{\pm
2\e_0;(0)}$, ${\bf
M}_{\mD^{\prime\pm}}=\psi_{\mD'}(x^\pm)T^{(0)}_{\pm 2;(0)}$ and
${\bf M}_{{\cal W}^{(\pm)}}=(1+\pi)\psi_{{\cal
W}^{(\pm)}}(x^+)T^{(0)}_{e_0^{(\pm)};(0)}$. Demanding $\pi({\bf
M})={\bf M}$ enforces ${\cal N}_{\mD^+}={\cal N}_{\mD^-}$ and
${\cal N}_{\mD^{\prime+}}={\cal N}_{\mD^{\prime-}}$, resulting in
\eq{decunity}. The final determination of ${\cal N}_{\mV}$ from
the condition $\1_{\cal M}\star\1_{\cal M}=\1_{\cal M}$ requires a
regularization of $\star$-product compositions of various special
functions. We shall address this issue elsewhere \cite{companion},
and continue here with the derivation of the embedding functions
and the expansion of Lorentz tensors in compact basis.


\scsss{Decomposition of Lorentz tensors}

The Lorentz tensor
$\ket{S_{(s+k,s)}}_{12}=S^{a(s+k),b(s)}\ket{T_{a(s+k),b(s)}}_{12}=S^{a(s+k),b(s)}T_{a(s+k),b(s)}(1)
\ket{\1}_{12}$ can be expanded in the compact basis as follows
\bea \ket{S_{(s+k,s)}}& = & \sum_{s+k\geqslant j_1\geqslant
s\geqslant j_2\geqslant 0} S^{r(j_1),t(j_2)}_{(s+k,s)}
\sum_{\mV=\mD,\mD',{\cal W}^{(\pm)}} {\cal
N}^{(s+k,s)}_{\mV;(j_1,j_2)}
\ket{{}^{(s+k,s)}_{(j_1,j_2)}}_{\mV;r(j_1),t(j_2)}\ ,
\label{embeddinggen}\eea
where: {(i)}
$S^{r(j_1),t(j_2)}_{(s+k,s)}=S^{0(s+k-j_1)\{r(j_1),t(j_2)\}0(s-j_2)}$
are the type-$(j_1,j_2)$ \emph{polarization tensors} contained in
$S^{a(s+k),b(s)}$; {(ii)} ${\cal N}^{(s+k,s)}_{\mV;(j_1,j_2)} $
are \emph{normalization constants}; {(iii)} the \emph{embedded
polarizations}
\bea \ket{{}^{(s+k,s)}_{(j_1,j_2)}}_{\mV}&=&
(1+\pi)\ket{{}^{(s+k,s)}_{(j_1,j_2)}}^+_{\mV}\ ,\qquad s+k\geqslant j_1\geqslant s\geqslant j_2\geqslant 0\ ,\\[5pt]
\ket{{}^{(s+k,s)}_{(j_1,j_2)}}^+_{\mV}& =&
\psi^{(s+k,s)}_{\mV;(j_1,j_2)}(x^+)
\ket{(s);e^{(\mV)}_0;(j_1,j_2)}^+_{\mV}\ ,\eea
where
$\ket{(s);e^{(\mV)}_0;(j_1,j_2)}^+_{\mV}=T^{(s)}_{e^{(\mV)}_0;(j_1,j_2)}(1)\ket{\1}_{12}$
and $e^{(\mV)}_0$ is the minimal positive energy compatible with
the embedding into $\mV$, which is given for $\mV=\mD,{\cal
W}^{(\pm)}$ by (see also \eq{ketj1j2mine})
\bea \label{ketj1j2} e^{(\mD)}_0&=&2\e_0+j_1+j_2\
,\qquad\mx{\{}{l}{e^{(\pm)}_0\in\{0,1\}\\ e^{(\pm)}_0+j_1+j_2\ =\
\ft12 (1\mp 1)\ }{.}\ ;\eea
{(iv)} the \emph{embedding functions}
$\psi^{(s+k,s)}_{\mV;(j_1,j_2)}(x^+)$ are determined by
$\psi^{(s+k,s)}_{\mV;(j_1,j_2)}(0)=1$ and the requirement that the
set
$\left\{\ket{{}^{(s+k,s)}_{(j_1,j_2)}}_{\mV}\right\}_{s+k\geqslant
j_1\geqslant s\geqslant j_2\geqslant 0}$ furnishes a
type-$(s+k,s)$ Lorentz tensor, \emph{i.e.}
\bea M_{0r}\ket{{}^{(s+k,s)}_{(j_1,j_2)}}_{\mV}&=&
M^{(s+k,s)}_{(j_1,j_2);(1,0)}\ket{{}^{(s+k,s)}_{(j_1+1,j_2)}}_{\mV}+
M^{(s+k,s)}_{(j_1,j_2);(-1,0)}\ket{{}^{(s+k,s)}_{(j_1-1,j_2)}}_{\mV}\nn\\[5pt]&&+
M^{(s+k,s)}_{(j_1,j_2);(0,1)}\ket{{}^{(s+k,s)}_{(j_1,j_2+1)}}_{\mV}+
M^{(s+k,s)}_{(j_1,j_2);(0,-1)}\ket{{}^{(s+k,s)}_{(j_1,j_2-1)}}_{\mV}\
,\label{ccoeff}\eea
for $M_{0r}=\ft12(L^+_r+L^-_r)$ and
$M^{(s+k,s)}_{(s+k,j_2);(1,0)}=M^{(s+k,s)}_{(j_1,s);(0,1)}=0$, and
$M^{(s+k,s)}_{(s,j_2);(-1,0)}=0$ for $j_2<s$. In ${\cal M}$ the
condition $j_1\geqslant s\geqslant j_2$ is obeyed identically, and
\eq{ccoeff} is equivalent to
\bea M_{0\{r}\ket{{}^{(s+k,s)}_{(s+k,0)}}_{\mV;r(s+k)\}}&=&0\
,\label{Morcond}\eea
providing a differential equation for
$\psi^{(s+k,s)}_{\mV;(s+k,0)}(x^+)$.

Since $\ket{S_{(0,0)}}_{12}=S_{(0,0)}\ket{\1}_{12}$ where
$S_{(0,0)}$ is a constant, it follows that \eq{embeddinggen} is
equivalent to \eq{decunity} for $s=k=0$. In this case \eq{ccoeff}
reads $(M_{ab}(1)+M_{ab}(2))\ket{{\bf M}_{\mV}}_{12}=0$, where the
\emph{reduced reflector}
\bea \ket{{\bf M}_{\mV}}_{12}&=& \ket{{}^{(0,0)}_{(0,0)}}_{\mV}\
=\
(1+\pi)\psi^{(0,0)}_{\mV;(0,0)}(x^+)T^{(0)}_{e^{(\mV)}_0;(0)}\ket{\1}_{12}\
. \label{bfmmv}\eea
Eq. \eq{Ef} can be written as $[E(1)-E(2),f(x^+)]_{\star}=0$ where
$x^+=2L^+_r(1)L^+_r(2)$, which implies that $(E(1)-E(2))\ket{{\bf
M}_{\mV}}_{12}=0$. Thus $(P_a(1)-P_a(2))\ket{{\bf
M}_{\mV}}_{12}=0$, and hence one may identify
\bea \ket{\1}_{12}&=&\sum_{\mV}{\cal
N}^{(0,0)}_{\mV;(0,0)}\ket{{\bf M}_{\mV}}_{12}\ ,\qquad {\cal
N}_{\mV}\ =\ {\cal N}^{(0,0)}_{\mV;(0,0)}\ ,\quad \psi_{\mV}\ =\
\psi^{(0,0)}_{\mV;(0,0)}\ .\eea
Using \eq{Lplusminus} and \eq{ll} the overlap conditions can also
be written as
\bea P_r(1)\ket{{\bf M}_\mV}_{12}&=&\ft{i}2L^+_r\ket{{\bf
M}_{\mV}}_{12}\ =\ -\ft{i}2 L^-_r\ket{{\bf M}_{\mV}}_{12}\
,\label{pr1}\\[5pt] M_{r0}(1)\ket{{\bf M}_{\mV}}_{12}&=&
-\ft{1}2(L^+_r(1)-L^+_r(2))\ket{{\bf M}_{\mV}}_{12}\
,\\[5pt]
M_{rs}(1)\ket{{\bf
M}_{\mV}}_{12}&=&(L^+_{[r}(1)L^+_{s]}(2)+L^-_{[r}(1)L^-_{s]}(2))\ket{{\bf
M}_{\mV}}_{12}\
,\\[5pt] E(1)\ket{{\bf M}_{\mV}}_{12}&=&
\ft{1}{2(D-1)}(L^+_r(1)L^+_r(2)-L^-_r(1)L^-_r(2))\ket{{\bf
M}_{\mV}}_{12}\ .\label{lhs2}\eea
It follows that $(P_{\{r_1}\cdots P_{r_k\}})(1)\ket{{\bf
M}_{\mV}}_{12}=(2i)^{-k}(-1)^{l} L^{+(l)}_{\{r(l)}
L^{-(k-l)}_{r(k-l)\}}\ket{{\bf M}_{\mV}}_{12}$ for
$l=0,1,\dots,k$. Combined with \eq{bfmmv} and $(P_{\{r_1}\cdots
P_{r_k\}})(1)\ket{{\bf
M}_{\mV}}_{12}=(1+\pi)\psi^{(k,0)}_{\mV;(k,0)}(x^+)\ket{(0);e_0^{(\mV)};(k,0)}_{\mV;r(k)}$
where
$\ket{(0);e_0^{(\mV)};(k,0)}_{\mV;r(k)}=f^{(\mV)}_{r(k)}\ket{(0);e_0^{(\mV)};(0)}$
with $f^{(\mV)}_{r(k)}$ being a monomial of degree $k$ in $L^+_r$
and $(L^+_r,L^-_r)$ for $\mV=\mD,\mD'$ and $\mV={\cal M}^{(\pm)}$,
respectively, it follows that ($k=0,1,2,\dots$)
\bea
\psi^{(k,0)}_{\mV;(k,0)}(x^+)&=&\psi^{(0,0)}_{\mV;(0,0)}(x^+)\ =\
\psi_{\mV}(x^+)\qquad \mbox{for $\mV=\mD,\mD'$}\ .\eea
The energy operator, on the other hand, acts as
$$E f(x^+)\ket{(s);e;(j_1,j_2)}\ =\
\left(2x^+\frac{d}{dx^+}+e\right)f(x^+)\ket{e;(j_1,j_2)}\ ,$$
which means that the functional forms of
$\psi^{(k,0)}_{\mV;(j_1,0)}(x^+)$ with $j_1<k$ differ from that of
$\psi_{\mV}(x^+)$ for $\mV=\mD,\mD'$. For example, from
$\ket{{}^{(k,0)}_{\mD;(k-1,0)}}\propto
E\ket{{}^{(k,0)}_{\mD;(k,0)}}$ it follows that
$\psi^{(k,0)}_{\mD;(k-1,0)}(x^+)=\ft1{2\e_0+k}(2x^+\frac{d}{dx^+}+2\e_0+k)
\psi^{(0,0)}_{\mD;(0,0)}(x^+)$.

Next we turn to closer look at the embedding functions in the
different sectors.


\scsss{Composite-massless sector}\label{Sec:comprefl}


A state $L^+_{r_1}\cdots L^+_{r_l} \ket{2\e_0+s;(s)}_{t(s)}\in
\mD(2\e_0+s;(s))$ at excitation level $l$ can be decomposed under
$\ms$ using $L^+_r\ket{2\e_0+s;(s)}_{rt(s-1)}=0$ with the result
\bea L^+_{r_1}\cdots L^+_{r_l}
\ket{2\e_0+s;(s)}_{t(s)}&=&\sum_{n=0}^{[l/2]}\sum_{p=0}^{\textrm{min}(s,l-2n)}
(x^+)^n\ket{(s);2\e_0+l-2n;(s+l-2n-p,p)}\ .\hspace{1.5cm}\eea
Thus, a state $\ket{(s);(j_1,j_2)}\in \mD(s+2\e_0;(s))$ of
type-$(j_1,j_2)$ is of the form
\bea \ket{(s);(j_1,j_2)}&=&\sum_{n=0}^\infty \psi_{(j_1,j_2);n}
(x^+)^n\ket{(s);e^{(\mD)}_0;(j_1,j_2)}\ ,\label{genexp}\eea
where $\psi_{(j_1,j_2);n}$ are arbitrary coefficients,
\bea \ket{(s);e^{(\mD)}_0;(j_1,j_2)}_{r(j_1),t(j_2)}\ =\
L^{+(j_1-s)}_{\{ r(j_1-s)} L^{+(j_2)}_{t(j_2)}
\ket{2\e_0+s;(s)}_{r(s)\}}\ ,\quad e^{(\mD)}_0=j_1+j_2+2\e_0\
,\hspace{1cm}\label{ketj1j2mine}\eea
is the type-$(j_1,j_2)$ state in $\mD(2\e_0+s;(s))$ of minimal
energy, and the series expansion does not collapse for any values
of $s,k,j_1$ and $j_2$ since
$(x^+)^n\ket{(s);e^{(\mD)}_0;(j_1,j_2)}$ are not null states. The
lemma \eq{lminusxn} implies
\bea M_{0\{r} (x^+)^n\ket{2\e_0+p;(p)}_{r(p)\}}&=& \ft12
(1+4n(n+\e_0-1))L^+_{\{r} (x^+)^{n-1}\ket{2\e_0+p;(p)}_{r(p)\}}\
,\eea
where $p=s+k$ and $\ket{2\e_0+p;(p)}_{r(p)}\equiv
L^{+(k)}_{\{r(k)}\ket{s+2\e_0;(s)}_{r(s)\}}$ has the property
$L^-_{\{r_1}\ket{2\e_0+p;(p)}_{r(p)\}}=0$. Hence, independently of
$s+k$, the embedding condition \eq{Morcond} implies
\bea \left(4x^+{d^2\over d(x^+)^2}+4\e_0{d\over
dx^+}+1\right)\psi^{(s+k,s)}_{\mD;(s+k,0)}(x^+)&=&0\
.\label{diffeqn}\eea
The transformation $\psi^{(s+k,s)}_{\mD;(s+k,0)}(x^+)=
y^{-\nu}J(y)$ with $\nu=\e_0-1$ and $x^+=y^2$ brings \eq{diffeqn}
to Bessel's differential equation \eq{Bessel} with index $\nu$ and
source $K=0$. Thus
\bea \psi^{(s+k,s)}_{\mD;(s+k,0)}(x^+)&=&
\psi^{(0,0)}_{\mD;(0,0)}(x^+)\ =\ \psi_{\mD}(x^+)\ =\
\C(\e_0)\left(\frac{y}{2}\right)^{1-\e_0} J_{\e_0-1}(y)
 \ ,\label{psi00D}\eea
as given also in \eq{psimD}. For even $D$ the embedding functions
are algebraic powers times trigonometric functions. In particular,
\bea D=4&:& \ket{{\bf M}_{\mD}}_{12} \ = \
(1+\pi)\cos(y)\ket{1;(0)}_{12}\ . \label{cos}\eea

Alternatively, the Flato-Fronsdal formula \eq{FFformula} yields
the following expression for the contribution from the $\mD^+$
sector to the right-hand side of the embedding formula
\eq{embeddinggen}:
\bea &&\sum_{s+k\geqslant j_1\geqslant s\geqslant j_2\geqslant
0}S^{r(j_1),t(j_2)}_{(s+k,s)}{\cal N}^{(s+k,s)}_{\mD;(j_1,j_2)}
\psi^{(s+k,s)}_{\mD;(j_1,j_2)}(x^+)
 f^{(j_1,j_2)}_{(s)} L^{+(j_1-s)}_{\{
r(j_1-s)}L^{+(j_2)}_{t(j_2)} f_{r(s)\}}(1,2) \ket{2\e_0;(0)}_{12}\
,\hspace{1.5cm} \label{embedding2}\eea
where $f^{(j_1,j_2)}_{(s)}$ are normalizations and
$L^+_r=L^+_r(1)+L^+_r(2)$. The left-hand side of \eq{embeddinggen}
decomposes under $\ms$ as
\bea \sum_{s+k\geqslant j_1\geqslant s\geqslant j_2\geqslant 0}
g^{(s+k,s)}_{(j_1,j_2)}
S^{r(j_1),t(j_2)}_{(s+k,s)}T_{0(s+k-j_1)\{r(j_1),t(j_2)\}0(s-j_2)}(1)\ket{\1}_{12}
\ ,\label{lhs}\eea
where $g^{(s+k,s)}_{(j_1,j_2)}$ are embedding coefficients and
$T_{0(s+k-j_1)\{r(j_1),t(j_2)\}0(s-j_2)}=(M_{rs})^{\star j_2}
\star (M_{r0})^{\star (s-j_2)} \star (P_r)^{\star(j_1-s)}\star
E^{\star(k-j_1+s)}$ plus trace corrections. (Anti)-symmetrizing
under $1\leftrightarrow 2$ using eqs. \eq{pr1}-\eq{lhs2}, and
rewriting the $\mD^+$ sector using \eq{bfmmv} and
\bea L^-_{[r}(1)L^-_{s]}(2)\psi_{\mD}(x^+)\ket{2\e_0;(0)}_{12}&=&
L^+_{[r}(1)L^+_{s]}(2)(D_M
\psi_{\mD})(x^+)\ket{2\e_0;(0)}_{12})\ ,\\[5pt]
L^-_{r}(1)L^-_{r}(2)\psi_{\mD}(x^+)\ket{2\e_0;(0)}_{12}&=& (D_E
\psi_{\mD})(x^+)\ket{2\e_0;(0)}_{12})\ ,\eea
where $D_M$ and $D_E$ are differential operators in $x^+$ with
coefficients that are analytic at $x^+=0$, the type-$(j_1,j_2)$
contribution to the $\mD^+$ sector of \eq{lhs} takes the form
\bea S^{r(j_1),t(j_2)}_{(s+k,s)}{\cal
N}_{\mD}(D^{(s+k,s)}_{(j_1,j_2)}\psi_{\mD})(x^+)
M_{r(j_1),t(j_2)}(1,2)\ket{2\e_0;(0)}_{12}\ ,\label{lhsd+}\eea
where $D^{(s+k,s)}_{(j_1,j_2)}$ are analytic differential
operators in $x^+$ (including the coefficients
$g^{(s+k,s)}_{(j_1,j_2)}$) and $M_{r(j_1),t(j_2)}(1,2)=(-1)^s
M_{r(j_1),t(j_2)}(2,1)$ is a normalized monomial in $L^+_r(1)$ and
$L^+_r(2)$ of degree $2j_2+s-j_2+j_1-s=j_1+j_2$. Equating
\eq{embedding2} and \eq{lhsd+} yields
$M_{r(j_1),t(j_2)}(1,2)=L^{+(j_1-s)}_{\{
r(j_1-s)}L^{+(j_2)}_{t(j_2)}
f_{r(s)\}}(1,2)$, and hence \\[-10pt]$$\psi^{(s+k,s)}_{\mD;(j_1,j_2)}(x^+)\ =\ {{\cal N}_{\mD}\over f^{(j_1,j_2)}_{(s)}{\cal N}^{(s+k,s)}_{\mD;(j_1,j_2)}} \left(D^{(s+k,s)}_{(j_1,j_2)}\psi_{\mD}\right)(x^+)\ .$$\\[-10pt]
In particular, from $M_{r0}=-\ft12(L^+_r+L^-_r)$,
$P_r=\ft{i}2(L^+_r-L^-_r)$ and
$(L^\pm_r(1)+L^\mp_r(2))\ket{\1}_{12}=0$ it follows that the
type-$(s+k,0)$ contribution to the $\mD^+$ sector of \eq{lhs} is
proportional to
$S^{r(s+k)}_{(s+k,0)}(L^+_r(1)+L^+_r(2))^k(L^+_r(1)-L^+_r(2))^s\psi^{(0,0)}_{\mD;(0,0)}(x^+)\ket{2\e_0;(0)}$,
which contains no factors of $M_{rs}$ and $E$, and hence
\bea \psi^{(s+k,s)}_{\mD;(s+k,0)}(x^+)&=&\psi_{\mD}(x^+)\ ,\qquad
D^{(s+k,s)}_{(s+k,s)}\ =\ {f^{(j_1,j_2)}_{(s)}{\cal
N}^{(s+k,s)}_{\mD;(j_1,j_2)}\over {\cal N}_{\mD}}\ .\eea
The trace corrections turns the binomial expansion of
$(L^+_r(1)-L^+_r(2))^s$ into the Flato-Fronsdal formula
\eq{FFformula}. For example, the embedding into
$\mD^+(2\e_0+2;(2))$ of a spin-2 Weyl tensor
$S^{a(2),b(2)}T_{a(2),b(2)}$, which decomposes under $\mso(D-1)$
into $(j_1,j_2)\in \{(2,0),(2,1),(2,2)\}$, contains the
type-$(2,0)$ contribution $S^{r(2)}T_{r(2),0(2)}(1)=\ft43
S^{r(2)}(M_{r_10}\star M_{r_20}+\ft1{2\e_0+1}P_{r_1}\star
P_{r_2})$ as can be seen using \eq{papa} and \eq{macmbc} to expand
\bea M_{\{a(2),b(2)\}_D} & = & \ft43 M_{a_1 b_1}\star M_{a_2
b_2}-\ft{4}{3(2\e_0+1)}\left( \eta_{a(2)}P_{b_1}\star
P_{b_2}-2\eta_{a_1b_1}P_{(a_2}\star
P_{b_2)}+\eta_{b(2)}P_{a_1}\star P_{a_2}\right)\nn\\[5pt]
&+&{4\e_0\over 3(2\e_0+1)}(\eta_{a(2)}\eta_{b(2)}-\eta_{a_1
b_1}\eta_{a_2 b_2})\ ,\nn\eea
and make contact with \eq{FFformula}:
$S^{r(2)}T_{r(2),0(2)}(1)\psi_{\mD}(x^+)\ket{2\e_0;(0)}=$
\bea & = & {2\e_0\over 3(2\e_0+1)}
S^{r(2)}\psi_{\mD}(x^+)\left(L^+_{r_1}(1)L^+_{r_2}(1)-{2(\e_0+1)\over
\e_0}L^+_{r_1}(1)L^+_{r_2}(2)+L^+_{r_1}(2)L^+_{r_2}(2)\right)\ket{2\e_0;(0)}_{12}
\nn\\[5pt]
  &=& {2\e_0\over 3(2\e_0+1)}
S^{r(2)}\psi_{\mD}(x^+)\ket{2\e_0+2; (2,0)}_{12;r_1 r_2}\ .\nn\eea


\scsss{Singleton factorization of composite-massless reflector}


The form \eq{psi00D} of the embedding function
$\psi_{\mD}=\psi^{(0,0)}_{\mD;(0,0)}$ can also be derived starting
from the compositeness relation \eq{envFF}, that is, the
isomorphism $\ket{2\e_0;(0)}_{12}\simeq \ket{\e_0;(0)}_{1}\otimes
\ket{\e_0;(0)}_{2}$. The reduced reflector $\ket{{\bf
M}_{\mD}}_{12}$ can thus be identified as the scalar-singleton
reflector
\bea \ket{{\bf M}_{\mD}}_{12}&=&{\cal
N}_{\mD_0}(1+\pi)\ket{\1_{\mD_0}}_{12}\ ,\qquad
\ket{\1_{\mD_0}}_{12}\ =\ R_2^{-1}(\1_{\mD_0})_{12}\
,\label{reducedrefl}\eea
where ${\cal N}_{\mD_0}$ is a normalization constant, $\1_{\mD_0}$
is the identity operator in $\mD_0$, and the reflection map $R$
induced from \eq{doubleton} and \eq{invrefl} is given by
\bea R(X\ket{\e_0;(0)}^\pm)\ =\ {}^\pm\bra{\e_0;(0)}(\tau\circ
\pi)(X)\ ,\quad R(X\star Y)\ =\ R(Y)\star R(X)\ ,\eea
for $X,Y\in {\cal A}$. It follows that $R(L^{\pm}_r)=-L^\mp_r$,
$R(E)=E$, $R(M_{rs})=M_{rs}$ and
$R(\ket{n}^\pm)=(-1)^n{}^\pm\bra{n}$, where we have defined the
following basis elements for $\mD_0$:
\bea \ket{n}^\pm_{r(n)}&=& \ket{\pm(n+\e_0);(n)}^\pm_{r(n)}\ =\
L^\pm_{r_1}\cdots
L^\pm_{r_n}\ket{\pm\e_0;(0)}^\pm\ ,\label{basisD0ket}\\[5pt]
{}^\pm\bra{n}_{r(n)}&=&{}^\pm\bra{\pm(n+\e_0);(n)}_{r(n)}\ =\
{}^\pm\bra{\e_0;(0)}L^\mp_{r_1}\cdots L^\mp_{r_n}\
.\label{basisD0bra}\eea
They are traceless as a consequence of the singular vector
\eq{singularsingleton} and normalized to
\bea {}^\pm\bra{m}^{r(m)}\ket{n}^\pm_{s(n)}&=&\d_{mn}{\cal
N}_n\d_{\{s(n)\}}^{\{r(n)\}} \ ,\qquad {\cal N}_n \ =\
2^n\,n!(\e_0)_n \ , \label{norm}\eea
as can be seen using $L^-_r\ket{n}_{s(n)} =
2n(n+\e_0-1)\d_{r\{s_1}\ket{n-1}_{s(n-1)\}}$. Defining the normal
order $\cross L^+_r L^-_s\cross=L^+_r L^-_s$ and $z=2L^+_r L^-_r$,
the unity $\1_{\mD_0}=\sum_{n=0}^\infty\, \left[{\cal
N}_n\right]^{-1} \ket{n}\bra{n}$ in $\mD_0$ assumes the form
\bea \1_{\mD_0} & = & \sum_{n=0}^\infty \cross
 {(L^+_r L^-_r)^n\over 2^n n!(\e_0)_n}\ket{\e_0;(0)}\bra{\e_0;(0)} \cross\ =\ {\C(\n+1) 2^{\nu}\over z^{\nu/2}}\cross
I_\n(\sqrt{z})\ket{\e_0;(0)}\bra{\e_0;(0)}\cross\ ,\eea
where $\nu=\e_0-1$ and the modified Bessel function $I_\n(w)=
e^{-\frac{i\pi\n}{2}}J_\n(e^{\frac{i\pi}{2}}w)$ for
$-\pi<\textrm{arg}\,w\leqslant\frac{\pi}{2}$ and
$I_\n(w)=e^{\frac{2i\pi\n}{3}}J_\n(e^{-\frac{3i\pi}{2}}w)$ for
$-\frac{\pi}{2}<\textrm{arg}\,w\leqslant\pi$. Applying
$R^{-1}:\mD_0^\ast\mapsto \mD_0$ to the $\bra{n}$ states in the
expansion of $\1_{\mD_0}$ yields $(-1)^n\ket{n}$. Using
$R^{-1}_2(z)=-2L^+_r(1)L^+_r(2)=-x$, which formally implies
$R^{-1}(\sqrt{z})=iy$, one finds
\bea \ket{\1_{\mD_0}}_{12}&=& {\C(\n+1)2^\nu\over(iy)^{\n}}
I_\n(iy)\ket{\e_0;(0)}_1\ket{\e_0;(0)}_2\ =\
\psi_{\mD}(x^+)\ket{\e_0;(0)}_1\ket{\e_0;(0)}_2\
,\label{leftidD0}\eea
with the embedding function given in \eq{psi00D} and in agreement
with \eq{reducedrefl}. In particular, in $D=4$ one has
\bea \1_{\mD_0}&=&\cross
\cosh{\sqrt{z}}\ket{\ft12;(0)}\bra{\ft12;(0)} \cross \ ,\qquad
\ket{\1_{\mD_0}}_{12}\ =\ \cos y
\ket{\ft12;(0)}_1\ket{\ft12;(0)}_2\ .\label{4Dsinglunity}\eea


\scsss{Remaining sectors and logarithmic contributions}


In the sector $\mD'$ the embedding condition \eq{ltilde1}, or
equivalently \eq{Morcond}, implies
\bea
D_2\psi_{\mD'}(x^+)&=&\left(4x^+\frac{d^2}{d(x^+)^2}+4(2-\e_0)\frac{d}{dx^+}+1\right)\psi_{\mD'}(x^+)
\ =\ 0 \ ,\eea
which upon $\psi_{\mD'}(x^+)=y^{-\nu'}J(y)$ with $\nu'=1-\e_0$ is
transformed to Bessel's differential equation \eq{Bessel} with
index $\nu'$ and source $K=0$. Taking into account also the
boundary condition $\psi_{\mD'}(0)=1$, it follows that
$\psi_{\mD'}(x^+)$ is given by the rescaled Bessel functions in
\eq{psimDprime} if $\nu'\neq-1,-2,\dots$ and the rescaled Neumann
functions in \eq{psimDprimelog} if $\nu'=-1,-2,\dots$. In the
latter case the lowest-weight space $\mD(2;(0))$ is the scalar
$p$-lineton in $D=5+2p=7,9,\dots$, $p=1,2,\dots$, and the
logarithmic tail in $\psi_{\mD'}(x^+)$ starting at order
$(x^+)^p\log x^+$ reflects the singular nature of the states
$(x^+)^p\ket{2;(0)}$. In $D=4$ the isomorphism
$\ket{2;(0)}_{12}\simeq \ket{1;(\ft12)}^i_1\otimes
\ket{1;(\ft12)}_{2,i}$ implies that the reduced reflector
$\ket{{\bf M}_{\mD'}}_{12}$ is proportional to the
spinor-singleton reflector
\bea D=4&:&\ket{{\bf M}_{\mD'}}_{12}\ =\ {\cal N}_{\mD_{1/2}}
(1+\pi) \ket{\1_{\mD_{\ft12}}}_{12}\ ,\qquad
\ket{\1_{\mD_{1/2}}}_{12}\ =\ R_{2}^{-1} \1_{\mD_{1/2}} ,\eea
where ${\cal N}_{\mD_{1/2}}$ is a normalization constant, the
spinor-singleton identity operator
\bea \1_{\mD_{1/2}}&=&\sum_{n=0}^\infty \left[{\cal
N}_{n+\ft12}\right]^{-1}\ket{n+\ft12}^{i,r(n)}\bra{n+\ft12}_{i,r(n)}\
,\eea
with $\gamma$-traceless basis states and normalizations given by
\bea \ket{n+\ft12}^\pm_{i,r(n)}&=& L^\pm_{r_1}\cdots L^\pm_{r_n}\ket{\pm 1;(\ft12)}^\pm_{i}\ ,\\[5pt]
{}^\pm\bra{n+\ft12}^{i,r(n)}\ket{m+\ft12}^\pm_{j,s(m)}&=&\delta_{mn}\d^i_j\delta^{\{r(n)\}}_{\{s(n)\}}\
,\quad {\cal N}_{n+\ft12}\ =\ 2^n n!(\ft32)_n\ ,\eea
as can be seen using $L^-_r\ket{n+\ft12}_{i,s(n)} =
2n(n+\ft12)\d_{r\{s_1}\ket{n-\ft12}_{i,s(n-1)\}}$ where
$\{\cdots\}$ denotes the $\gamma$-traceless projection, and the
action of the reflection map is defined by
\bea R(\ket{\pm 1;(\ft12)}^\pm_{i})&=& i{}^\pm\bra{\pm
1;(\ft12)}_i\ .\eea
One has the following agreement:
\bea D=4&:& \ket{\1_{\mD_{1/2}}}_{12}\ =\ i{\sin y\over y}
\ket{1;(\ft12)}^i_1~
\ket{1;(\ft12)}_{2,i}\quad\leftrightarrow\quad\psi^{(0,0)}_{\mD';(0,0)}(x^+)
\ =\ {\sin y\over y} \ .\label{spinrefl}\label{tpsi004}\eea

In the sector ${\cal W}^{(-)}$ the condition \eq{ltilde1} reads
\bea (\widetilde L^+_r+\widetilde
L^-_r)(1+\pi)\psi_{(-)}(\widetilde
x^+)T^{(0)}_{1;(0)}&=&(1+\pi)\widetilde L^+_r
(D_1\psi_{(-)})(\widetilde x^+)T^{(0)}_{1;(0)}\ =\ 0\ ,\eea
which is obeyed if
\bea D_1\psi_{(-)}(x^+)~=~
\left(4x^+\frac{d^2}{d(x^+)^2}+4(1-\e_0)\frac{d}{dx^+}+1+{2\e_0-1\over
x^+}\right)\psi_{(-)}(x^+) ~=~{2\e_0-1\over x_+} \
,\qquad\label{struve}\eea
since \eq{Ce} and \eq{tildeCe} implies that
\bea (1+\pi)\widetilde L^+_r {1\over \widetilde
x^+}T^{(0)}_{1;(0)}&=& {\cal C}_{-1}(1+\pi)\widetilde L^+_r
T^{(0)}_{-1;(0)}\ =\ {\cal C}_{-1}(1+{\cal C}'_1)\widetilde L^+_r
T^{(0)}_{-1;(0)}\ =\ 0\ .\qquad\eea
The transformation $\psi_{(-)}(x^+)=1+y^{\e_0} J(y)$, $x^+=y^2$,
brings \eq{struve} to Bessel's differential equation \eq{Bessel}
with index $\nu'=1-\e_0$ and source $K=-y^{-\e_0}\ =\
-y^{\nu'-1}$, whose particular solution is the Struve function if
$\nu'\neq -\ft12,-\ft32,\dots$ and the modified Struve function if
$\nu'=-\ft12,-\ft32,\dots$, as given in \eq{psiwminus} and
\eq{psiwminuslog}. In the latter case $D=4+2p=6,8,\dots$,
$p=1,2,\dots$, and the logarithmic tail in $\psi_{(-)}(x^+)$
starting at order $(x^+)^p\log x^+$ corresponds to the fact that
the states $(x^+)^p\ket{1;(0)}\in \mD(2\e_0;(0))\subset \mD$.

Finally, in the sector ${\cal W}^{(+)}$ the condition \eq{ltilde1}
reads
\bea (\widetilde L^+_r+\widetilde
L^-_r)(1+\pi)\psi_{(+)}(\widetilde
x^+)T^{(0)}_{0;(0)}&=&(1+\pi)\widetilde L^+_r
(D_0\psi_{(-)})(\widetilde x^+)T^{(0)}_{0;(0)}\ =\ 0\ ,\eea
which is obeyed if
\bea D_0\psi_{(-)}(x^+)&=&
\left(4x^+\frac{d^2}{d(x^+)^2}-4\e_0\frac{d}{dx^+}+1+{4\e_0\over
x^+}\right)\psi_{(-)}(x^+) \ =\ {4\e_0\over x_+}-1 \
,\label{psiplus1}\eea
since \eq{Ce} and \eq{tildeCe} implies that
\bea (1+\pi)\widetilde L^+_r \left({4\e_0\over \widetilde
x^+}-1\right)T^{(0)}_{0;(0)}~=~\lim_{e\rightarrow
0}\left({4\e_0\over \widetilde x^+}\widetilde L^+_r +{4\e_0\over
\widetilde x^-}\widetilde L^-_r\right)T^{(0}_{e;(0)}-(\widetilde
L^+_r +\widetilde L^-_r)T^{(0)}_{0;(0)}~=~0\ .\qquad\eea
The limiting procedure can be avoided by writing
$\psi_{(+)}(x^+)=1-{1\over 2\e_0} x^+ \chi_{(+)}(x^+)$, that is
\bea (1+\pi)\psi_{(+)}(\widetilde
x^+)T^{(0)}_{0;(0)}&=&(1+\pi)\left(T^{(0)}_{0;(0)}+
2\chi_{(+)}(\widetilde x^+)T^{(0)}_{2;(0)}\right)\ .\eea
The condition \eq{ltilde1} then reads
\bea (\widetilde L^+_r+\widetilde
L^-_r)(1+\pi)\left(T^{(0)}_{0;(0)}+ 2\chi_{(+)}(\widetilde
x^+)T^{(0)}_{2;(0)}\right)=2(1+\pi)\widetilde L^+_r
\left(T^{(0)}_{0;(0)}+ (D_2\chi_{(+)})(\widetilde
x^+)T^{(0)}_{2;(0)}\right)=0\ , &&\qquad\nn\eea
which is obeyed if
\bea D_2\chi_{(+)}(x^+)&=&
\left(4x^+\frac{d^2}{d(x^+)^2}+4(2-\e_0)\frac{d}{dx^+}+1\right)\psi_{(-)}(x^+)
\ =\ {4\e_0\over x_+} \ ,\label{chiplus}\eea
since \eq{Ce} and \eq{tildeCe} implies that ${4\e_0\over
\widetilde x^+}T^{(2)}_{0;(0)}=-T^{(0)}_{0;(0)}$, and
\eq{psiplus1} is indeed equivalent to \eq{chiplus}. By the
rescaling $\chi_{(+)}(x^+)=y^{-\nu'} J(y)$ brings \eq{chiplus} to
Bessel's differential equation \eq{Bessel} with index
$\nu'=1-\e_0$ and source $K=4\e_0y^{\nu'-2}$ with solution
\eq{psiwpluseven} if $\nu'\neq 0,1,\dots$. The logarithm starting
at order $x^+\log x^+$ corresponds to the lowest-weight nature of
$T^{(0)}_{2;(0)}$. If $\nu'=0$ the element $T^{(0)}_{2;(0)}$ is a
two-fold root to the lowest-weight condition leading to quadratic
logarithms $x^+(\log x^+)^2$. Likewise, if $D=5+2p=7,\dots$,
$p=1,2,\dots$ then the quadratic logarithms set in at order
$(x^+)^{p+1}(\log x^+)^2$. We leave the analysis of these cases
for future work.


\scss{Adjoint singleton-anti-singleton composites}\label{Sec:Adj}


As seen in Section \ref{sec:FT}, an element $Q\in\mho$ can be
mapped via the twisted reflector $\ket{\j1}_{12}$ to the state
$\ket{\widetilde Q}_{12}$ carrying the untwisted
$\mho$-representation $\ket{\adj_Q \widetilde Q'}_{12}
=Q\ket{\widetilde Q'}_{12}$. Since
$\ket{\j1}_{12}=k(2)\ket{\1}_{12}$ and $\ket{\1}_{12}$ contains
the reduced singleton-singleton reflector $\ket{{\bf
M}_{\mD}}_{12}$ given in \eq{leftidD0} and \eq{reducedrefl}, it
follows that $\ket{\j1}_{12}$ contains the reduced
singleton-anti-singleton reflector
\bea \ket{\widetilde {\bf M}_{\mD}}_{12}&=& k(2)\ket{{\bf
M}_{\mD}}_{12}\ =\ (1+\pi)\sum_{n=0}^\infty [{\cal
N}_n]^{-1}\ket{n}^+_{1;r(n)} \ket{n}^-_{2;r(n)}\
,\label{reducedrefltwiddle}\eea
where we use the basis \eq{basisD0ket}. One can show that
$(M_{AB}(1)+M_{AB}(2))\ket{\widetilde {\bf M}_{\mD}}_{12}=0$. The
$\ell$-th adjoint level, which is the lowest-and-highest-weight
space ($s=2\ell+2$)
\bea {\cal L}_\ell&\simeq & \mD(-(s-1);(s-1))\ =\ \bigoplus_{\tiny
\ba{c}e\in\integ,n\geqslant 0 \\ j_1\geqslant j_2\geqslant 0\\ |e|\leqslant j_1-j_2\\
j_1+n\leqslant s-1\ea}\Comp\otimes M_{e;(j_1,j_2);n}\
,\label{Lambdaell}\eea
with $M_{e;(j_1,j_2);n}$ given below \eq{Xs}, is thus reflected by
$\ket{\widetilde {\bf M}_{\mD}}_{12}$ to a bimodule containing the
lowest-weight and highest-weight states
\bea \ket{\mp(s-1);(s-1)}^\pm_{12;r(s-1)}&=& (L^\mp_{\{r_1}\cdots
L^\mp_{r_{s-1}\}})(1)\ket{\widetilde {\bf M}_{\mD}}_{12}\
,\label{adjlws}\eea
where we note that $L^\mp_r\ket{\mp(s-1);(s-1)}_{12}=0$ follows
from the $\mg$-invariance of $\ket{\widetilde {\bf
M}_{\mD}}_{12}$. Thus one has
\bea \ket{\mp(s-1);(s-1)}^\pm_{12;r(s-1)}&=&
(1+\pi)\sum_{n=0}^\infty {(-1)^n\over 2^n n!(\e_0)_n}
\ket{n+s-1}^\mp_{1;r(s-1)t(n)} \ket{n}^\pm_{2;t(n)}\ ,\eea
that is, the tensor product of a scalar singleton and a scalar
anti-singleton can be expanded in terms of finite-dimensional
adjoint levels, leading to the \emph{twisted Flato-Fronsdal
formula}:
\bea(\mD^+_0\otimes\mD^-_0)\oplus (\mD^-_0\otimes\mD^+_0)&=&
\bigoplus_{s=0}^\infty \mD(-(s-1);(s-1))\ .
\label{FFformulatwisted}\eea
%


\scs{Conclusions}


The unfolded formulation of dynamics provides a dual fiber
description of standard field equations in terms of functions in
the enveloping algebra of the underlying isometry group and
associated finite-dimensional as well as infinite-dimensional
representations. This approach is natural in the context of
higher-spin gauge theories. In this paper we have used it to
examine the harmonic expansion of simple bosonic higher-spin gauge
theories around constantly curved backgrounds. We have focused on
the case of $AdS_D$ with the aim of assessing the basic premises
(i)--(iv) listed at the end of Section \ref{Sec:Runaway}.

Indeed, we have confirmed properties (i)--(iii). In particular,
the factorization of ${\cal M}^{(+)}$ has been found to be given
in terms of the angleton ${\cal S}^{(+)}$ which is the lowest-spin
module given by the left action on the static ground state
$T^{(0)}_{0;(0)}$. Our nomenclature is motivated by the fact that
${\cal S}^{(+)}$ fills a wedge in compact weight space, as opposed
to the singleton $\mD_0$ which fills a single line. The relation
between ${\cal M}^{(+)}$ and the angletons (see eq. \eq{statsing})
involves an equivalence relation $\sim$, which is a form of
``gauging'' needed in order to avoid degeneracy, and that has no
analog in the Flato-Fronsdal formula. Moreover, although ${\cal
M}^{(+)}$ contains $(1+\pi)(\mD_0\otimes (\mD_0)^\ast)$, the
angleton does not contain $(1+\pi)\mD_0$ nor any of the states in
its negative-spin extension discussed in Appendix
\ref{App:negspin}. The field theoretic interpretation of the
angletons therefore poses an interesting problem, as does that of
the half-integer one-sided spins in ${\cal S}^{(0)}_{e;(0)}$ for
odd $e+D$ and $D>4$.

We have also found that the module ${\cal W}_{(0)}^{(+)}$, which
contains even runaway scalar fields, is unitarizable for all $D$.
The situation is intriguing since ${\cal W}_{(0)}^{(+)}$ forms an
indecomposable $\mg$-structure together with $\mD(2;(0))$ which is
known to be unitarizable only if $D\leqslant 7$. Moreover, the
twisted-adjoint $\mho$ action yields higher-spin lowest-spin
modules ${\cal W}^{(+)}_{(s)}$ containing elements with negative
rescaled trace norms such as the static ground state
$T^{(1)}_{0;(1,1)}$ in ${\cal W}^{(+)}_{(1)}$. It might be the
case, however, that the trace norm remains (negative or positive)
definite for fixed $s$, and a decisive analysis should take into
account additional phase-factors coming from the definitions of
real forms of the higher-spin generators and possible internal
sectors. In concluding, we also note that there exists an
inequivalent inner product on ${\cal M}$ induced from the
(non-definite) inner product on the angleton.

An important issue left out in this paper is related to the
property (v), namely whether ${\cal M}$ can be equipped with an
associative structure such that the completeness relation
$\1_{\cal M}\star\1_{\cal M}=\1_{\cal M}$ can be imposed on the
decomposition of the unity given in \eq{decunity}. This issue is
related to the question whether the linearized solutions in ${\cal
M}$ admit non-linear completions within Vasiliev's equations.
Physically speaking, such solutions would describe ``one-body''
systems whose extensions to ``two-body'' systems might provide an
effective potential in some limit. We plan to give more details on
this in \cite{companion}.

A related omitted topic is the inclusion of compact-weight
elements that are singular functions on the enveloping algebra. In
the case of $AdS_D$, each analytic compact $\mso(2)\oplus
\mso(D-1)$-type $T^{(s)}_{e;(j_1,j_2)}$ is accompanied by a dual
non-analytic element $\widetilde T^{(s)}_{e;(j_1,j_2)}$ forming a
dual module $\widetilde{\cal M}$. Since $T^{(s)}_{e;(j_1,j_2)}$
corresponds to the generalized spherical harmonic function that is
finite at $r=0$, its dual must correspond to the singular
fluctuation field that blows up at $r=0$. Moreover, for $D=4$ and
$s=0$, the duals of particles have logarithmic singularities at
$E=0$, while the duals of runaway solutions have poles at $E=0$.
Thus one might speculate that the $r\leftrightarrow 1/r$ symmetry
mentioned in Section \ref{Sec:Runaway} corresponds to that ${\cal
M}$ and $\widetilde{\cal M}$ have dual indecomposable structures
viewed as $\mho(D+1;\Comp)$-modules (as defined in Appendix
\ref{App:B}), and that this $\integ_2$ symmetry extends to the
full level. For example, on the Euclidean $S^D$ minus the
north-pole $N$ and the south-pole $S$, this symmetry would
exchange solitons/runaways centered at $N$ with runaways/solitons
centered at $S$.

Another direction to pursue is the extension of the phase-space
quantization program from finite-dimensional non-compact algebras
to infinite-dimensional non-compact algebras. In particular we
would like to realize the spectrally flowed multipleton vertex
operators of the subcritical WZW models discussed in
\cite{Engquist:2007pr} directly in terms of the Kac-Moody
currents, without resorting to free-field constructions, and then
study whether these constructions could be extended to the
angletons.

\vspace{1.5cm}

{\bf Acknowledgements}

It is a pleasure to thank N. Boulanger, P. P. Cook, F. A. Dolan,
J. Engquist, M. Gunaydin, J. Mourad, T. Poznansky, F. Riccioni, A.
Sagnotti, E. Sezgin, M. A. Vasiliev, and P. West for useful
discussions, and P. Rajan for collaboration at an early stage of
this work. This has been supported in part by the EU contracts
MRTN-CT-2004-503369 and MRTN-CT-2004-512194 and the NATO grant
PST.CLG.978785.

\vspace{1.5cm}


\begin{appendix}



\scs{Basic Properties of some infinite-dimensional
$\mso(D+1;\Comp)$ modules}\label{App:B}

This Appendix contains some basic properties of various
infinite-dimensional unitarizable representations of
$\mso(D+1;\Comp)$ that occur in the text. General treatises can be
found for example in \cite{Fuchs:1997jv,Humphreys:1980dw,Knapp}.

\scss{Harish-Chandra modules and Verma modules}


An infinite-dimensional irreducible $\mg$ module $\mR$ can be
``sliced'' in many different ways. The slicing under a Cartan
subalgebra $\mg^{(0)}$ yields the \emph{weight-space
representation} $\mR'=\bigoplus_{\l}{\rm mult}(\l)\mR_\l$ where
$\mR_\l=\Comp\otimes \ket{\l}$ with $\ket{\l}$ being an eigenstate
of the generators in $\mg^{(0)}$ with eigenvalues
$\l=(\lambda_1,\dots,\lambda_r)\in \Comp^{r}$, $r=\dim \mg^{(0)}$,
referred to as a weight vector, or just weight, and the
multiplicities ${\rm mult}(\l)\in\{1,2,\dots\}\cup\{\infty\}$. If
$\mR$ has finite dimension then $\mR\simeq \mR'$ while this need
not hold true if $\mR$ has infinite dimension. The slicing
$\mR|_{\mh}$ of $\mR$ under a subalgebra $\mh\subset \mg$ is said
to be \emph{admissible} if it contains only finite-dimensional
$\mh$ irreps, sometimes referred to as $\mh$-types, with finite
multiplicities, \emph{viz.}
\bea \mR|_{\mh}&=& \bigoplus_{\kappa} {\rm
mult}(\kappa)\mR_\kappa\ ,\qquad {\rm mult}(\kappa)
\dim\mR_\kappa<\infty\ ,\label{hcmodule}\eea
where $\k$ are referred to as the \emph{compact weights} of $\mh$
(we note that if $\mg^{(0)}\subset \mh$ then the further
decomposition of $\mR|_{\mh}$ under $\mg^{(0)}$ might yield
weights $\l$ with ${\rm mult}(\l)=\infty$). A remarkable theorem
(see for example \cite{Knapp}) states that if $\mR$ is a unitary
representation of a real form of $\mg$ with maximal compact
subalgebra $\mh$ then ${\rm mult}(\kappa)\leqslant \dim
\mR_\kappa$. If all $\mh$-types are generated by ${\cal U}[\mg]$
starting from a finite number of $\mh$-types then $\mR$ is
referred to as a \emph{Harish-Chandra module}. In particular, if
$\mR\simeq \mR|_{\mh}$ we shall refer to $\mR$ as a $(\mg|\mh)$
module.

If $\mR=\bigoplus_\kappa \mR_\k$ is a left $(\mg|\mh)$ module with
dual right module $\mR^\ast=\bigoplus_\kappa \mR^{\ast\k}$, then
the resulting matrix representation of $\mg$ in $\mR$ and
$\mR^\ast$ can be written as $\r(X)=\sum_{\k,\k'}v_\k \otimes
v^{\ast\k'} \rho^{\k}{}_{\k'}(X)$. Equipping $\mR$ with the
symmetric bilinear inner product $\eta_{\k\k'}\equiv
(v_\k,v_{\k'})|_{\mR}\equiv
\delta_{\k\k'}(v_\k,v_{\k'})_{\mR_\k}$, so that $\mR$ and
$\mR^\ast$ become isomorphic as vector spaces, makes $\mR^\ast$
equivalent to the \emph{dual} left module
$\widetilde\mR=\bigoplus_\kappa \widetilde\mR_\k$ with
representation matrix $\widetilde
\rho(X)=-(\eta\rho(X)\eta^{-1})^T$.

A reducible $\mg$ module $\mR$ is said to be \emph{indecomposable}
if it contains a proper ideal $\mI$ whose complement is not
invariant. This is denoted by $\mR=\mR_1\ssum \mR_2$ where
$\mR_1\simeq \mI$ and $\mR_2\simeq \mR/\mI$. Thus, the
representation of $\mg$ in $\mR$ is of the form $\rho=
\mx{[}{cc}{\r_1&\a_{12}\\0&\r_2}{]}$, where $\r_i$ are the
representations of $\mg$ in $\mR_i$, $i=1,2$, and the off-diagonal
piece $\a_{12}:\mR_2\rightarrow \mR_1$, referred to as a cocycle,
obeys $\r_1(X)\a_{12}(Y)+\a_{12}(X)\r_2(Y)-(X\leftrightarrow Y)\
=\ \a_{12}([X,Y])$ for all $X,Y\in \mg$. It follows that
$t\a_{12}$ with $t\in\Comp$ is a cocycle of a rescaled
indecomposable structure $\rho^t$, such that $\rho^0$ is
decomposable, and that the dual left module $\widetilde\mR=
\widetilde\mR_1\ssumr\widetilde\mR_2$ carries the dual
indecomposable structure
$\widetilde \r =\mx{[}{cc}{\widetilde \r_1&0\\
\widetilde\a_{21}&\widetilde \r_2}{]}$ with
$\widetilde\a_{21}(X)=-(\eta_1 \a_{12}(X)\eta_2^{-1})^T$.

A \emph{highest-weight} module $\mD(\L)$ is a $(\mg|\mg^{(0)})$
module in which the weights are bounded from above by a highest
weight $\L$ in the sense that the corresponding (unique)
highest-weight state $\ket{\L}\in \mD(\L)$ can be reached by
successive maximization of $\l_k$ keeping $\l_{k'}$, $k'<k$
maximal, inducing a weak three-grading $\mg=\mg^{(-)}\oplus
\mg^{(0)}\oplus \mg^{(+)}$ wherein $\mg^{(0)}$ is the Cartan
subalgebra, $\mg^{(+)}\ket{\L}=0$,
$[\mg^{(\pm)},\mg^{(\pm)}]\subseteq \mg^{(\pm)}$ and
$[\mg^{(0)},\mg^{(\pm)}]\subseteq \mg^{(\pm)}$. The corresponding
infinite-dimensional \emph{Verma module}, or cyclic
$(\mg|\mg^{(0)})$ module, $\mV(\L)\equiv {\cal
U}[\mg^{(-)}]\ket{\L}$ (where we recall that ${\cal U}[\mg]$
denotes the enveloping algebra of a Lie algebra $\mg$) is
irreducible for generic $\L$ and indecomposable for critical $\L$
in which case it contains at least one excited state that is
annihilated by $\mg^{(+)}$, referred to as a \emph{singular
vector}. The ${\cal U}[\mg^{(-)}]$ action on the singular vectors
generates a maximal ideal $\mN(\L)\subset \mV(\L)$, sometimes
referred to as the null module, and
\bea \mV(\L)&=& \mD(\L)\ssumr \mN(\L)\ ,\qquad \mD(\L)\ =\
{\mV(\L)\over \mN(\L)}\ .\label{A1quo}\eea

In the case of $\mg=\mso(D+1;\Comp)$ the finite-dimensional
representations are \emph{highest-and-lowest-weight} spaces. In
Euclidean signature $\eta_{AB}=(+,\dots,+)$ and with
$(M_{D+3-2k,D+2-2k}-\lambda_k)\mid \l\rangle=0$ for
$k=1,\dots,\nu=[(D+1)/2]$, these arise for highest weights obeying
\bea \L\in\integ^r\cup (\integ+\ft12)^r\ ,\qquad \L_1\geqslant
\L_2\geqslant\cdots\geqslant \L_\nu\geqslant 0\ ,\eea
which we shall refer to as positive integer or positive
half-integer highest weights, or more shortly, integer or
half-integer $\mg$-spins. They correspond to traceless tensors or
gamma-traceless tensor-spinors, respectively, in shapes with
$\L_k$ cells in the $k$th row (as can be seen by going to helicity
basis). In the tensorial case the lowest-weight state is
$\ket{-\L}$, thus obeying $\mg^{(-)}\mid -\L\rangle=0$.

\scss{Lowest-weight representations}\label{App:Blws}

Upon Wick-rotating the $D$ and $D+1$ directions into time-like $0$
and $0'$ directions and going to the helicity basis $V^\pm=V^0\mp
i V^{0'}$ and $V_\pm=\ft12(V_0\mp i V_{0'})$ with $\eta^{+-}=-2$,
one identifies the energy operator and the energy-raising and
energy-lowering ladder operators
\bea E&=&-M_{D+1,D}\ =\ M_{0'0}\ =\ i M_+{}^+\ =\ -2i M_{+-}\ ,\\[5pt]L^\pm_r&=& M_{0r}\mp iM_{0'r}\ =\ M_{0r}\mp iP_r\ =\ M_r{}^\pm\ =\ -2 M_{r\mp}\ ,\label{Lplusminus}\eea
with $P_a=M_{0' a}=(E,P_r)$. The resulting commutation rules read
\bea [L^-_r,L^+_s] & = & 2iM_{rs}+2\d_{rs}E \ ,\qquad
[E,L^{\pm}_r] \ =\ \pm L^{\pm}_r \ ,\qquad [M_{rs},L^\pm_t]\ =\
2i\d_{t[s}L^\pm_{r]}\ , \label{el}\label{ll}\label{ml}\eea
and $[M_{rs},M_{tu}]=4i\d_{[t|[s}M_{r]|u]}$, exhibiting a
three-grading whereby $\mg=\ml^{-}\oplus (\me\oplus
\ms)\oplus\ml^+$ such that $[E,\ml^\pm]=\pm \ml^\pm$ and hence
$[\ml^\pm,\ml^\pm]=0$. In this basis, the $\mg$-automorphism $\pi$
defined in \eq{pimap} acts as $\pi(L^\pm_r)= L^\mp_r$, $\pi(E)=-E$
and $\pi(M_{rs})=M_{rs}$, and extends to a map between
$(\mg|\me\oplus\ms)$ modules as the reflection
$\pi(\ket{e;\overrightarrow{s}})=\ket{-e;\overrightarrow{s}}$. The
$\mg$-antiautomorphism $\tau$ defined in \eq{taumap} extends to
the reflection
\bea \tau(\ket{e;\overrightarrow{s}})&=&
\varphi(e;\overrightarrow{s})\bra{-e;\overrightarrow{s}}\ ,\qquad
\tau^2\ =\ \left\{\ba{ll}\textrm{Id}, \ &\mbox{integer
spin}\\-\textrm{Id}, \ &\mbox{half-integer spin}\ea\right.\quad\
,\label{tau2}\eea
where ${}^\pm\bra{-e;\overrightarrow{s}}$ belong to dual
$(\mg|\me\oplus \ms)$ modules and $\varphi(e;\overrightarrow{s})$
are phase factors.

One particular type of infinite-dimensional representations of
$\mso(D+1;\Comp)$ are highest-weight spaces in which
$\overrightarrow{s}_0\equiv (\L_2,\dots,\L_\nu)$ is a spin of
$\ms\simeq\mso(D-1;\Comp)$, while the lowest eigenvalue $e_0\equiv
-\L_1$ of the energy operator $E=-M_{D+1,D}$ is no longer
quantized. Splitting $\mg^{(\pm)}=\ms^{(\pm)}\oplus \ml^\mp$,
where $\ms^{(\pm)}\in \ms\cap \mg^{(\pm)}$ and $\ml^\pm$ are the
``effective'' ladder operators obeying $[\ml^\pm,\ml^\pm]=0$, it
follows that $\mV(\L)={\cal U}[\ml^{+}]{\cal
U}[\ms^{(-)}]\ket{\L}$, where the $\ms$-submodule ${\cal
U}[\ms^{(-)}]\ket{\L}$ contains a null $\ms$-module
$\mN(\ms|\overrightarrow{s}_0)$ such that
$\mC_{e_0;\overrightarrow{s}_0}\equiv {{\cal
U}[\ms^{(-)}]\ket{\e;\overrightarrow{s}_0}\over
\mN(\ms|\overrightarrow{s}_0) }$ is the ($\me\oplus\ms$)-type with
(minimal) energy $e_0$. Factoring out the generic $\mg$ ideal
${\cal U}[\ml^{+}]\mN(\ms|\overrightarrow{s}_0)\subseteq\mN(\L)$
from $\mV(\L)$ yields the \emph{generalized Verma module}
\bea \mC(e_0;\overrightarrow{s}_0)&=& {\mV(\L) \over {\cal
U}[\ml^{+}]\mN(\ms|\overrightarrow{s}_0)}\ =\ {\cal
U}[\ml^{+}]\,\mC_{e_0;\overrightarrow{s}_0}\ =\
\bigoplus_{e\geqslant e_0;\overrightarrow{s}}{\rm
mult}(e;\overrightarrow{s})\,\mC_{e;\overrightarrow{s}}\
,\label{genverm}\eea
which is thus a $(\mg|\me\oplus\ms)$ module with energies bounded
from below by $e_0$. Thus we have a \emph{lowest-energy} state
$|e_0;\overrightarrow{s}_0\rangle^+\equiv
|(\L_1,\overrightarrow{s}_0)\rangle$ and a \emph{highest-energy}
state $|-e_0;\overrightarrow{s}_0\rangle^-\equiv
\pi(|e_0;\overrightarrow{s}_0\rangle$ obeying
\bea (E\mp e_0)\ket{\pm e_0;\overrightarrow{s}_0}^\pm\ =\ 0\
,\qquad L^\mp_r\ket{\pm e_0;\overrightarrow{s}_0}^\pm\ =\ 0\ ,\eea
and the restriction of $\mV(\L)$ and $\pi(\mV(\L))$, respectively,
to the subspace ${\cal U}[\ml^\pm]\ket{\pm
e_0;\overrightarrow{s}_0}^\pm$ yields
\bea \mC^\pm(\pm e_0;\overrightarrow{s}_0)&=&\bigoplus_{\pm
e\geqslant e_0;\overrightarrow{s}}\Comp\otimes
\ket{e;\overrightarrow{s}}^\pm\ ,\qquad
\ket{e_0+n;\overrightarrow{s}}=\mathbf P_{\overrightarrow{s}}
\left[(L^\pm)^n\ket{\pm e_0;\overrightarrow{s}_0}^\pm\right]\
,\quad\eea
where, in a slight abuse of notation, we are denoting
$\mC_{e;\overrightarrow s}$ by $\ket{e;\overrightarrow s}$. The
singular vectors are states
$\ket{e^{\prime}_0;\overrightarrow{s}'_0}^\pm$ with $\pm
e^{\prime}_0>e_0$ obeying
$L^\mp_r\ket{e^{\prime}_0;\overrightarrow{s}'_0}^\pm=0$. They
generate generalized Verma submodules forming a maximal ideal
$\mI^\pm(\pm e_0;\overrightarrow{s}_0)$ and
\bea \mD^\pm(\pm e_0;\overrightarrow{s}_0)&=&{\mC^\pm(\pm
e_0;\overrightarrow{s}_0)\over \mI^\pm(\pm
e_0;\overrightarrow{s}_0)}\ ,\eea
where thus $\mD^+(e_0;\overrightarrow{s}_0)=
\mD(\L_1,\overrightarrow{s}_0)$ and
$\mD^-(-e_0;\overrightarrow{s}_0)\equiv \pi(\mD(\L))$. For generic
$e_0$ the energy spectra are unbounded in
$\mD^+(e_0;\overrightarrow{s}_0)$ and
$\mD^-(-e_0;\overrightarrow{s}_0)$, that are then distinct albeit
isomorphic representations which we shall refer to as
lowest-weight and highest-weight spaces, respectively (although
they are actually of opposite type back in Euclidean signature).
In this context the finite-dimensional representations are
referred to as lowest-and-highest-weight spaces\footnote{The
finite-dimensional irreps arise for negative $e_0$. For example,
the $(D+1)$-plet $\mD^-(-1;(0))=\mC^-(-1;(0))/\mI(-1;(0))$. The
singular vector $L^+_{\{r}L^+_{s\}}\ket{-1;(0)}$ generates
$\mI(-1;(0))\simeq \mC(1;(2))$ containing \emph{all} states in
$\mC^-(-1;(0))$ with energy $e\geqslant 2$ since $L^-_r L^+_s
L^+_t L^+_t\ket{-1;(0)}= -6L^+_{\{r}L^+_{s\}}\ket{-1;(0)}$. Thus
$\mD(-1;(0))$ is spanned by $\ket{-1;(0)}$,
$\ket{0;(1)}=L^+_r\ket{-1;(0)}$ and $\ket{1;(0)}=L^+_r
L^+_r\ket{-1;(0)}$, and the highest-weight state is
$L^+_r\ket{1;(0)}\in \mI(-1;(0))$.}.

The lowest-weight and highest-weight representations are
unitarizable in two-time signature for $e_0\geqslant
e_0(\overrightarrow s_0)$ in such a way that
$\mD^\pm(e_0;\overrightarrow{s}_0)\simeq
\mC^\pm(-e_0;\overrightarrow{s}_0)$ for
$e_0>e_0(\overrightarrow{s}_0)$, where $e_0(\overrightarrow{s}_0)$
is a critical value at which at least one singular vector develops
while $E$ remains unbounded from above (two singular vectors arise
in conformal cases with $s_0\geqslant 1$). The critical cases of
interest to us are:
\bea \mbox{\emph{scalar and spinor singletons}}&:& e_0\ =\
s_0+\e_0\ ,\qquad s_0\ =\ 0,\ft12\
,\label{singletons}\\[5pt] \mbox{\emph{composite massless particles}}&:& e_0\ =\ s_0+2\e_0\ ,\qquad s_0\ =\
1,\ft32,2,\ft52,\dots\ ,\eea
where $\e_0=\ft12(D-3)$. The corresponding singular vectors are:
\bea \mbox{\emph{singletons}\hspace{1cm} $s_0=0$}:&&
\ket{\e_0+2;(0)}\ =\
L^+_r L^+_r|\e_0;(0)\rangle\ ,\\[5pt]s_0=\ft12:&&\ket{\e_0+\ft32;(\ft12)}_i\ =\
(\c_r)_i{}^j L^+_r|\e_0+\ft12;(\ft12)\rangle_j\ ,
\label{singularsingleton}\\[5pt]\mbox{\emph{massless}\hspace{1cm} $s_0=1,2,\dots$:}&& |s_0+2e_0+1;(s_0-1)\rangle_{r(s_0-1)}\ =\
L^+_t|s_0+2e_0;(s_0)\rangle_{t r(s_0-1)}\ ,\qquad \ ,\\[5pt]
s_0=\ft32,\ft52,\dots:&& |s_0+2e_0+1;(s_0-1)\rangle_{i,
r(s_0-\ft32)}\ =\ L^+_t|e_0;(s_0)\rangle_{i,tr(s_0-\ft32)}\ ,\eea
where the $\ms$ irreps of the ground states are given by
\bea M_{rs}\ket{e_0;(s_0)}_{t(s_0)}&=& 2is_0\d_{t_1[s
}\ket{e_0;(s_0)}_{r]t(s_0-1)}\ ,\\[5pt]
M_{rs}\ket{e_0;(s_0)}_{i,t(s_0-\ft12)}&=&
2i(s_0-\ft12)\d_{t_1[s}\ket{e_0;(s_0)}_{i,r]t(s_0-\ft32)}-\ft
i2(\c_{rs})_{i}{}^j \ket{e_0;(s_0)}_{j,t(s_0-\ft12)}\ ,\qquad\eea
with $\c_r$ and $\c_{rs}=\c_{[r}\c_{s]}$ being $\mso(D-1)$ Dirac
matrices, and the right-hand sides are automatically traceless and
$\gamma$-traceless, respectively. The phase factors in \eq{tau2}
can be chosen to be
\bea \mbox{\emph{scalar singletons}}&:& \tau(\ket{\pm\e_0;(0)}^\pm)\ =\ {}^\mp\bra{\mp\e_0;(0)}\ ,\\[5pt]
\mbox{\emph{massless bosons}}&:&
\tau(\ket{\pm(s_0+2\e_0);(s_0)}^\pm)\ =\ (-)^{s_0}\
{}^\mp\bra{\mp(s_0+2\e_0);(s_0)}\ ,\\[5pt]\mbox{\emph{$4D$ spinor singletons}}&:&
\tau(\ket{\pm 1;(\ft12)}^\pm)\ =\ i~{}^\mp\bra{\mp 1;(\ft12)}\ ,\\[5pt]
\mbox{\emph{$4D$ massless fermions}}&:&
\tau(\ket{\pm(s_0+1);(s_0)}^\pm)\ =\ (-)^{s_0-\ft12}i~\
{}^\mp\bra{\mp(s_0+1);(s_0)}\ .\eea

\scss{Lowest-spin modules}

In this paper, we refer to an infinite-dimensional
$(\mg|\me\oplus\ms)$ module ${\cal W}(e_0;\overrightarrow{s_0})$
as a \emph{lowest-spin module} if: (i) the energy spectrum is
\emph{unbounded} from both above and below (while the spins are
bounded from below at least by $(0\dots0)$ or $(\ft12\dots\ft12)$
by the admissibility assumption); and (ii) ${\cal
W}(e_0;\overrightarrow{s_0})={\cal
M}(e_0;\overrightarrow{s}_0)/{\cal I}(e_0;\overrightarrow{s}_0)$
where ${\cal M}(e_0;\overrightarrow{s}_0)={\cal U}[\ml^{-}]{\cal
U}[\ml^+]\ket{e_0;\overrightarrow{s}_0}$ with the \emph{static
ground state} $\ket{e_0;\overrightarrow{s}_0}$ being the
$(\me\oplus\ms)$-type with \emph{minimal}
$|e|+\dim(\overrightarrow{s}_0)$, and ${\cal
I}(e_0;\overrightarrow{s}_0)$ is the maximal ideal. The state
generation yields a canonical inner product $(\cdot,\cdot)_{{\cal
W}(e_0;\overrightarrow{s}_0)}$ (following the prescription under
\eq{inner}). We say that ${\cal W}(e_0;\overrightarrow{s}_0)$ is
\emph{unitary} if $(\cdot,\cdot)_{{\cal
W}(e_0;\overrightarrow{s}_0)}$ is positive definite for the real
form of ${\cal W}(e_0;\overrightarrow{s}_0)$ compatible with the
$\mg$ action. We note that $L^\pm_r\ket{e_0;\overrightarrow{s}_0}$
are non-vanishing, and that ${\cal M}(e_0;\overrightarrow{s}_0)$
may contain an infinite number of $(\me\oplus\ms)$-types with
given energy $e$, in which case its decomposition under the
$\mg^{(0)}\subset \me\oplus\ms$ contains infinitely degenerate
weights.

\scss{Twisted-adjoint $(\mg|\mm_1)$ modules}

The twisted-adjoint $(\mg|\mm_1)$ modules (where we follow the
notation of Sections \ref{assocquot} and \ref{adjtwadj})
\bea \mR(s)&=&\bigoplus_{k\geqslant 0} \mR_{(s+k,s)}\ ,\qquad {\rm
mult}(s+k,s)\ =\ 1\ ,\qquad s=0,1,\dots\ ,\eea
with representation matrices given by
\bea M_{mn}
\ket{{}^{(s,s)}_{(s_1,s_2)}}_{p(s_1),q(s_2)}\!\!\!&=&\!\!\!
2is_1\eta_{p_1[n}\ket{{}^{(s,s)}_{(s_1,s_2)}}_{m]p(s_1-1),q(s_2)}+2is_2\eta_{q_1[n|}
\ket{{}^{(s,s)}_{(s_1,s_2)}}_{p(s_1),|m]q(s_2-1)}\ ,\label{Mmn}\\[5pt]
P_m \ket{{}^{(s,s)}_{(s+k,s)}}_{n(s+k),p(s)}\!\!\!&=&\!\!\!
2\D_{s+k,s}\ket{{}^{(s,s)}_{ (s+k+1,s)}}_{m\{n(s+k),p(s)\}}+2
\l_k^{(s)}\eta_{m\{n_1}\ket{{}^{(s,s)}_{
(s+k-1,s)}}_{n(s+k-1),p(s)\}}\ ,\qquad\quad\label{TPa}\eea
where the traceless type-$(s+k,s)$ projection implies that
$\D_{s+k,s}={(k+2)(k+s+1)\over (k+1)(k+s+2)}$ and\footnote{To
compute $\D_{s+k,s}$ one may use $\mathbf P(P_{c}
T_{a(s+k),b(s)})=\D_{s+k,s}\mathbf PT_{c\{
a(s+k),b(s)\}}=\D_{s+k,s} T_{c a(s+k),b(s)}$ with $\mathbf P\equiv
\mathbf{P}_{\langle ca(s+k),b(s)\rangle }$, or equivalently
$T_{a(s+k+1),b(s)}=\D_{s+k,s}T_{a\langle a(s+k),b(s)\rangle}$
where
\bea T_{a\langle a(s+k),b(s)\rangle}&=& {(s+k)! s!\over
(s+k+1)\cdots (k+2)(k)\cdots 1\times s!} \sum_{n=0}^s
(-1)^n{s\choose n}T_{ a(s+k+1-n)b(n),a(n)b(s-n)}\ .\nn\eea
Using $T_{(a(s+k+1-n)b(n),a_1)a(n-1)b(s-n)}=0$ to cycle the
$b$-indices yields $T_{a(s+k+1-n)b(n),a(n)b(s-n)}=-{n\over
s+k-n+2}T_{a(s+k-n+2)b(n-1),a(n-1)b(s-n+1)}={(-1)^n\over
{s+k+1\choose n}}T_{a(s+k+1),b(s)}$. Thus
$\left[\D_{s+k,s}\right]^{-1}={k+1\over s+k+1}\sum_{n=0}^s
{{s\choose n}\over {s+k+1\choose n}}={(k+1)(s+k+2)\over
(k+2)(s+k+1)}$ as a consequence of the lemma $\sum_{n=0}^s
{{s\choose n}\over {s+k+1\choose n}}={1\over
(s+1)_{k+1}}\sum_{n=0}^s(s+1-n)_{k+1}$.}
\bea \eta_{ m \{ n_1}
\ket{{}^{(s,s)}_{(s+k-1,s)}}_{n(s+k-1),p(s)\}}\!\!\!&=&\!\!\!
\eta_{ mn_1}
\ket{{}^{(s,s)}_{(s+k-1,s)}}_{n(s+k-1),p(s)}+\a_{s+k,s}\eta_{n(2)}\ket{{}^{(s,s)}_{(s+k-1,s)}}_{n(s+k-2)m,p(s)}+\nn\\[5pt]
&&\!\!\!\!\!\!\!\!\!\!\!\!\!\!\!\!+\b_{s+k,s}\eta_{n(2)}\ket{{}^{(s,s)}_{(s+k-1,s)}}_{n(s+k-2)p,mp(s-1)}+\gamma_{s+k,s}
\eta_{np}T_{n(s+k-1),mp(s-1)}\ ,\qquad\qquad\label{hooked}\eea
with $\a_{s+k,s}=-{1\over 2}{s+k-1\over s+k+\e_0-\ft12}$,
$\b_{s+k,s}={1\over 2}{(s+k-1)s\over
(s+k+\e_0-\ft12)(2s+k+2\e_0-1)}$ and $\c_{s+k,s}= -{s\over
2s+k+2\e_0-1}$, while $\l_k^{(s)}$ is fixed by the closure
relation $[P_m,P_n]=i\e M_{mn}$, $\e=\pm1$. One solution is given
by the dimensional reduction of \eq{AcMAB} which takes the
form\footnote{Acting on a Lorentz tensor the mass operator
$\nabla^2=M^2\equiv -P^2=\e(C_2[\m_1]-C_2[\mg])$. The value of
$C_2[\mg]$ in \eq{lambda} yields the critical masses for composite
massless Weyl tensors and scalars as examined in Appendix
\ref{App:T}.}
\bea \lambda_k^{(s)}&=& {\e\over 8}{k(k+s+1)(k+2s+2\e_0-1)\over
k+s+\e_0+\ft12}\ ,\quad C_2[\mg|(s)]\ =\ C_2[\ell]\ ,\quad
s=2\ell+2\ .\label{lambda}\eea
For $s\geqslant 1$ this is the unique solution while $s=0$ admits
the \emph{massive} deformation\footnote{The finite-dimensional
highest-weight representation $\mD(\ell)$, containing the scalar
spherical harmonics, \emph{i.e.} the Killing-normalizable
solutions to $(\nabla^2_{S^D}+C_2[\mso(D+1)]|(\ell)])\phi=0$,
arises inside $\mR(0)$ for
$C_2[\mso(D+1)|(\ell)]=\ell(\ell+2(\e_0+1)))$ and $\e=-1$, leading
to $\l_k^{(0)}=\ft18 {k\over
k+\e_0+\ft12}(\ell+1-k)(\ell+k+2\e_0+1)$, which are positive for
$k=0,\dots,\ell$, vanish for $k=\ell+1$, and are negative for
$k=\ell+2,\ell+3,\dots$, so that
$\mI(0)\equiv\bigoplus_{k=\ell+1}^\infty \mR_{(k)}$ is an
(non-unitarizable) ideal and
$\mD(\ell)\equiv\mR(0)/\mI(0)=\bigoplus_{k=0}^\ell\mR_{(k)}$ a
unitarizable quotient.}
\bea \lambda_k^{(0)}&=&{\e\over 8}{k(k^2+2\e_0k-2\e_0-1-\e
M^2_0)\over k+\e_0+\ft12}\ ,\quad C_2[\mg|(0)]\ =\ \e M^2_0\
.\label{lambdas=0}\eea
This can be seen by expanding $[P_m,P_n]\ket{(k)}_{p(k)}$ as
\bea &&2P_m \ket{(k+1)}_{np(k)}+2\l_k^{(0)}\eta_{n\{p_1|}P_m
\ket{(k-1)}_{|p(k-1)\}}-(m\leftrightarrow
n)\label{intermediate4}\\[5pt]
&=&4\l_{k+1}^{(0)} \eta_{m\{n}\ket{(k)}_{p(k)\}}+4\l_{k}^{(0)}
\eta_{n\{p_1}\ket{(k)}_{p(k-1)\}m}+4\l_k^{(0)}\l_{k-1}^{(0)}\eta_{n\{p_1|}\eta_{m|p_2}\ket{(k-2)}_{|p(k-2)\}}-(m\leftrightarrow
n)\ ,\nn\eea
where the anti-symmetrization on $m$ and $n$ removes the last
term\footnote{Explicitly, using \eq{symmtraceless} one finds that
$\eta_{n\{p_1|}\eta_{m|p_2}\ket{(k-2)}_{|p(n-2)\}}-(m\leftrightarrow
n)$ equals $$\a_{k-1}\eta_{np_1}\eta_{p(2)}\ket{(k-2)}_{p(k-3)m}
+\a_k\eta_{p(2)}\left(\eta_{m
(p_3}\ket{(k-2)}_{p(k-3)n)}+\a_{k-1}\eta_{(p(2)}
\ket{(k-2)}_{p(k-4)n)m}\right)-(m\leftrightarrow n)$$$$=\
2\left(\a_{k-1}- {k-2\over k-1}\a_k+{2\a_k\a_{k-1}\over
k-1}\right)\eta_{p(2)}\eta_{p_3[n}\ket{(k-2)}_{m]p(k-3)}\ =\ 0\
.$$ }, leaving
\bea &&4\l_{k+1}^{(0)} \left(\eta_{m(n}
\ket{(k)}_{p(k))}+\a_{k+1}\eta_{(np_1}\ket{(k)}_{p(k-1))m}\right)+
\nn\\[5pt]&&+4\l_{k}^{(0)}\left(\eta_{np_1}\ket{(k)}_{p(k-1)m}+\a_k\eta_{p(2)}\ket{(k)}_{p(k-2)nm}\right)
-(m\leftrightarrow n)\nn\\[5pt]
&=&8\left({k\l_{k+1}^{(0)}\over k+1}-{2\a_{k+1}\l_{k+1}^{(0)}\over
k+1}-\l_k^{(0)}\right)\eta_{p_1[m}\ket{(k)}_{n]p(k-1)}\ \equiv\
2k\eta_{p_1[m}\ket{(k)}_{n]p(k-1)}\ ,\eea
which is an inhomogeneous first-order recursion relation for
$\l^{(0)}_k$ with initial datum $\l^{(0)}_0=0$ whose general
solution is given by \eq{lambdas=0}. The massively deformed
twisted-adjoint spinor $\mR(\ft12)=\bigoplus_{k=0}^\infty
\mR_{(\ft12+k,\ft12,\dots,\ft12)}$ has the representation matrices
\bea M_{mn}\ket{(k+\ft12)}_{p(k)}&=&2ik\eta_{p_1[n}\ket{(k+\ft12)}_{m]p(k-1)}-\ft{i}2 \c_{mn}\ket{(k+\ft12)}_{p(k)}\ ,\\[5pt]
P_m\ket{(k+\ft12)}_{n(k)}\!\!\!&=&\!\!\!2\left(\ket{(k+\ft32)}_{m\{n(k)\}}+
\l_k^{(\ft12)}\eta_{m\{n_1}\ket{(k-\ft12)}_{n(k-1)\}}
+i\m^{(\ft12)}_k\c_m\ket{(k+\ft12)}_{\{n(k)\}}\right)\
,\qquad\quad\label{TPa1/2}\eea
where the Dirac matrices obey $\{\c_m,\c_n\}=2\eta_{mn}$; the
$\c$-traceless projections
\bea \eta_{m\{n_1}\ket{(k-\ft12)}_{n(k-1)\}}\!\!\!&=&\!\!\!
\eta_{mn_1}\ket{(k-\ft12)}_{n(k-1)}+A_k\eta_{n(2)}\ket{(k-\ft12)}_{n(k-2)m}+
B_k\c_{mn_1}\ket{(k-\ft12)}_{n(k-1)}\ ,\qquad\quad\\[5pt]
\c_m\ket{(k+\ft12)}_{\{n(k)\}}&=&
\c_m\ket{(k+\ft12)}_{n(k)}+C_k\c_{n_1}\ket{(k+\ft12)}_{n(k-1)m}\
,\eea
with $A_k=-{k-1\over 2(k+\e_0)}$, $B_k={1\over 2(k+\e_0)}$ and
$C_k=-{k\over k+\e_0+\ft12}$; and the closure is solved by
\bea \l^{(\ft12)}_k&=& {\e\over 8}{k(k+\e_0)((k+\e_0+\ft12)^2-\e M^2_{1/2})\over (k+\e_0+\ft12)^2}\ ,\nn\\[5pt] \mu^{(\ft12)}_k&=&{\sqrt{\e}\over 4}{M_{1/2}\over k+\e_0+\ft32}\ ,\qquad C_2[\mg|(\ft12)]\ =\ \e M^2_{1/2}-\ft12(\e_0+1)(\e_0+\ft32)\ .\label{lambdas=1/2}\eea

\scss{The conformal twisted-adjoint modules}

The conformal twisted-adjoint $(\mg|\mm_1)$ modules, namely
$\mR(s,s)\simeq \mD(s+1;(s,s))$ for $s\geqslant 1$ in $D=4$, and
$\mR(0)\simeq \mD(\e_0+\ft12;(0))$ and $\mR(\ft12)\simeq
\mD(\e_0+1;(\ft12))$ in $D\geqslant 3$ with the conformal masses
given in \eq{mconf}, are singleton lowest-weight
spaces\footnote{Conversely, the $\mg$ singletons remain
irreducible under $\mm\subset \mg$. Thus the $\mg$ singletons are
zero-momentum representations of $\miso(D-1,1)$, \emph{i.e.}
infinite-dimensional unitary representations of $\mso(D-1,1)$ with
vanishing momentum. In particular, the massless UIRs of
$\miso(3,1)$ are conformal and as such composite in terms of
$Di\oplus Rac$, although in a non-tensorial split (see comment in
Appendix \ref{Sec:4D}). This suggests a manifestly
$\mso(3,1)$-covariant unfolded realization of the singletons in
the singular light-like geometry of the boundary of 4D Minkowski
spacetime.} of the $\mso(D+2;\Comp)$ generated by
$M_{\unA\unB}=(M_{AB},R_B)$, $\eta_{\unA\unB}=(\eta_{AB},+)$,
where the conformal translations $R_A=(R_m,R)$, which obey
$[M_{AB},R_C]= 2i\eta_{C[B}R_{A]}$ and $[R_A,R_B]=-iR_{AB}$, can
be represented for integer $s$ by
\bea \frac12 R_m \ket{{}^{(s,s)}_{(s+k,s)}}_{n(s+k),p(s)}\!\!&=&\!\! \D_{s+k,s}\ket{{}^{(s,s)}_{ (s+k+1,s)}}_{m\{n(s+k),p(s)\}}- \l_k^{(s)}\eta_{m\{n_1}\ket{{}^{(s,s)}_{ (s+k-1,s)}}_{n(s+k-1),p(s)\}}\ ,\qquad\qquad\label{TRa}\\[5pt]
R
\ket{{}^{(s,s)}_{(s+k,s)}}_{n(s+k),p(s)}\!\!&=&\!\!i\D_k^{(s)}\ket{{}^{(s,s)}_{
(s+k,s)}}_{n(s+k),p(s)}\ ,\qquad \D^{(0)}_k\ =\ \e(k+\e_0+\ft12)\
.\qquad\label{TR0}\eea
The closure relation $[R_m,R_n]=-iM_{mn}$ holds for all
$\l^{(s)}_k$, while $[R_m,P_n]=i\eta_{mn} R$ requires
conformality. For example, for $s=0$ the contributions to
$[R_m,P_n]\ket{{(k)}}_{p(k)}$ of type
$\eta_{\{p_1|(m}\ket{{(k)}}_{n)|p(k-1)\}}$ cancel iff
$M^2_0=M^2_0({\rm conf})$, leaving $8i{\l^{(0)}_{k+1}({\rm
conf})\over k+1}\eta_{mn}\ket{{(k)}}_{p(k)}$ to be identified with
the action of $R$. We note that the conformal representation
matrices of $\mg$ simplify, \emph{viz.}
\bea \l^{(s)}_k|_{\rm conf}&=&{\e\over 8}k(k+2s)\ ,\qquad \mbox{for $s\geqslant 1$ in $D=4$}\ ,\\[5pt] \l^{(0)}_k|_{\rm conf}&=&{\e\over 8} k(k+\e_0-\ft12)\ ,\qquad
\l^{(\ft12)}_k|_{\rm conf}\ =\ {\e\over 8} k(k+\e_0)\ ,\quad
\mu^{(\ft12)}_k|_{\rm conf}\ =\ 0\ .\label{conflambda}\eea
The conformal embedding of $\miso(D;\Comp)={\rm
span}_\Comp\{M_{mn},\Pi_m\}$ reads
\bea \Pi_m&=&\ft1{\sqrt{2}}(P_m+R_m)\ ,\quad \Sigma_m\ =\
\ft1{\sqrt{2}}(P_m-R_m)\ ,\quad [\Pi_m,\Sigma_n]\ =\
iM_{mn}+i\eta_{mn}R\ ,\eea
with representation matrices
$\Pi\ket{{}^{(s,s)}_{(s+k,s)}}=2\sqrt{2}\ket{{}^{(s,s)}_{(s+k+1,s)}}$
and
$\Sigma\ket{{}^{(s,s)}_{(s+k,s)}}=2\sqrt{2}\l_k^{(s)}\eta\ket{{}^{(s,s)}_{(s+k-1,s)}}$,
leading to the identifications $\Pi_m=\ft1{\sqrt{2}}L^+_m$,
$\Sigma_m=\ft1{\sqrt{2}}L^-_m$ and $E=R$ (whose hermiticity
properties are discussed in \cite{Gunaydin:1999jb}, for example),
and to the identification of the smallest Lorentz tensor with the
lowest-weight state. Thus, the light-likeness condition
$\Pi^m\Pi_m\approx 0$ can be written $\mso(D+2;\Comp)$-covariantly
as the hyperlight-likeness condition
\bea V_{\underline A\underline B}&\equiv &\ft12 M_{\{\underline
A}{}^{\underline C}M_{\underline B\}\underline C}\ \approx\ 0\
.\eea
%


\scs{Details of the Lorentz-covariant factorization of ${\cal
I}[V]$}\label{App:VAB}


This Appendix contains details of the procedure of factoring out
the ideal ${\cal I}[V]$ defined in \eq{idealV} from the enveloping
algebra ${\cal U}[\mg]$. The constraints $V_{AB}\approx 0$ and
$V_{ABCD}\approx 0$, defined by \eq{VAB} and \eq{VABCD}, decompose
under $\mm$ into
\bea V_{\sharp\sharp}&=&\ft12(\s P^a\star P_a -\mu^2)\ \approx 0\ ,\qquad V_{\sharp a}\ =\ \ft14\{M_a{}^b,P_b\}_\star\ \approx\ 0\ ,\label{v0a}\\[5pt]
V_{ab}&=& \ft12 (M_{(a}{}^c\star M_{b)c}-\s P_{(a}\star
P_{b)}+\mu^2\eta_{ab})\ \approx 0\ ,\label{vab}\\[10pt]
V_{abcd}&=&M_{[ab}\star M_{cd]}\approx 0\ ,\qquad V_{\sharp abc}\
=\ -P_{[a}\star M_{bc]}\ =\ 0\ ,\label{v0abc}\eea
with $\mu^2\equiv -{2C_2[{\cal S}]\over D+1}$. The
$\mg$-irreducibility of $V_{AB}\approx 0$ implies that $V_{\sharp
a}\approx0$ and $V_{ab}\approx 0$ follow from
$V_{\sharp\sharp}\approx 0$. Similarly, the constraint
$V_{abcd}\approx 0$ follows from $V_{\sharp abc}\approx0$. More
explicitly, using that $\mu^2$ is a commuting element one can show
that $V_{\sharp\sharp}\approx 0$ implies that
\bea P^a\star M_{ab}\approx M_{ba}\star P^a\approx i(\e_0+1)P_b\
,\label{pamab2}\eea
from which \eq{v0a} follows immediately (alternatively one may
compute $V_{\sharp a}=-\ft{i\sigma}4 [P_a,P^b\star
P_b]_\star\approx-\ft{i}4 [P_a,\mu^2]_\star=0$). Next, eq.
\eq{pamab2} and $P^a\star P_a\approx \s\mu^2$ imply
\bea M_{(a}{}^c\star M_{b)c}&\approx & \s P_{(a}\star
P_{b)}-\mu^2\eta_{ab}\ ,\label{macmbc}\eea
that is $V_{ab}\approx 0$. Similarly,
$[P_{[a},V_{0'bcd]}]_\star\propto V_{abcd}$, so that
$V_{abcd}\approx 0$, \emph{i.e.} $M_{[ab}\star M_{cd}]\approx 0$,
follows from $V_{0'abc}\approx 0$, \emph{i.e.} $P_{[a}\star
P_b\star P_{c]}\approx 0$. Finally, the value of $\mu^2$ is
determined from $P^a\star P_{[a}\star P_b\star
P_{c]}\approx\ft{i}6(\mu^2-\e_0) M_{bc}$.

Next, the $\mm$-covariant basis elements defined in \eq{Tambn2}
can be expanded in traces in the first row. For example, for $m=0$
one has
\bea T_{a(n)}&=& P_{\{a_1}\cdots P_{a_n\}}\ = \ P_{(a_1}\star
\cdots\star P_{a_n)}+ \k_{n,0;1}\eta_{(a_1 a_2}P_{a_3}\star\cdots
\star P_{a_n)}+{\cal O}(\eta^2)\ ,\label{lemmaApp}\eea
with $\k_{n,0;1}= -\sigma {(n+1)n(n-1)(n+4\e_0-2)\over
48(n+\e_0-\ft12)}$. Let us use the contraction rules \eq{papa},
\eq{pamab2} and \eq{macmbc} to compute $\k_{n}\equiv \k_{n,0;1}$
by demanding the right-hand side to be traceless. To this end, we
first use \eq{papa} to expand
$$\eta^{bc}P_{(b}\star P_c\star P_{a_1}\cdots\star P_{a_{n-2})}
\ \approx\ \e_0P_{a_1}\cdots\star P_{a_{n-2}}$$$$ +{2\over
n(n-1)}\sum_{1\leqslant i<j\leqslant n} P_{a_1}\cdots\star
P_{a_{i-1}}\star [P_b,P_{a_{i}}\star\cdots \star
P_{a_{j-2}}]_\star\star P^b\star P_{a_{j-1}}\star \cdots \star
P_{a_{n-2}}\ ,$$
where the summand can be evaluated modulo trace parts, which only
affect the higher traces in \eq{lemmaApp}. Thus, for $i=1$ and
$j=n$ we find using \eq{pamab2} that $[P_b,P_{a_{1}}\star\cdots
\star P_{a_{n-2}}]_\star\star P^b$
\bea &=&iM_{ba_1}\star P_{a_2}\star\cdots \star P_{a_{n-2}}\star
P^b +P_{a_1}\star (iM_{ba_2})\star \cdots \star P_{a_{n-2}}\star
P^b+\cdots\nn\\[5pt] &\approx&(\e_0+1) P_{a_1}\star\cdots \star
P_{a_{n-2}}+\eta_{ba_2}P_{a_1}\star P_{a_3}\star \cdots\star
P_{a_{n-2}}\star P^b\nn\\[5pt] &&+P_{a_2}\star (\eta_{ba_3}P_{a_2})\star
P_{a_4}\star \cdots\star P_{a_{n-2}}\star P^b+P_{a_2}\star
P_{a_3}\star (\eta_{ba_4}P_{a_3})\star P_{a_5}\star
\cdots\star P_{a_{n-2}}\star P^b+\cdots\qquad\qquad\nn\\[5pt]&&+
(\e_0+1) P_{a_1}\star\cdots \star P_{a_{n-2}}+P_{a_2}\star
(\eta_{ba_3}P_{a_2})\star P_{a_4}\star
\cdots\star P_{a_{n-2}}\star P^b+\cdots\nn\\[5pt]&&+\cdots+{\cal O}(\eta)\ =\ \ft12(n-2)(n+2\e_0-1)P_{a_1}\star\cdots
\star P_{a_{n-2}}+{\cal O}(\eta)\ .\nn\eea
The contributions from $j=n$ and $i=1+k$ for $k=0,\dots,n-2$ are
obtained by letting $n\rightarrow n-k$, and are given modulo trace
parts by $P_{a_1}\star\cdots \star P_{a_{n-2}}$ times
$\sum_{k=0}^{n-2} k(\e_0+1+\ft12(k-1))= \ft16(n-1)(n-2)(n+3\e_0)$.
Letting $n\rightarrow n-k$ yields the contributions from $j=n-k$
for $k=0,\dots,n-2$, which are thus given modulo trace parts by
$P_{a_1}\star\cdots \star P_{a_{n-2}}$ times $\ft16
\sum_{k=0}^{n-2}(n-1-k)(n-2-k)(n-k+3\e_0)=\ft1{24}
n(n-1)(n-2)(n+4\e_0+1)$. Hence the trace of the first term of the
right-hand side of \eq{lemmaApp} is given by
\bea \eta^{bc}P_{(b}\star P_c\star P_{a_1}\cdots\star
P_{a_{n-2})}&\approx&
\left(\e_0+\ft1{12}(n-2)(n+4\e_0+1)\right)P_{a_1}\star\cdots\star
P_{a_{n-2}}+ {\cal O}(\eta)\nn\\[5pt]&=& \ft1{12}(n+1)(n+4\e_0-2)P_{a_1}\star\cdots\star
P_{a_{n-2}}+ {\cal O}(\eta)\ .\eea
Tracing the second term of the right-hand side of \eq{lemmaApp}
yields $\k_n$ times
\bea \eta^{bc}\eta_{(bc}P_{a_1}\star\cdots\star P_{a_{n-2})}&=&
{4(n+2\e_0-\ft12)\over n(n-1)}P_{a_1}\star \cdots\star
P_{a_{n-2}}+ {\cal O}(\eta)\ .\eea
The tracelessness of the right-hand side of \eq{lemmaApp} thus
requires
\bea \ft1{12}(n+1)(n+4\e_0-2)+{2(2n+2\e_0-1)\over n(n-1)}\k_n&=&0\
,\eea
which yields the value of $\k_n$ given below \eq{lemmaApp}.

The contraction rules \eq{papa} and \eq{pamab2} can be verified
against \eq{TPa} which implies
\bea \ac_{P_c} (T_{a(s+k),b(s)})&=&2\Delta_{s+k,s}T_{c\{
a(s+k),b(s)\}}+ 2\lambda_{k}^{(s)} \eta_{ c \{ a}
T_{a(s+k-1),b(s)\}}\ ,\eea
where $\D_{s+k,s}$ and $\{\cdots\}$ are given below \eq{TPa}, and
$\lambda_k^{(s)}$ in \eq{lambda}. For $s=0$ one has
\bea \ac_{P_a} T_{b(n)}&=& \{P_a,T_{b(n)}\}_\star\ =\ 2T_{ab(n)}+2
\l^{(0)}_n\eta_{a\{b_1}T_{b(n-1)\}}\ ,\label{TPa0}\eea
with $\lambda^{(0)}_n={n(n+2\e_0-1)(n+1)\over 8(n+\e_0+\ft12)}$
and
\bea \eta_{a\{b_1}T_{b(n-1)\}}&\equiv &
\eta_{a(b_1}T_{b(n-1))}+\alpha_n\eta_{(b_1 b_2}T_{b(n-2))a}\
,\quad \alpha_n\ =\ \alpha_{n,0}\ =\ -{n-1\over 2(n+\e_0-\ft12)}\
,\qquad \label{symmtraceless}\eea
which means that $\l_n^{(0)}$ is determined by the trace condition
\bea \{P^a, T_{ab(n-1)}\}_\star&=& \l_n^{(0)}
\eta^{ac}\eta_{a\{b_1}T_{b(n-2)c\}}\label{trcond}\ .\eea
The right-hand side can been simplified using \eq{symmtraceless},
which yields
\bea
\eta^{ac}\eta_{a\{b_1}T_{b(n-2)c\}}&=&{(n+\e_0+\ft12)(n+2\e_0)\over
n(n+\e_0-\ft12)} T_{b(n-1)}\ .\label{intermediate2}\eea
On the left-hand side we first use the $\tau$-map to show that
$P^a\star T_{ab(n-1)}=T_{ab(n-1)}\star P^a$. We then use
\eq{lemmaApp} and the contraction rules \eq{papa} and \eq{pamab2}
to compute $P^a\star T_{ab(n-1)}$ as
\bea &&{1\over n}P^a\star\sum_{i=1}^n P_{b_{1}}\star\cdots
P_{b_{i-1}}\star P_a\star P_{b_{i}}\star\cdots\star P_{b_{n-1}}+
{2\k_n\over n}P_{b_1}\star\cdots \star P_{b_{n-1}}+{\cal O}(\eta)\nn\\[5pt]&\approx&
\left(\e_0+\ft12(\e_0+1)(n-1)+\ft16(n-1)(n-2)+{2\k_n\over
n}\right)P_{b_1}\star\cdots
\star P_{b_{n-1}}+{\cal O}(\eta)\nn\\[5pt]
&=& {(n+2\e_0)(n+2\e_0-1)(n+1)\over
8(n+\e_0-\ft12)}T_{b(n-1)}+{\cal O}(\eta)\ .\label{trcond2}\eea
Substituting \eq{intermediate2} and \eq{trcond2} into \eq{trcond}
then yields $\l_n^{(0}$ in agreement with \eq{lambda}.

Another basic $\star$-product in the $\mm$-covariant basis is
\bea \ac_{M_{ab}}(T_{c(n)})&=& 2
T_{c(n)[a,b]}+2\rho_n\eta_{[a|\{c_1}T_{c(n-1)\},|b]}\ ,\quad
\rho_n\ =\ -{\s\over4}{(n-1)n(n+1)\over n+\e_0+\ft12}\
,\label{MabT}\eea
with $\D_n={2(n+1)\over n+2}$,
$\eta_{[a|\{c_1}T_{c(n-1)\},|b]}=\eta_{c_1[a|}T_{c(n-1),|b]}-{n-1\over
2(n+\e_0-\ft12)} \eta_{c_1 c_2}T_{c(n-2)[a,b]}$, and we note the
alternative form
$\ac_{M_{ab}}(T_{c(n)})=2M_{ab}T_{c(n)}+2\r_n\eta_{[a|\{c_1}
T_{c(n-2)}M_{c_n\}|b]}$.

Finally, the $\Tr$ norms of the basis elements defined in
\eq{TAnBn} and \eq{Tambn2} read
\bea \Tr[M_{A(n),B(n)}\star M^{C(m),D(m)}]&=& \delta_{mn}
\delta_{\{A(n),B(n)\}}^{\{C(n),D(n)\}}{\cal
N}_{(n,n)_{D+1}}\ , \label{trtABtCD}\\[5pt]
\Tr[T_{a(n),b(m)}\star T^{c(n'),d(m')}]&=& \delta_{n,n'}
\delta_{m,m'}\delta_{\{a(n),b(m)\}}^{\{c(n),d(m)\}}{\cal
N}_{(n,m)_D}\ ,\label{trtabtcd}\eea
%
where the normalizations are given by
\bea {\cal N}_{(n,n)_{D+1}}&=& \l_n\l_{n-1}\cdots \l_1\ =\
(-2)^{-n} {n!(n+1)!(\e_0)_n\over
(\e_0+\ft32)_n} \ ,\label{calNn}\\[5pt] {\cal N}_{(n)_D}&=& \l_n^{(0)}\l_{n-1}^{(0)}\cdots \l_1^{(0)}\ =\ (8\s)^{-n}
{n!(n+1)!(2\e_0)_n\over (\e_0+\ft32)_n} \ ,\label{calNn0}\eea
as can be seen by making repeated use of \eq{AcMAB} and \eq{TPa}.
For example,
\bea &&\Tr[T_{a(n)}\star T_{b(n)}]\ =\
\Tr[P_{\{a_1}\star\cdots\star P_{a_n\}}\star
T_{b(n)}]\nn\\[5pt]&&=\
\Tr[P_{\{a_1}\star\cdots\star P_{a_{n-1}}\star\left(
T_{n+1}+T_n+\l_n^{(0)}\eta_{a_n\}\{b_1}T_{b(n-1)\}}\right)]\nn\\[5pt]&&=\
\Tr[P_{\{a_1}\star\cdots\star P_{a_{n-1}}\star
\l_n^{(0)}\eta_{a_n\}\{b_1}T_{b(n-1)\}}]\ ,\eea
where $T_n$ denote traceless symmetric $\star$-products of $n$
transvections, and we have used the fact that $\Tr[T_{n+1}\star
T_{n-1}]=\Tr[T_n\star T_{n-1}]=0$. In particular, using
\bea \d^{\{a(n)\}}_{\{b(n)\}}&=& \sum_{k=0}^{[n/2]}
t_{k}(\eta^{a(2)}\eta_{b(2)})^k\delta^{a(n-2k)}_{b(n-2k)}\ ,\qquad
t_k\ =\ {(-n)_{2k}\over 4^k k!(-n-\e_0+\ft12)_k}\ ,\eea
from which it follows that
$\d^{\{\sharp'(n)\}}_{\{\sharp'(n)\}}=\sum_{k=0}^{[n/2]}
{(-n)_{2k}\over 4^k k!(-n-\e_0+\ft12)_k}=2^{-n}{(2\e_0+1)_n\over
(\e_0+\ft12)_n}$, we obtain
\bea \Tr[T_{\sharp'(n)}\star T_{\sharp'(n)}]&=& (-\sigma')^n
4^{-2n}{n!(n+1)!(2\e_0)_n(2\e_0+1)_n\over
(\e_0+\ft12)_n(\e_0+\ft32)_n}\ .\label{TrEnEn}\eea


\scs{Quadratic and quartic Casimir operators}\label{App:Cas}


The values $C_{2n}[\mg|(e_0;\overrightarrow{s}_0)^\pm]$ of
$C_{2n}[\mg]=\ft12 M_{A_1}{}^{A_2}\star
M_{A_2}{}^{A_3}\star\cdots\star M_{A_{2n}}{}^{A_1}$ in
$\mD^\pm(e_0;\overrightarrow{s}_0)$,
$\overrightarrow{s}_0=(m_1,\dots,m_{\nu-1})$, are given for $n=1$
and $n=2$ by
\bea C_2[\mg|(e_0;\overrightarrow{s}_0)^\pm]&=& x_0^\pm
+C_2[\mathfrak{s}|\overrightarrow{s}_0]\ ,\label{C2lhws}\\[5pt]
C_4[\mg|(e_0;\overrightarrow{s}_0)^\pm]&=&
x_0^\pm\left(x_0^\pm+\D_0\right)+C_4[\ms|\overrightarrow{s}_0]-C_2[\ms|\overrightarrow{s}_0]\
,\label{C4lhws}\eea
where $x_0^\pm=e_0(e_0\mp(D-1))$ and $\D_0=\ft12 (D-1)(D-2)$, and
the values of $C_2[\ms]=\ft12 M^{rs}\star M_{rs}$ and
$C_4[\ms]=\ft12 M_r{}^s \star M_s{}^t \star M_t{}^u\star M_u{}^r$
in the $\overrightarrow{s}_0$-plet are given by
\bea C_2[\ms|\overrightarrow{s}_0] &=& \sum_{k=1}^{\nu-1}x_k\
,\qquad C_4[\ms|\overrightarrow{s}_0] \ =\
\sum_{k=1}^{\nu-1}x_k\left(x_k+\D_k\right)\ ,\eea
with $x_k=m_k(m_k+D-1-2k)$ and $\D_k=\ft12 (D-1-2k)(D-2-2k)+1-k$.
The $\ell$th level adjoint level ${\cal L}_\ell$ defined in
\eq{Lell} has lowest weight $(-(2\ell+1);2\ell+1)$, which yields
\eq{c2ell} and \eq{c4ell} (with $s=2\ell+2$). The composite
massless lowest-weight and highest-weight spaces
$\mD^\pm(\pm(s+2\e_0);(s))$ have the same values of $C_2[\mg]$ and
$C_4[\mg]$.

To compute $C_2[\mg]$ in the twisted-adjoint representation we use
$P^a\star P_a=\s\e_0$ to write
\bea&& \widetilde{\adj}_{C_2[\mg]}(S)\ =\
\adj_{C_2[\mm]}(S)-\s\{P^a,\{P_a,S\}_\star\}_\star\nn\\[5pt]&&=\
\adj_{C_2[\mm]}(S)-2\s(\e_0 S+P^a\star S\star P_a)\ =\
\adj_{C_2[\mg]}(S_\ell)+4\s P^a\star S\star P_a\ ,\eea
where $C_2[\mm]=\ft12 M_{ab}M^{ab}$, from which $\s P^a\star
S\star P_a$ can be eliminated, which yields
\bea
\widetilde{\adj}_{C_2[\mg]}(S)&=&2\adj_{C_2[\mm]}(S)-\adj_{C_2[\mg]}(S_\ell)-4\e_0
S\ .\eea
An element $S_\ell\in{\cal T}_\ell$, defined by \eq{Tell}, carries
the highest weights $(s+k,s)$ and $(s+k,s+k)$ with $s=2\ell+2$ and
$k=0,1,\dots$ of the adjoint $\mm$ and $\mg$ actions,
respectively, and for all $k$
\bea
\widetilde{\adj}_{C_2[\mg]}(S_\ell)&=&\left(2C_2[\mm|(s+k,s)]-C_2[\mg|(s+k,s+k)]-4\e_0\right)S_\ell
\nn\\[5pt]&=&\left(
2C_2[\mm|(s,s)]-C_2[\mg|(s,s)]-4\e_0\right)S_\ell\ =\
C_2[\mg|\ell] S_\ell\ .\eea
Similarly,
\bea
\widetilde{\adj}_{C_4[\mg]}(S)&=&\adj_{C_4[\mm]}(S)+\nn\\
&&+\ft{\s}2[M_a{}^b,[M_b{}^c,\{P_c,\{P^a,S\}_\star\}_\star]_\star]_\star
+\ft{\s}2[M_a{}^b,\{P_b,\{P^c,[M_c{}^a,S]_\star\}_\star\}_\star]_\star\nn\\&&
+\ft{\s}2\{P_a,\{P^b,[M_b{}^c,[M_c{}^a,S]_\star]_\star\}_\star\}_\star
+\ft{\s}2\{P^a,[M_a{}^b,[M_b{}^c,\{P_c,S\}_\star]_\star]_\star\}_\star\nn\\&&
+\ft12 \{P_a,\{P^b,\{P_b,\{P^a,S\}_\star\}_\star\}_\star\}_\star +
\ft12
\{P^a,\{P_b,\{P^b,\{P_b,S\}_\star\}_\star\}_\star\}_\star\nn\\[5pt]
&=&\adj_{C_4[\mm]}(S)+C_+(S)+C_-(S)\ =\
\adj_{C_4[\mg]}(S)+2C_-(S)\ ,\eea
where $C_{+}(S)$ and $C_-(S)$ are the terms with an even and odd
number of translation generators standing to the right of $S$,
respectively. Eliminating $C_-(S)$ leads to
\bea
\widetilde{\adj}_{C_4[\mg]}(S)&=&2\adj_{C_4[\mm]}(S)-\adj_{C_4[\mg]}(S)+2C_+(S)\
.\eea
The quantity $C_+(S)$ can be calculated using \eq{pamab2},
\eq{macmbc}, and $M_{ab}\star S\star
M^{ab}=-\adj_{C_2[\mm]}(S)+\{M^{ab}\star M_{ab},S\}_\star$, with
the result
\bea C_+(S)&=&\adj_{C_2[\mm]}(S)-2\e_0(2\e_0^2-\e_0+1)S\ .\eea
Thus, using the $\mg$-adjoint and $\mm$-adjoint highest weights
for $S_\ell$, we find that
\bea \widetilde{\adj}_{C_4[\mg]}(S_\ell)&=&\left(C_4[\mm|(s+k,s)]-
C_4[\mg|(s+k,s+k)]\right)S_\ell\nn\\&&+
\left(C_2[\mm|(s+k,s)]-4\e_0(2\e_0^2-\e_0+1)\right)S_\ell\ =\
C_4[\mg|\ell] S_\ell\ .\eea


\scs{Critical and conformal masses for the Weyl
zero-forms}\label{App:T}


In this Appendix we give some details related to
\eq{DPhicomponents} and the mass formula \eq{msk}. Let us begin
with the case of $s=0$, where the linearized master-field
constraint reads $\nabla\Phi_{(0)}-i
e^a\{P_a,\Phi_{(0)}\}_\star=0$, with $\Phi_{(0)}$ given by
\eq{Phis} and $P_a$ in the presentation \eq{TPa0}. The constraint
takes the component form
\bea \nabla_b \Phi_{a(n)}-2n
\eta_{b\{a_1}\Phi_{a(n-1)\}}+{2\l_n^{(0)}\over
n+1}\Phi_{ba(n)}&=&0\ ,\label{DPhicomponents0}\eea
with $\eta_{b\{a_1}\Phi_{a(n-1)\}}$ given by \eq{symmtraceless}.
The symmetric and traceless part of \eq{DPhicomponents0} yields
\eq{Phiauxiliary} for $s=0$ while its trace part leads to the
masses in \eq{msk}. To this end, contraction with $\nabla^{b}$
yields
\bea \nabla^2\Phi_{a(n)}-2n\eta^{bc}\left(\eta_{b
a_1}\nabla_c\Phi_{a(n-1)}+\a_n
\eta_{a_1a_2}\nabla_c\Phi_{a(n-2)b}\right)+{2\l^{(0)}_{n+1}\over
n+1}\nabla^b\Phi_{ba(n)}&=&0\ .\eea
Elimination of $\nabla_c\Phi_{a(n-1)}$ and $\nabla_c\Phi_{a(n1)}$
using \eq{Phiauxiliary} followed by $\{a(n)\}$-projection leads to
\bea
\nabla^2\Phi_{a(n)}+4\l_n^{(0)}\Phi_{a(n)}+4\l^{(0)}_{n+1}\eta^{bc}
\left(\eta_{c(b}\Phi_{a(n))}+\a_{n+1}\eta_{(ba_1}\Phi_{a(n-1))c}\right)&=&0\
.\eea
Performing the traces one ends up with the following expression
for the critical mass:
\bea M^2_{0,n}&=&
-4\l_n^{(0)}-4\l_{n+1}^{(0)}{(n+2\e_0+1)(n+\e_0+\ft32)\over
(n+1)(n+\e_0+\ft12)}\ .\label{masses0}\eea
Inserting \eq{lambda} leads to the critical mass
$M^2_{0,n}=-(n^2+(2\e_0+1)n+4\e_0)\s$ in agreement with \eq{msk}.
For general $s$ we use \eq{TPa} and \eq{Phis} to expand
$\nabla_c\Phi-i\{P_c,\Phi\}_\star$ as
\bea&&\sum_{s,n}{i^n\over n!} \left(T_{a(s+n),b(s)} \nabla_c
-2i(\D_{n+s,s}T_{ca(s+n),b(s)}+ \l_n^{(s)} \eta_{c\{
a}T_{a(s+n-1),b(s)\}})\right)\Phi^{a(s+n),b(s)}\ .\qquad\eea
The component form \eq{DPhicomponents} follows by rewriting the
middle term as $T_{ca(s+n),b(s)}
\Phi^{a(s+n),b(s)}=T^{a(s+n+1),b(s)} \eta_{c\{
a}\Phi_{a(s+n),b(s)\}}$ and last term as
\bea \eta_{c\{ a}T_{a(s+n-1),b(s)\}} \Phi^{a(s+n),b(s)}&=&
\left(\eta_{ca}T_{a(s+n-1),b(s)} +\mbox{($\eta_{aa}$ and
$\eta_{ab}$ traces)}\right)\Phi^{a(s+n),b(s)}\nn\\[5pt]&=&
T^{a(s+n-1),b(s)}\Phi_{c\{ a(s+n-1),b(s)\} }\ .\eea
Contracting \eq{DPhicomponents} by $\nabla^c$ yields
\bea \nabla^2\Phi_{a(s+n),b(s)}&=&
2n\D_{s+n-1,s}\eta^{cd}\nabla_d\eta_{c\{ a}\Phi_{a(s+n-1),b(s)\}}-
{2\l_{n+1}^{(s)}\over n+1} \eta^{cd}\nabla_d\Phi_{c\{
a(s+n),b(s)\}}\ ,\label{intermediatemasses}\eea
with $\eta_{c\{ a}\Phi_{a(s+n-1),b(s)\}}$ given by \eq{hooked},
and where the gradients on the right-hand side are to be
eliminated using \eq{DPhicomponents}. In the first term one finds
\bea &&-4\D_{s+n-1,s}\l_{n}^{(s)}\eta^{cd} \left(\eta_{c a}\Phi_{d
\langle a(s+n-1),b(s)\rangle} +\a_{s+n,s}\eta_{a(2)}
\Phi_{d\langle
a(s+n-2)c,b(s)\rangle}\right.\nn\\[5pt]&&\left.+\beta_{s+n,s}\eta_{a(2)}\Phi_{d\langle
a(n+s-2)b,cb(s-1)\rangle}+\c_{s+n,s}\eta_{ab}\Phi_{d\langle
a(s+n-1),cb(s-1)\rangle}\right)\nn\\[5pt]&=&
-4\D_{s+n-1,s}\l_{n}^{(s)}\Phi_{a\langle a(s+n-1),b(s)\rangle}\ =\
-4\l_{n}^{(s)}\Phi_{a(s+n),b(s)}\ ,\label{appmass1}\eea
where the $\langle\cdots\rangle$ Young projections are imposed
\emph{prior} to the final symmetrization on $a$ and $b$ indices
and $\Phi_{a\langle
a(s+n-1),b(s)\rangle}=\left(\D_{s+n-1,s}\right)^{-1}\Phi_{a(s+n),b(s)}$,
which follows from the definition of $\D_{s+n,s}$ in \eq{TPa}. In
the second term
\bea &&- 4\l_{n+1}^{(s)} \D_{s+n,s}\eta^{cd}\eta_{c\{ d}\Phi_{\{
a(s+n),b(s)\}\}}\nn\\[5pt]&=&
- 4\l_{n+1}^{(s)}\D_{s+n,s}{1\over
n+s+1}\left(s+n+2\e_0+3+2\a_{s+n+1,s}-{2\beta_{s+n+1,s}\over
s+n}+\c_{s+n+1,s}\right)\Phi_{a(s+n),b(s)}\nn\\[5pt]&=&-4\l_{n+1}^{(s)}\D_{s+n,s}
{(n+s+2\e_0)(n+s+\e_0+\ft32)(n+2s+2\e_0+1)\over (n+s+1)
(n+2s+2\e_0)(n+s+\e_0+\ft12)}\Phi_{a(s+n),b(s)}\
.\label{appmass2}\eea
Combining \eq{appmass1} and \eq{appmass2} the critical mass can be
identified with \eq{msk}.

Unfolding a scalar field $\phi$ obeying $(\nabla^2-M^2_0)\phi=0$
yields the master-field equation $(\nabla-ie^a
P_a)\ket{\Phi_{(0)}}=0$ where $\ket{\Phi_{(0)}}=\sum_{n=0}^\infty
{i^n\over n!} \Phi^{a(n)}\ket{{(n)}}_{a(n)}$ belongs to $\mR(0)$
defined by \eq{TPa} with $\e=\s$ and $\mm_1=\mm$. From
$[\nabla_a,\nabla_b]V_c=2\s\eta_{c[b}V_{a]}$ it follows that the
auxiliary $0$-forms obey
\bea (\nabla^2-M^2_{0,n})\Phi_{a(n)}\ =\ 0\ ,\qquad M^2_{0,n}\ =\
\s(M^2_0-(n+2\e_0+1)n)\ ,\eea
for non-critical masses obeying \eq{masses0} with $\l_n^{(s)}$
given by \eq{lambdas=0}. Similarly, unfolding a spinor $\psi$
obeying $(\c^a\nabla_a+\e' M_{1/2})\psi=0$ yields the master-field
equation $(\nabla-i e^a P_a)\ket{\Psi_{(1/2)}}=0$ where
$\ket{\Psi_{(1/2)}}=\sum_{n=0}^\infty {i^n\over n!}
\overline\Psi^{a(n)}\ket{{(n+\ft12)}}_{a(n)}$ belongs to
$\mC(\ft12)$ defined by \eq{TPa1/2} with $\e=\s$ and $\e'$ is the
sign in $\overline\psi^\b(\c_a)_\b{}^\a=\e'(\c_a)^{\a\b}\psi_\b$.
The conformal masses in the maximally symmetric $D$-dimensional
geometry with $R_{ab,cd}=-2\s \eta_{a[c}\eta_{d]b}$ are given by
\bea M^2_0|_{\rm conf}&=&{(D-2)\eta^{ac}\eta^{bd}R_{ab,cd}\over
4(D-1)}\ =\ -\s(\e_0+\ft12)(\e_0+\ft32)\ ,\qquad M^2_{1/2}|_{\rm
conf}\ =\ 0\ ,\label{mconf}\eea
We note that $M^2_0|_{\rm conf}=M^2_0|_{\rm crit}$ iff $D=4$ or
$D=6$ and that $M^2_{1/2}|_{\rm conf}=M^2_{1/2}|_{\rm crit}$ iff
$D=4$.


\scs{Indecomposable negative-spin extension of scalar
singletons}\label{App:negspin}


The scalar singleton $\mD^+_0=\mD^+(\e_0;(0))$, which consists of
compact $(\me\oplus\ms)$ weights $(\e_0+j;(j))$ with
$j=0,1,\dots$, can be embedded together with the scalar
anti-singleton $\mD_0^-=\mD^-(-\e_0;(0))$ into the indecomposable
representation
\bea \mM_0&=& \mW_0\ssumr (\mD^+_0\oplus\mD^-_0)\ ,\qquad \mW_0\
=\ \bigoplus_{(e;(\nu))\in \L(\e_0)}\mW_{e;(\nu)}\ , \eea
where
$\L(\e_0)=\left\{(\pm([\e_0]+p);(-(\e_0+[\e_0]+p)))\right\}_{p=0}^\infty$
and ${\rm mult}(e;(\nu))=1$. The negative spins $(\nu)$ label
$\ms$-irreps $\mW_{e;(\nu)}|_{\ms}\simeq \mW(\nu)$ given by the
quotient submodule sitting in the dual $\widetilde \mR_\ms(0)$ of
the twisted-adjoint $\ms$-module $\mR_\ms(0)$ defined by \eq{Mmn},
\eq{TPa} and \eq{lambda} with $D\rightarrow D-2$ and
$C_2[\ms|(0)]= \nu(\n+2\e_0)=p^2-\e^2_0$. This twisted-adjoint
module is irreducible for $p\leqslant \e_0-[\e_0]-1$ and
indecomposable for $p\geqslant \e_0-[\e_0]$, in which case its
dual $\widetilde \mR_\ms(0)$ contains $\mD(j)$ with
$j=p-\e_0+[\e_0]$ as an invariant subspace. The representation
matrix of $\widetilde \mR_\ms(0)$ is given below in \eq{Rrprime}
and \eq{tilderho}. The decomposition $\mM_0|_{\mg'}={\cal
M}^{(+)}_{(0)}\oplus {\cal M}^{(-)}_{(0)}$ under
$\mg'=\mso(D-2,2)\subset \mg$ is the twisted-adjoint
compact-weight space of the harmonic expansion of a conformal
scalar field\footnote{Alternatively, on $S^1\times S^{D-2}$ with
metric $ds^2=-dt^2+d\Omega^2_{S^{D-2}}$, the harmonic expansion
reads $\phi=\sum_{\o,\nu,\a}\phi_{\o,\nu,\a} e^{i\omega
t}D_{\nu,\a}(\widehat n)$ where $|\omega|=-\nu-\e_0$ and
$(\nabla^2_{S^{D-2}}+\nu(\nu+2\e_0)) D_{\nu,\a}=0$ for
$\nu=-(\e_0+[\e_0]+p)$ with $p=0,1,\dots$. For fixed
$\nu\equiv-2\e_0-\ell$, the generalized spherical harmonics
$\{D_{\nu,\a}\}\simeq \mW(\nu)$, which is irreducible if $\ell<0$
and indecomposable if $\ell\geqslant 0$ with ideal consisting of
the standard spherical harmonics $Y_{\ell,\a}$.} in $AdS_{D-1}$,
such that $\mW_0|_{\mg'}={\cal W}^{(+)}_{(0)}\oplus {\cal
W}^{(-)}_{(0)}$ and $\mD^\pm_0|_{\mg'}=\mD^\pm(\pm\e_0;(0))\oplus
\mD^\pm(\pm(\e_0+1);(0))$, with
\bea {\cal
M}^{(\pm)}_{(0)}&=&\bigoplus_{\tiny\ba{c}e\in\integ+[\e_0]\,,\
j'\in\{0,1,\dots\}\\ e+j'=[\e_0]+\ft{1\mp 1}2~\mbox{mod
$2$}\ea}\Comp\otimes \ket{e;(j')}\ =\ {\cal W}^{(\pm)}_{(0)}\ssumr
\left[(1+\pi)\mD^+(\e_0+\d_\pm;(0))\right]\ ,\eea
where $\d_\pm=\ft12(1\mp(-1)^{\e_0-[\e_0]})$. We note that the
action of $\mg'$ is unitarizable in $\mM_0$, while the action of
$\mg$ is only unitarizable in $\mD^\pm_0$.

Explicitly, letting $\mm'=\mso(D-2,1)\subset \mso(D-2,2)$, the
harmonic map from the conformal twisted-adjoint $(\mg'|\mm')$
module $\mR(0)|_{\mg'}$ (given by \eq{TPa},\eq{TRa}, \eq{TR0} and
\eq{conflambda} for $D\rightarrow D-1$) to ${\cal M}_0$ takes the
form
\bea \ket{{}^{(0){\rm c}}_{e;(j')}}_{r'(j')}&=&\sum_{n=0}^\infty
f^{(0){\rm c}}_{e;(j');n}
\ket{{}^{(0)}_{(n+j')}}_{0(n)\{r'(j')\}}\ ,\label{wnu}\eea
where $\ket{{}^{(0)}_{(n)}}_{a'(n)}\in \mR(0)|_{\mg'}$
($a'=0,1,\dots,D-2$) and $\ket{{}^{(0){\rm c}}_{e;(j')}}_{r'(j')}$
($r'=1,\dots,D-2$) is a type-$(j')$ tensor of $\ms'=\mso(D-2)$
with energy $(E-e)\ket{{}^{(0){\rm c}}_{e;(j')}}_{r'(j')}=0$. The
latter condition yields $\theta(n-1)f^{(0){\rm
c}}_{e;(j');n-1}-\frac{e}2 f^{(0){\rm c}}_{e;(j');n}-\frac
1{16}(n+1)(n+2(j'+\e_0))f^{(0){\rm c}}_{e;(j');n+1}=0$ for
$n\geqslant 0$, where $\theta(x)$ equals $1$ for $x\geqslant 0$
and $0$ for $x<0$. Equivalently, the generating function
$f^{(0){\rm c}}_{e;(j')}(z)=\sum_{n=0}^\infty z^n f^{(0){\rm
c}}_{e;(j');n}$ obeys
\bea \left(\frac{z}{16}{d^2\over dz^2}+\frac{j'+\e_0}{8}{d\over
dz}+\frac{e}2-z\right)f^{(0){\rm c}}_{e;(j')}(z)&=&0\ ,\eea
whose solutions that are analytic at $z=0$ can be written as the
closed contour integrals\footnote{The line integral $\int_a^b ds
f(s)$, where $a,b\in\Real$ and $f(s)$ is analytic in a
neighborhood of $[a,b]$ can be rewritten as the closed contour
integral
\bea \int_a^b ds f(s)&=& \oint_\gamma {ds\over 2\pi i}
\log\left({s-a\over s-b}\right) f(s)\ ,\eea
where $\gamma$ encircles $[a,b]$. For example, if $n=0,1,\dots$
then
\bea \int_0^1 ds s^n&=& \oint_\gamma {ds\over 2\pi i}
\log\left({s\over s-1}\right) s^n\ =\ \oint_0 {ds\over 2\pi i}
\log\left({1\over 1-s}\right) s^{-n-2}\ =\ {1\over n+1}\ .\eea
This lemma generalizes to the case where $f(s)$ has branch-cut
along $[a,b]$ provided $f(s)\sim (s-s_0)^{\eta_{s_0}}$ with
$\eta_{s_0}>-1$ as $s\sim s_0\in\{a,b\}$, so that the real line
integral is finite.}
\bea f^{(0){\rm c}}_{e;(j')}(z)&=&{\cal C}^{(0){\rm
c}}_{e;(j')}\oint_C {ds\over 2\pi
i}\d_{e,j'}(s)(1-s)^{\e_0+j'+e-1}(1+s)^{\e_0+j'-e-1}e^{4sz}\
,\label{confgenf}\eea
where ${\cal C}^{(0){\rm c}}_{e;(j')}$ is a normalization constant
chosen such that $f^{(0){\rm c}}_{e;(j')}(0)=1$; $\d_{e,j'}(s)=1$
for $|e|\geqslant \e_0+j'$ and $\d_{e'j'}(s)=\log\frac{s+1}{s-1}$
for $|e|\leqslant \e_0+j'-1$; and $C$ is a closed contour
encircling the branch cut from $[-1,1]$. The integral collapses on
residues at $s=\pm1$ for $|e|\geqslant \e_0+j'$, \emph{i.e.} for
elements in the lowest-weight and highest-weight spaces, while it
collapses on the (logarithmic) branch cut and turns into a real
line integral from $s=-1$ to $s=+1$ for $|e|\leqslant \e_0+j'-1$,
\emph{i.e.} for elements in the lowest-spin spaces.

The non-compact $\ms$ module $\mW(\nu)$ thus consists of the
states $\ket{{}^{(0){\rm c}}_{e;(j')}}$ on which the $\ms$ action
is represented by $M_{r's'}$, generating the $\ms'$ subalgebra of
$\ms$, and the conformal translations
\bea \frac12 R_{r'}\ket{{}^{(0){\rm c}}_{e;(j')}}_{s'(j')}&=&
\widetilde \r^{(e)}_{j'} \ket{{}^{(0){\rm
c}}_{e;(j'+1)}}_{r's'(j')}-\widetilde\l^{(e)}_{j'}\d_{r'\{}\ket{{}^{(0){\rm
c}}_{e;(j'-1)}}_{s'(j'-1)\}}\ ,\label{Rrprime}\\ \widetilde
\r^{(e)}_{j'}&=& {(j'+e+\e_0)(j'-e+\e_0)\over
(j'+\e_0)(j'+\e_0+\ft12)}\ ,\qquad \widetilde\l^{(e)}_{j'}\ =\
\frac18 j'(j'+\e_0-1)\ ,\label{tilderho}\eea
as can be seen by acting on \eq{wnu} with $R_{r'}$ and using
\eq{TRa}; one first obtains $\ft12R_{r'}\ket{{}^{(0){\rm
c}}_{(n+j')}}_{s'(j')0(n)}=\ket{{}^{(0){\rm
c}}_{(n+j'+1)}}_{s'(j')0(n)}-
\ft18(n+j')(n+j'+\e_0-1)\eta_{r'\{s'_1}\ket{{}^{(0){\rm
c}}_{(n+j'-1)}}_{s'(j')0(n)\}_{D-1}}$, which yields $$\ft12
R_{r'}\ket{{}^{(0){\rm c}}_{(n+j')}}_{\{s'(j')\}0(n)}=
\sum_{p=0,2}\left(\widetilde\r^{n,j'}_p\ket{{}^{(0){\rm
c}}_{(n+j'-p)}}_{\{r's'(j')\}0(n-p)}-
\widetilde\l^{n,j'}_p\d_{r'\{s'_1}\ket{{}^{(0){\rm
c}}_{(n+j'+p)}}_{s'(j'-1)\}0(n+p)}\right)$$ with
$\widetilde\rho^{n,j'}_0=1$, $\widetilde\rho^{n,j'}_2=-\ft1{16}
n(n-1)$, $\widetilde\l^{n,j'}_0=\ft18
j'(n+j'+\e_0-1)+\ft1{32}{n(n-1)j'\over j'+\e_0-\ft12}$ and
$\widetilde\l^{n,j'}_2=-\ft12{j'\over j'+\e_0-\ft12}$, and where
$\{\cdots\}$ denotes the traceless type-$(j')$ projection. The
coefficients in \eq{tilderho} are then given by
$\widetilde\rho^{(e)}_{j'}=\widetilde\rho^{0,j'}_0+\widetilde\rho^{2,j'}_2
f^{(0){\rm c}}_{e;(j');2}$ and
$\widetilde\l^{(e)}_{j'}=\widetilde\l^{0,j'}_0$.

The Flato-Fronsdal construction raises the issue of what is the
additional content of the direct product $\mM_0\otimes \mM_0$. It
cannot be identified with ${\cal M}_{(0)}$, which has already the
factorization given in \eq{statsing} in terms of two angletons.
Indeed, in odd dimensions the shadow $\ket{2;(0)}_{12}$, which
obey $(E(\x)-1)\ket{2;(0)}_{12}=0$ and
$L^-_r(\x)\ket{2;(0)}_{12}=0$ ($\x=1,2$), cannot be realized in
$\mM_0\otimes \mM_0$ except in the trivial case of $D=5$.


\scs{Some properties of the $T^{(0)}_{e;(j)}$
elements}\label{App:T0}


To analyze the elements $T^{(0)}_{e;(j)}$ defined by \eq{TE}, we
start from \eq{TPa} which for $\s=1$ implies
\bea \widetilde P_0 T_{0(n)\{r(j)\}}&=& 2\left(T_{0(n+1)\{ r(j)\}}-{1\over 16}\l_{(j);n}T_{0(n-1)\{ r(j)\}}\right)\ ,\label{TP0r}\\[5pt]\l_{(j);n}&=& {n(n+j+1)(n+j+2\e_0-1)(n+2j+2\e_0)\over (n+j+\e_0+\ft12)(n+j+\e_0-\ft12)}\ .\eea
Thus, the coefficients $f^{(0)}_{e;(j);n}$ obey the recursion
relation
\bea \theta(n-1) f^{(0)}_{e;(j);n-1}-{e\over 2}
f^{(0)}_{e;(j);n}-{1\over 16} \l_{(j);n+1}
f^{(0)}_{e;(j);n+1}&=&0\ , \label{Erec}\eea
with initial condition $f^{(0)}_{e;(0);0}=1$ and . If $j=0$ one
has
\bea E\star E^n&=& E^{n+1}-{1\over 16} \l_{(0);n} E^{n-1}\ ,\qquad
\l_{(0);n}\ =\ {n(n+1)(n+2\e_0-1)(n+2\e_0)\over
(n+\e_0-\ft12)(n+\e_0+\ft12)}\ ,\label{estaren}\eea
where $E^n\equiv T_{0(n)}$, and the recursion relation can be
rewritten as
\bea 2e(n+\e_0+\ft12)\widetilde f_n&=&
(n+1)(n+2\e_0)\left(\widetilde f_{n-1}-\widetilde f_{n+1}\right)\
,\qquad f_n\ =\ {4^n (\e_0+\ft32)_n\over
(n+1)!(2\e_0+1)_n}\widetilde f_n\ .\qquad\eea
If $e=0$ then $f^{(0)}_{0;(0);2p+1}=0$ and
\bea f^{(0)}_{0;(0);2p}&=&
4^{2p}{\left(\e_0+\ft32\right)_{2p}\over (2)_{2p}(2\e_0+1)_{2p}}\
=\ 2^{2p}{\left(\ft{2\e_0+5}4\right)_p
\left(\ft{2\e_0+3}4\right)_p\over p!\left(\ft32\right)_p
\left(\e_0+1\right)_p \left(\e_0+\ft12\right)_p}\ ,\eea
and one may write the generating function as
\bea f^{(0)}_{0;(0)}(z)&=& {1\over 2E}\int_0^E dz \left({}_1F_1(\e_0+\ft32;2\e_0+1;4z)+{}_1F_1(\e_0+\ft32;2\e_0+1;-4z)\right)\nn\\[5pt]
&=&{}_2 F_3\left({2\e_0+3\over 4},{2\e_0+5\over 4};
\ft32,\e_0+\ft12,\e_0+1;4z^2\right)\ .\eea
Since $T^{(0)}_{e;(0)}$ obeys $(E-\frac{e}2)\star
T^{(0)}_{e;(0)}=0$ it can be represented as
\bea T^{(0)}_{e;(0)}&=& {\cal
C}^{(0)}_{e;(0)}\oint_{C}{d\a(s)\over 2\pi i} \d_e(s)\exp
(-\frac{e\a(s)}2) \,g(s;E)\ ,\qquad g(s;E)\ =\ \exp_\star[\a(s)E]\
,\eea
where $g(s;E)$ is the Weyl-ordered form of the group element; $C$
is a closed and bounded contour; $\d_e(s)$ equals $1$ if there are
poles or branch cuts in the remaining part of the integrand, which
is the case for generic $e$, and $\log[(s-a)/(s-b)]$ for suitable
$a$ and $b$ for special values of $e$; and ${\cal
C}^{(0)}_{e;(0)}$ a normalization chosen such that
$f^{(0)}_{e;(0)}(0)=1$. In $D=4,6$ the composite massless scalars
are conformal (see Appendix \ref{App:T}), with simpler
representations matrices, \emph{viz.} $\l^{(0)}_n={\s\over
8}n(n+\e_0-\ft12)$ and $\l_{(0);n}=n(n+2\e_0)$, and the group
element can be written as
\bea g(s;E)&=& {1\over \cosh^{2\e_0+1}\ft{\a(s)}2} \exp[4E\tanh
\ft{\a(s)}4]\ .\eea
Specifically, in $D=4$, one has $E\star T(E)=\left(1-{1\over
16}{d^2\over dE^2}\right)E T(E)$, implying $\left({1\over
16}{d^2\over dz^2} +\frac{e}{2z}-1\right)zf^{(0)}_{e;(0)}(z)=0$,
and $f^{(0)}_{e;(0)}(z)=e^{-4z}{}_1 F_1(1-e;2;8z)$ for
$e\in\Comp$. These functions can be represented via the Laplace
transformations
\bea e\neq 0&:& f^{(0)}_{e;(0)}(z)\ =\ -{1\over 2e} \oint_{\gamma}{ds\over 2\pi i} \left({s-1\over s+1}\right)^e e^{4sz}\ ,\label{eneq0}\\[5pt]
e=0&:& f^{(0)}_{0;(0)}(z)\ =\ {1\over 2} \oint_{\gamma}{ds\over
2\pi i} \log \left({s+1\over s-1}\right) e^{4sz}\ ,\label{e=0}\eea
where the closed contour $\gamma$ encircles the interval $[-1,+1]$
which is a branch cut except for $e\in\{\pm 1,\pm2,\dots\}$. In
the latter case, the contour encloses the pole at $-{\rm sign}(e)$
with residue the rescaled Laguerre polynomial
\bea e\in\{\pm 1,\pm2,\dots\}&:& f^{(0)}_{e;(0)}(z)\ =\ {e^{-4{\rm
sign}(e)z}\over |e|} L^1_{|e|-1}(8{\rm sign}(e)z)\ .\eea
The integral \eq{eneq0} approaches \eq{e=0} as $e\rightarrow 0$,
since $\left(\frac{s-1}{s+1}\right)^{e}=1+e\log
\frac{s-1}{s+1}+{\cal O}(e^2)$, and \eq{e=0} can be rewritten as
the real line integral
\bea e=0&:& f^{(0)}_{0;(0)}(z)\ =\ \ft12 \int_{-1}^1 ds\, e^{4sz}\
=\ {\sinh 4z\over 4z}\ .\eea
In $D=6$ the functions take on a similar form:
\bea e\neq 0,\pm1 &:& f^{(0)}_{e;(0)}(z)\ =\ {3\over 4e(e^2-1)} \oint_{\gamma}{ds\over 2\pi i} (1-s^2)\left({s-1\over s+1}\right)^{e} e^{4sz}\ ,\label{eneq0D6}\\[5pt]
e=0,\pm 1&:& f^{(0)}_{0;(0)}(z)\ =\ {3\over 4(1+e^2)}
\oint_{\gamma}{ds\over 2\pi i} (1-s^2)\left({1-s\over
s+1}\right)^{e}\log \left({s+1\over s-1}\right) e^{4sz}\
.\label{e=0D6}\eea
If $e=2$ or $e=2\e_0$ it is possible to impose the lowest-weight
condition
\bea e=2,2\e_0&:& \widetilde L^-_r T^{(0)}_{e;(0)}\ =\ 0\
,\label{Lminuscondapp}\eea
which by means of
\bea \widetilde L^-_r T_{0(n)}&=& i\left(nT_{r0(n-1)}+2T_{r0(n)}+\ft18 \l'_{(0);n}T_{r0(n-2)}\right)\ ,\\[5pt]\l'_{(0);n}&=&{(n-1)n(n+1)(n+2\e_0-1)\over (n+\e_0-\ft12)(n+\e_0+\ft12)}\ ,\eea
amounts to the recursive relation ($n\geqslant 0$)
\bea e=2,2\e_0&:& f^{(0)}_{e;(0);n}+{n+1\over
2}f^{(0)}_{e;(0);n+1}+{1\over 16}\l'_{(0);n+2}f^{(0)}_{e;(0);n+2}\
=\ 0\ .\label{Lrec}\eea
Subtracting \eq{Erec} yields the first-order difference equation
($n\geqslant 0$)
\bea &&(n+1+e)f^{(0)}_{e;(0);n+1}+{1\over 8} (\l'_{(0);n+2}+\l_{(0);n+2})f^{(0)}_{e;(0);n+2}\ =\ 0\ ,\\[5pt] && \l'_{(0);n+2}+\l_{(0);n+2}\ =\ 2{(n+2)(n+3)(n+2\e_0+1)\over n+\e_0+\ft52}\ ,\eea
whose solution can be shown to obey \eq{Erec} iff $e=2,2\e_0$, in
which case $f^{(0)}_{2\e_0;(0);n}=(-4)^n{(\e_0+\ft32)_n\over
n!(2)_n}$ and $f^{(0)}_{2;(0);n}=(-4)^n{(\e_0+\ft32)_n\over
n!(2\e_0)_n}$ corresponding to the generating functions given in
\eq{scalarf020plus}.

Eq. \eq{mue} implies that $L^-_r\star T^{(0)}_{e;(0)}$ can vanish
only if $2(\e_0+1)e=e^2+4\e_0$, that is $e=2\e_0$ or $e=2$.
Similarly, from \eq{Cslemma} it follows that $M_{rs}\star
T^{(0)}_{e;(0)}$ can vanish only if $e=\pm 2\e_0$. Indeed, using
\eq{MabT} to derive the lemmas
\bea \ac_{M_{0r}} T_{0(n)}&=&2M_{0r} T_{0(n)}+{1\over 8} {(n-1)n(n+1)(n+2\e_0-1)\over (n+\e_0-\ft12)(n+\e_0+\ft12)}M_{0r}T_{0(n-2)}\ ,\\[5pt]
\ac_{M_{rs}} T_{0(n)}&=& 2M_{rs}T_{0(n)}-{1\over
8}{(n-1)n^2(n+1)\over
(n+\e_0-\ft12)(n+\e_0+\ft12)}M_{rs}T_{0(n-2)}\ ,\label{MrsT0n}\eea
one can then show that
\bea L^-_r\star T^{(0)}_{2\e_0;(0)}&=& L^-_r\star T^{(0)}_{2;(0)}\
=\ 0\ ,\qquad M_{rs}\star T^{(0)}_{2\e_0;(0,0)}\ =\ 0\
.\label{Tleft}\eea
%


\scs{Oscillator Realizations}\label{App:Osc}


In this Appendix we collect some basic properties of oscillator
algebras, and some particular properties of the spinor-oscillator
realization of 4D higher-spin representations.


\scss{On traces and projectors in oscillator
algebras}\label{App:G}


We first discuss traces, inner products and projectors in the
phase-space (two-sided) and Fock-space (one-sided) representations
of oscillator algebras. To further illustrate ideas we also
discuss generalized projectors and fermionic oscillators.

\begin{center}{\it Phase-Space Trace and Supertrace}\end{center}

The complexified Heisenberg algebra $u\star v-v\star u=1$
generates the associative algebra of Weyl-ordered, \emph{i.e.}
symmetrized, functions $f(u,v)$ with product
\bea f\star g&=& \int_{\Comp\times \Comp}~{d\xi d\bar\xi d\eta
d\bar\eta\over \pi^2} ~e^{2i(\bar\xi\eta+\bar\eta\xi)}
f(u+\xi,v+\bar\xi)g(u+i\eta,v-i\bar\eta)\ ,\eea
where $d\xi d\bar \xi=2d({\rm Re}\xi)d({\rm Im}\xi)$. The algebra
admits the two inequivalent hermitian conjugations
\bea u^\dagger&=&v\ ,\quad v^\dagger\ = u\ ,\qquad u^\ddag\ =\ -v\
,\qquad v^\ddag\ =\ -u\ ,\label{ddag}\eea
and two associated inequivalent traces, namely the cyclic trace
and the graded-cyclic supertrace
\bea \mathrm{Tr}_+(f)&=& \int_{\Comp} {du d\bar u\over 2\pi}
f(u,\bar u)\ ,\qquad \mathrm{Tr}_-(f)\ =\ {f(0,0)\over 2}\
,\label{trplus}\eea
obeying $\Tr_+(f\star g)=\Tr_+(fg)=\Tr_+(g\star f)$ up to boundary
terms and $\mathrm{Tr}_-(f\star g)=
(-1)^{\e(f)\e(g)}\mathrm{Tr}_-(g\star f)$ for functions $f$ and
$g$ with definite parity defined by $f(-u,-v)=(-1)^{\e(f)}f(u,v)$
\emph{idem} $g$. The two traces are related as follows:
\bea \Tr_\pm(f)&=&\Tr_\mp((-1)^{N}_\star\star f)\ ,\qquad N\ =\
v\star u\ ,\label{thm}\eea
where we use the notation $x^A_\star=\exp_\star (A\ln x)$ with
$\exp_\star A= \sum_{n=0}^\infty {A^{\star n}\over n!}$ and
$A^{\star n}= \underbrace{A\star\cdots \star A}_{\mbox{$n$
times}}$~. Eq. \eq{thm} is a consequence of the Weyl-ordering
formula
\bea \exp_\star (\a w)&=& {\exp({2w\tanh\ft\a2})\over
\cosh{\a\over 2}}\ ,\qquad w\ =\ N+\ft12\ =\ uv\
,\label{lemma}\eea
for $\a\in\Comp\backslash\{\pm i\pi,\pm 3i\pi,\dots\}$. This
formula follows by acting with $\partial/\partial\a$ and using
\bea w\star f(w)&=&\left(w-{1\over 4} {\partial\over \partial
w}-{1\over 4} w{\partial^2\over \partial w^2}\right)f(w)\
.\label{wstarf}\eea
Thus, setting $\exp_\star \a w =r(\a)\exp(s(\a) w)$, one finds
$r'=-rs/4$ and $s'=1-s^2/4$ subject to $r(0)=1$ and $s(0)=0$, with
the solution $r^{-1}=\cosh(\a/2)$ and $s=2\tanh(\a/2)$. Eq.
\eq{thm} then follows from
\bea \exp_\star(i(\pi+\e)N)\sim -i {\exp{2iuv\over\eta}\over
\eta}\qquad \mbox{as $\eta=-\sin(\e/2)\rightarrow 0$}\ ,\eea
which together with $Tr_+(f\star g)=\Tr_+(fg)$ implies
\bea \lim_{\e\rightarrow 0}\Tr_+(\exp_\star(i(\pi+\e)N)\star f)\
=\ -i\lim_{\eta\rightarrow 0}\int_{\Comp}{dud\bar u\over 2\pi}
{\exp{2iu\bar u\over\eta}\over \eta} f(u,\bar u)\ =\ {f(0,0)\over
2}\ .\eea
Moreover, from $\exp_\star (\a N)\star \exp_\star(\b
N)=\exp_\star((\a+\b)N)$, it follows that
\bea \lim_{\e\rightarrow 0}
\left[\exp_\star(i(\pi+\e)N)\right]^{\star 2}
&=&\lim_{\e\rightarrow
0} \exp_\star(2i(\pi+\e)N)\\[5pt]& =&\lim_{\e\rightarrow 0}{\exp(2\tanh
(i(\pi+\e))N -i(\pi+\e))\over \cosh i(\pi+\e)}\ =\ 1\ ,\eea
in agreement with $(-1)_\star^{N}\star
(-1)_\star^N=(-1)_\star^{2N}=1$.

\begin{center}{\it Fock-Space Inner Products and Projectors}\end{center}

The standard Fock space ${\cal F}=\bigoplus_{n=0}^\infty
\Comp\otimes \ket{n}$, where $\ket{n}={v^n\over \sqrt{n!}}\ket{0}$
and $u\ket{0}=0$, has two inequivalent inner products
$I_\pm(\ket{\Psi},\ket{\Psi'})\equiv
{}_\pm\langle\Psi|\Psi'\rangle$ defined by
\bea I_\pm (\m \ket{m},\n \ket{n})&=& \bar\mu \nu (\pm
1)^m\d_{mn}\ ,\qquad \mu,\nu\in \Comp\ ,\eea
and related to the traces $\Tr_\pm$ by
\bea I_\pm (\ket{m},\ket{n})&=& \Tr_\pm(P_{n,m})\ ,\qquad P_{n,m}\
=\ {1\over \sqrt{m!n!}}~v^n\star P_{0,0}\star u^m\ ,\qquad
P_{0,0}\ =\ 2e^{-2w}\ .\qquad\label{Pmn}\label{relation}\eea
One may identify $P_{n,m}\leftrightarrow
\ket{n}{}_+\bra{m}=\ket{n}{}\bra{m}$, where $\bra{\Psi}\equiv
{}_+\bra{\Psi}$, since the projector algebra $P_{m,n}\star
P_{p,q}=\d_{np}P_{m,q}$ follows from $P_{0,0}\star
P_{0,0}=P_{0,0}$ and $u\star P_{0,0}=P_{0,0}\star v=0$, and
\eq{relation} follows from the cyclicity properties of $\Tr_\pm$
which imply $\Tr_\pm(P_{n,m})=(\pm
1)^n\d_{mn}\Tr_\pm(P_{0,0})=(\pm 1)^n\d_{mn}$ using
$\Tr_\pm(P_{0,0})=1$. Moreover, it follows from \eq{thm} that
${}_\pm\langle\Psi|\Psi'\rangle ={}_\mp\langle\Psi|
(-1)_\star^N\vert \Psi'\rangle$, which indeed induces the
hermitian conjugation rules \eq{ddag}:
\bea I_\pm(f
\ket{\Psi},\ket{\Psi'})&=&I_\pm(\ket{\Psi},g\ket{\Psi'})\ ,\qquad
g\ =\ \left\{\ba{ll}
f^\dagger&\mbox{for $I_+$}\\[5pt] f^\ddag&\mbox{for $I_-$}\ea\right.\quad\ .\eea
Next, writing the projectors on even and odd states as
$P_{(\pm)}=\sum_{n=0}^\infty \ft12(1\pm(-1)^n) \vert
n\rangle\langle n\vert=\ft12:(e^N\pm e^{-N})\vert 0\rangle\langle
0\vert:$ and using $1=P_{(+)}+P_{(-)}=:e^{N}\vert 0\rangle\langle
0\vert:$, yields the useful lemma:
\bea (-1)_\star^N&=& :e^{-2N}:\ ,\qquad \vert 0\rangle\langle
0\vert\ =\ :e^{-N}:\ ,\qquad P_{(\pm)}\ =\ \ft12 (1\pm:e^{-2N}:)\
.\eea
To verify $\Tr_\pm(P_{n,n})=(\pm 1)^n$, one may compute
$P_{n,n}=\ket{n}\bra{n}=2(-1)^n e^{-2w}L_n(4w)$ by either direct
evaluation of the $\star$-products in \eq{Pmn}, or by using $:e^{a
u+ b v}:= e^{au}_\star\star e^{bv}_\star=e_\star^{a u+ bv+\ft12
ab}=e^{au+bv+\ft 12 ab}$ followed by Fourier
transformation\footnote{The Laguerre polynomials $L_n(x)={1\over
n!}e^{x}{d^n\over dx^n} (e^{-x} x^n)=\sum_{p=0}^n{n\choose n-p}
{(-1)^p\over p!} x^p$ obey $\sum_{n=0}^\infty (-z)^n
L_n(2x)={e^{2xz\over 1+z}\over 1+z}=\sum_{n=0}^\infty\sum_{p=0}^n
{n\choose n-p}(-2)^pL_p(x) z^n$.}:
\bea P_{n,n}&=&\ket{n}\bra{n}\ =\ {1\over n!} :v^n e^{-vu} u^n:\
=\ \int {dkd\bar k\over 2\pi}:e^{-i(\bar k u+k v)-\bar k
k}:L_n(\bar k k)\nn\\[5pt]
&=&\sum_{p=0}^n{n\choose n-p} {1\over
p!}(\partial_u\partial_v)^p\int {dkd\bar k\over 2\pi}e^{-i(\bar k
u+k v)-\ft12 \bar k k}\ =\ 2\sum_{p=0}^n{n\choose n-p} (-2)^p e^{-2w}L_p(2w)\nn\\[5pt]&=&
2(-1)^n e^{-2w} L_n(4w)\ .\eea

\begin{center}{\it Anti-Fock Space and (Anti-)Automorphisms}\end{center}

The anti-Fock space is defined by
\bea {\cal F}^- &=& \bigoplus_{n=0}^\infty \Comp\otimes \ket{n}^-\
,\qquad \ket{n}^-\ =\ \frac{u^n}{\sqrt{n!}}~\ket{0}^-\ , \quad
v\ket{0}^-\ =\ 0 \ .\eea
Its two inequivalent inner products are defined by
\bea {}_\pm^-\bra{m}n\rangle^-&=& (\mp 1)^m\d_{mn}\ =\
\Tr_\pm(P^-_{n,m})\ ,\eea
with $P^-_{n,m}=\ket{n}^-{}^-\bra{m}={1\over \sqrt{n!m!}}~u^n\star
P^-_{0,0}\star v^m$ and $P^-_{0,0}=2e^{2w}$. It follows that
$P^-_{m,n}\star P^-_{p,q}=(-1)^n\delta_{np}P^-_{m,q}$ and that
\bea P^-_{n,n}&=&2 (-1)^ne^{2w}L_n(-4w)\ .\eea
For uniformity we define ${\cal F}^+={\cal F}$,
$\ket{n}^+=\ket{n}$, ${}^+\bra{n}=\bra{n}$ and
$P^+_{n,m}=P_{n,m}$, so that
$w\ket{n}^\pm=\pm(n+\ft12)\ket{n}^\pm$ and
${}^\pm\bra{n}w=\pm(n+\ft12){}^\pm\bra{n}$. The oscillator algebra
has the automorphism $\pi$ and anti-automorphism $\tau$ given by
\bea \pi(f(u,v))&=& f(i v,iu)\ ,\qquad \tau(f(u,v))\ =\ f(i u,i
v)\ ,\eea
which exchange the $P^\pm$ projectors, \emph{viz.}
$\pi(P^\pm_{n,m})= \tau(P^\pm_{n,m})=i^{m+n}P^\mp_{m,n}$, and with
compositions
\bea \pi\circ \pi&=&\tau\circ \tau\ =\ \Ad_{(-1)_\star^{N'}}\
,\qquad N'\ =\ \sum_{n=0}^\infty\left((w-\ft12)\star
P^+_{n,n}+(w+\ft12)\star P^-_{n,n}\right)\ ,\label{squares}\eea
where $\Ad_X(Y)=X\star Y\star X^{-1}$, and
\bea R&=&\pi\circ \t\ ,\qquad R(f(u,v)) =\ f(-v,-u)\ ,\qquad
R\circ R\ =\ \mathrm{Id}\ .\eea
The action of the discrete maps can be extended to the Fock
spaces, such that
\bea \pi&:&{\cal F}^\pm\oplus {\cal F}^{\star \pm}\rightarrow
{\cal F}^\mp\oplus{\cal F}^{\star\mp}\ ,\qquad \tau\ :\ {\cal
F}^{\pm}\rightarrow {\cal F}^{\star\mp}\ ,\qquad R\ :\ {\cal
F}^\pm\rightarrow{\cal F}^{\star\pm}\ ,\eea
upon defining
\bea \pi(\ket{0}^\pm)&=& \ket{0}^\mp\ ,\qquad
\pi({}^\pm\bra{0})\ =\ {}^\mp\bra{0}\ ,\\[5pt] \tau(\ket{0}^\pm)&=&
{}^\mp\bra{0}\ ,\qquad \tau({}^\pm\bra{0})\ =\ \ket{0}^\mp \ ,
\eea
and $\Ad_X(\ket{\Psi})=X\ket{\Psi}$ and
$\Ad_X(\bra{\Psi})=\bra{\Psi}X^{-1}$, such that \eq{squares}
remains valid. It follows that
\bea R(\ket{0}^\pm)&=&{}^\pm\bra{0}\ ,\qquad R({}^\pm\bra{0})\ =\
\ket{0}^\pm\ ,\eea
that is, $R(\cdot)\leftrightarrow (\cdot)^\ddagger$ in the real
basis $\{\ket{n}^\pm,{}^\pm\bra{n}\}$ and where $\ddagger$ is
defined in \eq{ddag}. In Section \ref{Sec:4D} the discrete maps
acting on pseudo-real $\msu(2)$-doublets are defined analogously
with conventiones preserving $SU(2)$ quantum numbers.

\begin{center}{\it Generalized projectors and their composition}\end{center}

To illustrate the distinction between operators in Fock spaces and
more general classes of phase-space functions, we consider
analytic functions $M^C_{\kappa}(w)$ obeying
\bea (w-\kappa)\star M^C_{\kappa}&=&0\ ,\qquad \kappa\in\Comp\
,\label{Mkappa}\eea
where $C$ indicize a basis of linearly independent solutions. One
may consider
\bea M^C_\kappa&=& {\cal N}^C_\kappa\oint_{C} {d\alpha\over 2\pi
i} g^{(\kappa)}(\alpha)\ =\ {\cal N}^C_\kappa\oint_\Gamma
{ds~g^{(\kappa)}(\alpha(s)) \over 2\pi i (1-\ft{s^2}4)} \ =\ {\cal
N}^C_\kappa\oint_{\Gamma'}{d\lambda~g^{(\kappa)}(\a(\l))\over 2\pi
i(1+\lambda)} \ ,\eea
where $C$, $\C(C)$ and $\C'(C)$ are contours in the $\a$, $s$ and
$\l$ planes; $g^{(\kappa)}(\alpha)=e_\star^{\alpha(w-\kappa)}$
with Weyl-ordered and normal-ordered forms
\bea g^{(\kappa)}(\alpha(s))&=&
(1+\ft{s}2)^{\ft12-\kappa}(1-\ft{s}2)^{\ft12+\kappa}e^{s w}\
,\qquad
s\ =\ 2\tanh\ft{\alpha} 2\ ,\\[5pt]
g^{(\kappa)}(\alpha(\l))&=& (1+\lambda)^{\ft12-\kappa}:e^{\lambda
w}:\ ,\qquad \lambda\ =\ {s\over 1-\ft s2}\ =\ e^\a-1\ ;\eea
and ${\cal N}^C_\kappa\in\Comp$ are normalizations such that
$M^C_\kappa(0)=1$. In view of \eq{wstarf}, eq. \eq{Mkappa} holds
if $[g^{(\kappa)}(\alpha(s))]|_{\partial C}=0$. In particular, if
$C=C_{i\pi}$ is a small closed contour encircling $i\pi$
clockwise, then its image $\Gamma_{[2,-2]}\equiv s(C_{i\pi})$ is a
large contour encircling $[-2,2]$ counterclockwise\footnote{Other
interesting choices of closed contours are $C=i[-\pi,\pi]$ and
$C=\Real$ leading to integrals over $U(1)$ and $GL(1;\Real)$,
respectively.}. Enlarging $C_{i\pi}$ to the ''box''
$\{i\e+x:-L\leqslant x\leqslant L\}\cup \{L+ix:\e\leqslant
x\leqslant 2\pi-\e\}\cup\{i(2\pi-\e)-x:-L\leqslant x\leqslant
L\}\cup\{-L+i(2\pi-x):\e\leqslant x\leqslant 2\pi-\e\}$, its image
$\Gamma_{[-2,2]}$ shrinks to a ''dogbone'' containing $[-2,2]$.
For $\kappa=\pm(n+1/2)$, there is no branch cut,
$\oint_{\Gamma_{[-2,2]}}\rightarrow \oint_{\pm 2}$, and one finds
\bea M^{C_{i\pi}}_{\pm(n+\ft12)}&=& 2(\mp 1)^n e^{\mp 2w}L_n(\pm
4w)\ =\ P^\pm_{n,n}\quad\mbox{for ${\cal
N}^{C_{i\pi}}_{\pm(n+\ft12)}=(\pm 1)^{n+1}$}\ .\eea
Their $\star$-products can be computed using the composition rule
\bea g^{(\kappa)}(\a)\star g^{(\kappa)}(\a')&=&
g^{(\kappa)}(\a+\a')\ =\ g^{(\kappa)}(\a(s''))\ =\
g^{(\kappa)}(\a(\l''))\ ,\label{comprule}\eea
with $s''={s+s'\over 1+\ft {ss'}4}$ and $\l''=\l+\l'+\l\l'$. This
rule, which is equivalent to $e^{sw}\star e^{s' w}= {1\over
1+\ft{ss'}4}e^{s'' w}$ and $:e^{\l w}:\star
:e^{\l'w}=:e^{\l''w}:$, holds by analytical continuation for all
$s$ and $s'$ such that $ss'\neq -4$. A change of variables and
analytical contour deformation then yield
\bea (M^{C_{i\pi}}_{\pm(n+\ft12)})^{\star 2}&\!\!=\!\!&
\left[{\cal N}^{C_{i\pi}}_{\pm(n+\ft12)}\oint_{\mp 2}{ds\over 2\pi
i (1-\ft{s^2}4)}\right] M^{C_{i\pi}}_{\pm(n+\ft12)}\!\!\ =\!\!\
\pm {\cal
N}^{C_{i\pi}}_{\pm(n+\ft12)}M^{C_{i\pi}}_{\pm(n+\ft12)}\!\!\
=\!\!\ (-1)^n M^{C_{i\pi}}_{\pm(n+\ft12)}\ ,\qquad\qquad\eea
in agreement with $(P^\pm_{n,n})^{\star 2}=(-1)^n P^\pm_{n,n}$.
For $\kappa\notin(\integ+\ft12)$, the branch cut along $[-2,2]$
prevents $\Gamma_{[-2,2]}$ from collapsing, and
$(M^{C_{i\pi}}_{\k})^{\star 2}$ is computed with $s$ and $s'$
lying on large $\Gamma_{[-2,2]}$-contours; the change of variables
now yields\footnote{The functions $M^{C_{i\pi}}_0$ are related to
the dressing functions $M$ appearing in the perturbative
weak-field expansion of the oscillator formulations of higher-spin
gauge theories
\cite{SS5,Sezgin/Sundell-7,Vasiliev:2003ev,Sagnotti:2005ns}, which
are the analytical solutions to $K\star M=0$ for $K$ belonging to
the ``internal'' gauge group of the oscillator algebra, which is
$U(1)$, $SU(2)$ and $Sp(2)$, respectively, for 5D spinor, 7D
spinor and $D$-dimensional vector oscillators.}
\bea M_\kappa^{\star 2}&=& \left[{\cal N}^{C_{i\pi}}_\kappa
\oint_{\Gamma_{[-2,2]}}{ds\over 2\pi i (1-\ft{s^2}4)}\right]
M^{C_{i\pi}}_{\kappa}\ =\ 0\ \mbox{for
$\kappa\notin(\integ+\ft12)$}\ .\eea

\begin{center}{\it Fermionic Oscillators}\end{center}

In the case of fermionic oscillators, there is less distinction
between Fock-space and phase-space formulations. The complexified
Clifford algebra $\{\c,\d\}_\star =1$ has isomorphic Fock and
anti-Fock spaces, generated by $\ket{0}$ and $\ket{1}=\d\ket{0}$
obeying $\c\ket{0}=0$. The inequivalent inner products $I_\pm$ are
related by ${}_\pm\langle\Psi|\Psi'\rangle ={}_\mp\langle\Psi|
(-1)_\star^F\star \Psi'\rangle$ with $F\equiv \d\star\c$, and
induce the hermitian conjugation rules
\bea \c^\dagger&=&\d\ ,\quad \d^\dagger\ = \ \c\ ,\qquad \c^\ddag\
=\ -\d\ ,\quad \d^\ddag\ =\ -\c\
,\label{ddagfer}\label{dagfer}\eea
via
$I_+(f\star\ket{\Psi},\ket{\Psi'})=I_+(\ket{\Psi},f^\dagger\star\ket{\Psi'})$
and
$I_-(f\star\ket{\Psi},\ket{\Psi'})=I_-(\ket{\Psi},f^\ddag\star\ket{\Psi'})$.
The phase-space formulation uses the Weyl-ordered product
$\c\star\d=\c\d+1/2$, with $\c\d=[\c,\d]_\star/2=-\d\c$, and the
trace operations
\bea \Tr_+(f) & = & 2\,f(0,0) \ ,\qquad \Tr_-(f) \ =\ -\int d\c
d\bar{\c}\,f(\c,\bar{\c}) \ ,\label{ftrace2}\label{ftrace1}\eea
where $f=f(\c,\d)$ is Weyl-ordered and $\int d\c d\bar{\c}~\bar\c
\c=1$, and we note the interchanged role of $\Tr_+$ and $\Tr_-$ in
comparison to bosons. For example, in terms of the projectors
$P_0=\ket{0}\bra{0}=1-\d\star\c=\ft12 (1-2\d\c)$ and
$P_1=\ket{1}\bra{1}=\d\star\c=\ft12(1+2\d\c)$ one has
$\Tr_+(P_0)=\Tr_+(P_1)=1$ and $\Tr_-(P_0)=-\Tr_-(P_1)=1$. One can
show that
\bea \Tr_\pm((-1)_\star^F\star f) \ =\ \Tr_\mp(f) \
.\label{ftraces}\eea
To this end, one uses $\int d\c d\bar{\c}\,f\star g=\int d\c
d\bar{\c}\,fg$, and $F=P_1$ which implies that
\bea (-1)_\star^F\ =\ \exp_\star(i\pi F)\ = \ 1+\sum_{n=1}^\infty
\frac{(i\pi)^n}{n!}P_1 \ = \ 1+(e^{i\pi}-1)P_1 \ = \ 1-2P_1 \ =\
-2\bar{\c}\c\ ,\eea
which indeed implies
$(-1)_\star^F\star(-1)_\star^F=(-2\bar{\c}\c)\star(-2\bar{\c}\c)=1$.
Finally, using $\bar{\c}\c=\d(\bar \c)\d(\c)$, one finds
\bea \Tr_-((-1)_\star^F\star f)&=&-\int d\c
d\bar{\c}\,(-2\bar{\c}\c)f(\c,\bar{\c})\ =\ 2f(0,0)\ =\ \Tr_+(f)\
.\eea


\scss{4D Spinor-oscillator realizations}\label{Sec:4D}


In $D=4$ the algebra ${\cal A}$ is isomorphic to the space of
Weyl-ordered even arbitrary polynomials
\bea f(y,\yb) &=& \!\!\!\!\sum_{\tiny \ba{c}n,m\\\mbox{$n+m$
even}\ea}f^{\a(n),{\ad}(m)}T_{\a(n),{\ad}(m)} \ ,\qquad
T_{\a(n),{\ad}(m)}\ =\ {1\over n!m!}y_{\a_1}\cdots
y_{\a_n}\yb_{\ad_1}\cdots \yb_{\ad_m}\ ,\qquad\eea
in an $\msl(2;\Comp)_L\oplus \msl(2;\Comp)_R$-quartet
$(y_\a,\yb_{\ad})$ obeying
\bea y_\a\star y_\b&=&y_\a y_\b+i\epsilon_{\a\b}\ ,\qquad \bar
y_{\dot\a}\star \bar y_{\dot\b}\ =\ \bar y_{\dot\a} \bar
y_{\dot\b}+i\epsilon_{\dot\a\dot\b}\ ,\qquad y_{\a}\star \bar
y_{\dot\b}\ =\ \bar y_{\dot\b}\star y_{\a} \ = \ y_{\a}\bar
y_{\dot\b}\ , \qquad\label{osc1}\eea
where juxtaposition denotes Weyl-ordered products and
$T_{\a(n),\ad(m)}$ has $\msl(2;\Comp)_L\oplus \msl(2;\Comp)_R$
spin $(j_L,j_R)=\ft12(n,m)$ and Lorentz spin
$(s_1,s_2)=(\frac{n+m}{2},\frac{|n-m|}2)$. The $\star$-product
reads\footnote{The composition rules in more general classes of
functions require separate definitions; the case of the oscillator
realization of the compact basis elements defined in \eq{calMs}
shall be discussed in \cite{companion}.}
\bea f(y,\bar y)~\star~ g(y,\bar y)&=&\ \int \frac{d^2\xi d^2\eta
d^2\bar\xi d^2\bar\eta}{(2\pi)^4}~ e^{i\eta^\a\xi_\a+
i\bar\eta^{\dot\a}\bar\x_{\dot\a}} ~ f(y+\xi,\bar y+\bar \xi)~
g(y+\eta,\bar y+\bar \eta)\ .\label{star}
 \eea
The anti-automorphism $\t$ and automorphism $\pi$ can be taken to
be
\bea \t(f(y,\yb)) \ = \ f(iy,i\yb)\ ,\qquad \pi(f(y,\yb)) \ = \
f(-y,\yb)\ , \qquad \pb(f(y,\yb)) \ = \ f(y,-\yb)\ .\eea
The realization of the $\ell$th levels of the bosonic higher-spin
algebra $\mho(5;\Comp)$ and its twisted-adjoint representation
${\cal T}(5;\Comp)$ read
\bea Q_\ell&=&
\sum_{n+m=4\ell+2}Q_{\alpha_{1}\dots\alpha_{n}\dot{\alpha}_{1}\dots\dot
{\alpha}_{m}}T^{\a(n),\ad(m)}\ ,\qquad \ell=-\ft12,0,\ft12,1,\dots\ ,\\[5pt]
S_\ell&=& \sum_{|n-m|=4\ell}S^{\a(n),\ad(m)} T_{\a(n),\ad(m)}\
,\qquad \ell=-1,-\ft12,0,\ft12,1,\dots\ ,\eea
and the minimal algebra is obtained by truncating to integer
$\ell$.

The hermitian conjugation is defined in various signatures by
\cite{Iazeolla:2007wt}
\bea (f)^\dagger &=& \iota ((f)^{\dagger_{\rm osc}})\ ,\eea
where: {(i)} $\dagger_{\rm osc}$ acts on the oscillators in
different signature of the real form of $\mm$ as follows:
\bea \msu(2)_L\oplus \msu(2)_R&:&\quad (y^\a)^{\dagger_{\rm osc}}\
=\ y^{\dagger}_{\a}\ ,\quad (\bar y^{\ad})^{\dagger_{\rm osc}}\ =\
\bar
y^{\dagger}_{\ad}\ ,\label{su2}\\[5pt]
 \msl(2;\Comp)_{\rm diag}&:&\quad (y^\a)^{\dagger_{\rm osc}}\ =\ \bar y^{\ad}\
,\label{sl2}\\[5pt]
\msp(2;\Real)_L\oplus \msp(2;\Real)_R&:&\quad (y^\a)^{\dagger_{\rm
osc}}\ =\ y^{\a}\ ,\quad (\bar y^{\ad})^{\dagger_{\rm osc}}\ =\
\bar y^{\ad}\ ;\label{sp2} \eea
and {(ii)} $\iota$ is an oscillator-algebra isomorphism that acts
in different signatures of the real forms of $\mg$ and $\mm$ as
follows:
\bea \mso(5)\supset \mso(4)&:& \iota\ =\ \rho\ ,\\[5pt]
\mso(1,4)\supset \mso(4)&:& \iota\ =\ \pi\rho\ ,\\[5pt]
\mso(1,4)\supset \mso(1,3)&:& \iota\ =\ \pi\ ,\\[5pt]
\mso(2,3)\supset \mso(1,3)&:& \iota\ =\ {\rm Id}\ ,\\[5pt]
\mso(2,3)\supset \mso(2,2)&:& \iota\ =\ {\rm Id}\ ,\eea
where in Euclidean signature the $SU(2)$ doublets are pseudo real
(\emph{i.e.} $(y_\a)^{\dagger_{\rm osc}}=-y^{\dagger\a}$
\emph{idem} $\bar y_{\ad}$), and $(y_\a,\yb_{\ad})$ and
$(y^{\dagger}_\a,\yb^{\dagger}_\a)$ generate equivalent oscillator
algebras with isomorphism
\bea \rho(f(y^{\dagger}_\a,\bar y^{\dagger}_\a))&=&f(y_\a,\bar
y_\a)\ . \label{iso}\eea
The hermitian conjugation obeys
$((f)^\dagger)^\dagger=\iota((\iota((f)^{\dagger_{\rm
osc}}))^{\dagger_{\rm osc}})=f$, which relies on $\pi\bar\pi(f)=f$
for the real form $\mso(1,4)$. Thus, the real forms of the
generators $M_{AB}$ obey $(M^\Real_{AB})^\dagger=\iota
((M^\Real_{AB})^{\dagger_{\rm osc}})=M^\Real_{AB}$, and have the
oscillator realizations\footnote{ Raising and lowering of
two-component indices follow the convention
$y^\a=\epsilon^{\a\b}y_\b$ and $y_\a=y^\b\epsilon_{\b\a}$ where
$\e^{\a\b}\e_{\c\d}=2 \d^{\a\b}_{\c\d}$ \emph{idem} $\bar y_{\ad}$
and $\e_{\ad\bd}$. The van der Waerden symbols
$(\s^a)_{\a\ad}=(\bar \s^a)_{\ad\a}$ obey
  \bea
   (\s^{a})_{\a}{}^{\ad}(\sb^{b})_{\ad}{}^{\b}&=&
\y^{ab}\d_{\a}^{\b}\
  +\ (\s^{ab})_{\a}{}^{\b} \ ,\qquad
  (\sb^{a})_{\ad}{}^{\a}(\s^{b})_{\a}{}^{\bd}\ =\
\y^{ab}\d^{\bd}_{\ad}\
  +\ (\sb^{ab})_{\ad}{}^{\bd} \ ,\label{so4a}\w2
  \ft12 \e_{abcd}(\s^{cd})_{\a\b}&=& \left\{\ba{ll}
  (\s_{ab})_{\a\b}\ ,&\mbox{$(4,0)$ and $(2,2)$ signature\ ,}\\[5pt]i(\s_{ab})_{\a\b}
  \ ,&\mbox{$(3,1)$ signature\ ,}\ea\right.\label{so4b}\eea
and the reality conditions
\bea
\left(\e_{\a\b},(\s^a)_{\a\bd},(\s^{ab}_{\a\b})\right)^\dagger\ =\
\left\{
\begin{array}{ll}
\left(\e^{\a\b},-(\sb^a)^{\bd\a},(\s^{ab})^{\a\b}\right)
&\mbox{for $SU(2)$\ ,}
\\[5pt]
\left(\e_{\ad\bd},(\sb^a)_{\ad\b} ,
(\sb^{ab})_{\ad\bd}\right)&\mbox{for
$SL(2,\Comp)$\ ,} \\[5pt]
\left(\e_{\a\b},(\sb^a)_{\bd\a} ,(\s^{ab})_{\a\b}\right)
&\mbox{for $Sp(2)$\ .}
\end{array}
\right.\eea
One may verify the conventions using $\e=i\s^2$ and for $SU(2)$,
$SL(2;\Comp)$ and $SL(2;\Real)$, respectively, $[\s^a,\sb^a]$
given by $[(i,\s^i),(-i,\s^i)]$,
$[(-i\s^2,-i\s^i\s^2),(-i\s^2,i\s^2\s^i)]$ and
$[(1,\ts^i)),(-1,\ts^i)]$ where $\ts^i = (\s^1,i\s^2,\s^3)$.}
\bea M_{ab}&=& -\frac18 \left[~ (\s_{ab})^{\a\b}y_\a y_\b+
(\sb_{ab})^{\ad\bd}\tilde y_{\ad}\yb_{\bd}~\right]\ ,\qquad P_{a}\
=\ \frac{\sqrt{\l}}{4} (\s_a)^{\a\bd}y_\a \yb_{\bd}\ .\label{mab2}
\eea
The Fock-space and anti-Fock-space submodules in ${\cal A}$ are
exhibited by going from $(y_\a,\yb_{\ad})$ to a
$\mathfrak{u}(2)$-covariant basis $(a_i,\bar a^{i})$, $i=1,2$,
obeying
\bea [a_i,\bar a^{j}]_{\star} \ = \ \d^j_i\ ,\qquad
[a_i,a_j]_\star\ =\ [\bar a^{i},\bar a^j]_\star\ =\ 0\
,\label{acomm}\eea
after which the $\mso(5;\Comp)$ generators can be expressed as
\bea E & = & \frac{1}{2}(\bar a^{i}a_i+1)\ , \qquad M_{rs} \ =
\ \frac{i}{2}(\sigma_{rs})_i{}^j \bar a^{i}a_j \ , \\[5pt]
L^+_r & = & \frac{i}{2}(\sigma_r)_{ij}\bar a^{i}\bar a^{j} \ ,
\qquad L^-_r=\frac{i}{2}(\sigma_r)^{ij}a_{i}a_{j} \ .\eea
The Fock and anti-Fock spaces
\bea \cF^\pm&=&\bigoplus_{n=0}^\infty \Comp\otimes \ket{n}^\pm\
,\qquad \ket{n}^+\ =\ \bar a^{i_1}\cdots \bar a^{i_n}\ket{0}\
,\quad \ket{n}^-\ =\ a^{i_1}\cdots a^{i_n}\ket{0}^-\ ,\eea
can be identified as the $\mosp(4|1)$ supersingleton
\bea {\cal F}^\pm\ =\ \mD^\pm_0\oplus \mD^\pm_{1/2}\ ,\eea
where $\mD^\pm_0$ and $\mD^\pm_{1/2}$ consist of the even and odd
states, respectively, and $\ket{\ft12;(0)}=\ket{0}$ and
$\ket{1;(\ft12)}^i=\ \bar a^{i}\ket{0}$. The lowest-weight states
of the composite-massless scalars now read\footnote{The conformal
algebra $\mso(4,2)$ can be realized in ${\cal F}(1)\otimes {\cal
F}(2)$ as $M_{AB}=\ft18\sum_{\x=1,2} \bar Y(\xi)\C_{AB} Y(\xi)$,
that act in a tensorial split, and $R_a=\ft14 \bar Y(1)\C_A Y(2)$,
that act in a non-tensorial split, where $Y_\a(\x)=\bar Y_\a(\x)$
are Majorana spinors obeying $Y_\a(\x)\star Y_\b(\eta)=Y_\a(\x)
Y_\b(\eta)+i\d_{\x\y}C_{\a\b}$. The conformal generators can be
re-written as $M_{AB}=\ft14 \bar u \C_{AB} u$ and $R_A=-\ft{i}2
\bar u\C_A u$ where $u_\a=\ft1{\sqrt{2}}(Y_\a(1)+Y_\a(2))$ and
$\bar u_\a=\ft1{\sqrt{2}}(Y_\a(1)-Y_\a(2))$ are Weyl spinors
obeying $u_\a \star \bar u_\b=u_\a \bar u_\b+i C_{\a\b}$.}
\bea \ket{1;(0)}_{12}&=&\ket{\ft12;(0)}_1\ket{\ft12;(0)}_2\ =\ \ket{0}_1\ket{0}_2\ ,\label{oscd10}\\[5pt] \ket{2;(0)}&=&\ket{1;(\ft12)}^i_1\ket{1;(\ft12)}_{2;i} \ =\ -y\ket{1;(0)}_{12}\ ,\label{oscd20}\eea
where
\bea y&\equiv & \sqrt{x^+}\ =\ \sqrt{(L^+(1)+L^+(2))^2}\ =\ \bar
a_i(1) \bar a^{i}(2) \ .\eea
To make contact with \eq{4Dsinglunity} and \eq{spinrefl} we define
the reflection by $R(f\star g)=R(g)\star R(f)$ and
\bea R(\ket{0}^\pm) & = & {}^\pm\bra{0} \ , \qquad R(\bar a^{i})\
=\ ia^i \ , \quad R(a^{i}) \ = i\bar a^{i} \ .\label{su2refl}\eea
It follows that $R(\ket{n})=i^n \bra{0}a^{i_1}\cdots a^{i_n}$ and
$R(\ket{n}^-)=i^n {}^-\bra{0}\bar a^{i_1}\cdots \bar a^{i_n}$. We
also define
\bea \pi(\ket{0}^\pm)&=&\ket{0}^\mp\ ,\qquad \tau(\ket{0}^\pm)\ =\
{}^\mp\bra{0}\ ,\qquad R\ =\ \pi\circ\t\ .\eea
Using $z=2L^+_rL^-_r=:N^2:$ where $N=\bar a^{i}a_i$ and $:\bar a^i
a^j:=\bar a^i a^j$, we have $\1_{{\cal
F}}=\1_{\mD_0}+\1_{\mD_{\ft12}}$ with
\bea \1_{\mD_0}&=& \sum_{\mbox{n even}}\ket{n}\bra{n} \ = \
\sum_{\mbox{n even}}\frac{1}{n!}\bar a^{i_1}\ldots \bar
a^{i_n}\ket{0}\bra{0}a_{i_1}\ldots a_{i_n} \ = \ :\cosh \sqrt{z}
\ket{0}\bra{0}: \ , \label{cosh}\\[5pt]
\1_{\mD_{1/2}}&=& \sum_{\mbox{n odd}}\ket{n}\bra{n} \ = \
\sum_{\mbox{n odd}}\frac{1}{n!}\bar a^{i_1}\ldots \bar
a^{i_n}\ket{0}\bra{0}a_{i_1}\ldots a_{i_n}\ =\ :\sinh \sqrt{z}
\ket{0}\bra{0}: \ . \label{sinh}\eea
Thus $\1_{\cal F}=:e^{\sqrt{z}}\ket{0}\bra{0}:$ which implies
$\ket{0}\bra{0}=:e^{-N}:$, and hence
$\1_{\mD_0}=\frac{1}{2}(1+\C)$ and $\1_{\mD_{1/2}}=
\frac{1}{2}(1-\C)$ with $\C=:e^{-2N}:$ obeying $\C\star\C=1$.
Applying $(R_2)^{-1}$ to $\1_{\cal F}$ thus yields the
superreflector
\bea \ket{\1_{\cal F}}_{12} \ = \ e^{i y}\ket{0}_1\ket{0}_2 \ =\
\cos y \ket{1;(0)}_{12}-i{\sin y\over y}\ket{2;(0)}_{12}\
,\label{reflId} \eea
where we have used \eq{oscd10} and \eq{oscd20}. The two composite
trace operations on ${\cal F}^+$ read
\bea \Tr_\pm(f) &=&\Tr_{\mD_0}(f)\pm\Tr_{\mD_{1/2}}(f)\ ,\eea
where $\Tr_{\mD_0}(f) =\sum_{\tiny\mbox{$n$ even}}\bra{n}f\ket{n}$
and $\Tr_{\mD_{1/2}}(f) =\sum_{\tiny\mbox{$n$
odd}}\bra{n}f\ket{n}$. As shown in Appendix \ref{App:G}, the odd
composite trace coincides with the supertrace
\cite{Vasiliev:1986qx}, that is\footnote{The normalization can
also be derived by from $\Tr_{\mD_0}(\1)=\sum_{k=0}^\infty (2k+1)$
and $\Tr_{\mD_{1/2}}(\1)=\sum_{k=0}^\infty (2k+2)$ which imply
that $\Tr_{-}(\1)=\lim_{s\rightarrow
-1}\sum_{n=0}^\infty\,s^n\,(n+1) =\frac{1}{4}$.}
\bea \Tr_-(f)&=&\Tr_+((-1)^N_\star\star f)\ =\ \ft14 f(0,0)\ =\
\ft14 \Str(f)\ .\eea
Thus, the appearance of the spinor singleton in $D=4$ as an extra
solution to $V\approx 0$ leads to the identification \eq{D4strace}
of the non-composite trace $\Tr$ defined in \eq{Trprime} with the
supertrace.


\end{appendix}


\end{document}